\documentclass[12pt, preprint]{aastex}

\usepackage{wasysym} 
\usepackage{amssymb}
\usepackage{lscape}
\usepackage{color}

\begin{document}

\title{The \mbox{X-Ray} Variability of a Large, Serendipitous Sample of Spectroscopic Quasars}

\author{Robert R. Gibson\altaffilmark{1}, W. N. Brandt\altaffilmark{2,3}}
\email{rgibson@astro.washington.edu}

\altaffiltext{1}{Department of Astronomy, University of Washington, Box 351580, Seattle, WA 98195, USA}
\altaffiltext{2}{Department of Astronomy and Astrophysics, The Pennsylvania State University, 525 Davey Laboratory, University Park, PA 16802, USA}
\altaffiltext{3}{Institute for Gravitation and the Cosmos, The Pennsylvania State University, University Park, PA 16802, USA}

\shorttitle{Quasar \mbox{X-ray} Variability}
\shortauthors{Gibson et al.}

\hbadness=10000

\clearpage

\begin{abstract}We analyze the \mbox{X-ray} variability of 264 Sloan Digital Sky Survey spectroscopic quasars using the {\it Chandra} public archive.  This data set consists of quasars with spectroscopic redshifts out to $z \approx 5$ and covers rest-frame time scales up to $\Delta t_{sys} \approx$2000~d, with 3~or more \mbox{X-ray} observations available for 82~quasars.  It therefore samples longer time scales and higher luminosities than previous large-scale analyses of AGN variablity.  We find significant ($\gtrsim$3$\sigma$) variation in $\approx$30\% of the quasars overall; the fraction of sources with detected variability increases strongly with the number of available source counts up to $\approx$70\% for sources with $\ge$1000~counts per epoch.  Assuming the distribution of fractional variation is Gaussian, its standard deviation is $\approx$16\% on $\gtrsim$1~week time scales, which is not enough to explain the observed scatter in quasar \mbox{X-ray-to-optical} flux ratios as due to variability alone.  We find no evidence in our sample that quasars are more variable at higher redshifts ($z > 2$), as has been suggested in previous studies.  Quasar \mbox{X-ray} spectra vary similarly to some local Seyfert AGN in that they steepen as they brighten, with evidence for a constant, hard spectral component that is more prominent in fainter stages.  We identify one highly-variable Narrow Line Seyfert~1-type spectroscopic quasar in the {\it Chandra} Deep Field-North.  We constrain the rate of kilosecond-timescale flares in the quasar population using $\approx$8~months of total exposure and also constrain the distribution of variation amplitudes between exposures; extreme changes ($> 100$\%) are quite rare, while variation at the 25\% level occurs in $<$25\% of observations.  [\ion{O}{3}]~$\lambda$5007\AA\ emission may be stronger in sources that are identified as \mbox{X-ray} variable; if confirmed, this would represent an additional link between small-scale (corona) and large-scale (narrow line region) AGN properties.
\end{abstract}


\section{INTRODUCTION}\label{introSec}

The variation observed in spectra of active galactic nuclei (AGN) is governed by physical processes that we do not fully understand, but are integral to disk/corona emission and absorbing outflows.  The short time scales over which variation can be observed indicate that, in many cases, it is occurring in relatively small structures near to the supermassive black hole (SMBH).  AGN variability studies provide new temporal constraints for accretion and outflow models, particularly concerning size scales; black hole masses and accretion rates; ionization structure; emission mechanisms; and relations between various structures such as the disk, corona, jet, broad line region (BLR), and narrow line region (NLR).  Variability studies also attempt to map out how accretion and outflows may depend on luminosity, black hole mass, accretion rate, and redshift, in order to identify the fundamental factors that influence AGN structure, black hole growth, and galaxy evolution.

The AGN identified in optical surveys such as the Sloan Digital Sky Survey \citep[SDSS;][]{y+00} can be efficiently studied in other wavebands using publicly-archived data from observatories such as {\it Chandra}.  Because the SDSS spectroscopic quasar catalog is primarily optically-selected, the \mbox{X-ray} properties of these AGN have had little influence on the AGN selection process.  High-quality SDSS spectra also provide secure redshifts, allow us to distinguish highly-absorbed broad absorption line (BAL) AGN, and even support estimates of black hole masses.  Furthermore, the flux limits of the SDSS spectroscopic quasar catalog \citep{s+10} are well-matched to archived {\it Chandra} observations, with a very high \mbox{X-ray} detection rate for non-BAL AGN \citep[e.g.,][]{gbs08}.

In this study, we describe the \mbox{X-ray} variability properties of hundreds of SDSS spectroscopic quasars that have been observed multiple times by {\it Chandra}.  Most of the {\mbox X-ray} observations were serendipitous, in the sense that the quasar was not targeted by {\it Chandra}.  As a result, the sample is also relatively free of biases that could arise from selecting sources for targeting in the \mbox{X-rays}.  Although most sources were observed only two or three times, the large sample size permits us to quantify variability in ensembles out to redshift $z \approx 5$ and covering time scales of tens-of-minutes to years.

One of the primary goals of our analysis is to characterize quasar \mbox{X-ray} variability as a function of time scale, redshift, luminosity, and optical spectral properties.  We measure the extent of \mbox{X-ray} variation to determine whether the scatter in \mbox{X-ray-to-optical} flux ratios is dominated by variation or is largely intrinsic \citep[e.g.,][]{gbs08, vtta10}.  We also provide a new test of claims that variability increases at higher redshifts \citep{alsebggsg00, mal02, psgkg04}, and examine its dependence on luminosity and black hole mass.  Another goal of this study is to compare the variation of high-luminosity quasars to current models derived from intensive monitoring of local Seyfert AGN.  Using hardness ratios and model fits, we determine how spectral shape evolves as sources brighten and fade, and test simple models of spectral variation against the data.  Finally, we examine optical spectra for features that differ between ensembles of significantly-variable and non-variable sources.

Because the most detailed studies of variation in individual AGN have been conducted by monitoring local Seyfert AGN, we will be drawing from that body of research to guide our current study and to suggest directions for future analyses.  \mbox{X-ray} variation time scales in local Seyfert AGN are observed to depend on black hole mass and accretion rate \citep[e.g.,][]{onpt05, m10}.  The correlation between \mbox{X-ray} spectral steepness and brightness has been previously modeled in detail \citep[e.g.,][]{tum03, vf04, m10}.  The relationship between \mbox{X-ray} and optical variability is complex, with at least two different mechanisms modulating the emission.  On shorter time scales (days), optical emission can lag \mbox{X-ray} emission, suggesting that variability is partly caused by \mbox{X-rays} being reprocessed down to lower energies.  Variation on longer time scales (years) may be driven by changes in the accretion rate \citep[e.g.,][]{aukblm08, aulbmc09}.  The complexity of the temporal relationship between optical and \mbox{X-ray} emission may be due to differences in the geometry of the regions that emit at these wavelengths.  For example, an AGN with a larger black hole mass ($M_{BH}$) will generally have a cooler accretion disk, so that the region of the disk that emits in the optical may be relatively closer to, and subtend a greater solid angle of, a central \mbox{X-ray} emitting corona.  Such mass-dependent geometric effects could, for example, influence the effectiveness of reprocessing and Compton scattering \citep[e.g.,][]{m10}.  This example demonstrates one scenario in which variability studies could determine that high-luminosity quasars are not simply a ``scaled-up'' version of local Seyfert AGN.

In contrast to Seyfert AGN, much of our understanding of the temporal \mbox{X-ray} properties of distant, luminous quasars is derived from lower-sensitivity sky surveys and a limited number of resource-intensive targeted observations.  Previous \mbox{X-ray} variability analyses using {\it ROSAT} observations of radio-quiet quasars by \citet[][hereafter A00]{alsebggsg00} and \citet[][hereafter M02]{mal02}, as well as an analysis of {\it Chandra} Deep Field--South \citep[CDFS;][]{grtnhnbbggz01, lbbalsbcffghhkmprssssv08, xlbblbsabcfghhklmprrssstv11} AGN by \citet[][hereafter Pao04]{psgkg04}, have found an intriguing tendency for AGN at higher redshifts ($z > 2$) to have larger \mbox{X-ray} variability amplitudes than would be expected from an extrapolation of the properties of lower-redshift AGN.  A related study using {\it XMM-Newton} observations of the Lockman Hole region by \citep[][hereafter Pap08]{pcamg08} found that variability decreased with increasing redshift in the sample overall, but for a given luminosity range of AGNs, variability increased out to $z \sim 1$, then remained constant at higher redshifts.  Pao04 also estimated that a large fraction ($> 90$\%) of AGNs likely exhibit \mbox{X-ray} variability, and sources for which spectral variability could be measured exhibited a tendency to soften spectrally as they brighten.

Our current sample allows us to expand on these previous studies in several ways.  It is constructed using high-quality optical spectra that can be used to unambiguously identify quasars and determine their redshifts.  By contrast, AGN were estimated to account for only $\sim$80\% of the CDFS sample of Pao04.  Photometric redshifts were used for the CDFS sources, and some of the less-luminous ($L_X < 10^{42}$~erg~s$^{-1}$) sources may have been contaminated by emission from their host galaxies.  Drawing on the large area of sky covered by the SDSS survey, our sample includes a large number of highly-luminous quasars, and extends to luminosity levels $\sim$10~times higher than even the A00 and M00 {\it ROSAT} samples.  (See \S\ref{chandraRedSec} for further description of our sample properties.)  Compared to earlier {\it ROSAT} studies, {\it Chandra}'s sensitivity to hard \mbox{X-rays} permits us to measure hardness ratios and spectral shapes at energies $>$2~keV, where absorbing material (if present) has a weaker effect on spectral shape.  {\it Chandra's} spatial resolution also resolves away background contaminants.  The data in our serendipitous sample cover long time scales, with {\it rest-frame} times between epochs up to 5.4~yr.  By contrast, the Deep {\it ROSAT} Survey \citep{sgspbg91} used for A00 and much of the M02 sample spans about two weeks in the {\it observed} frame, while the Lockman Hole observations of Pap08 covered under two months and the CDFS observations used by Pao04 were collected over about 15~months.  Our sample also has a larger number of sources at higher redshift ($z > 2$), although larger samples are still needed in this regime.  For these reasons, we especially focus on the dependence of variability on redshift, luminosity, and time scale, and also examine how spectral properties are related to \mbox{X-ray} variation.

Although our current analysis is focused on \mbox{X-ray} variation, we note that the temporal emphasis of upcoming deep-wide surveys such as Pan-STARRS \citep{k+02} and LSST \citep{i+08} will greatly enhance variability studies by selecting large new samples of bright AGN and also by extending our understanding of the temporal properties of AGN in optical wavebands.  Optical and \mbox{X-ray} views are complementary because the processes that generate optical and \mbox{X-ray} emission are strongly related \citep[e.g.,][]{zhmtsalsswsc81, sbsvv05, ssbaklsv06, gbs08}.  Growing \mbox{X-ray} and optical data sets will support increasingly sophisticated AGN research that incorporates time-domain information across the spectrum.  The SDSS has measured two epochs of photometry for large numbers of spectroscopic quasars, permitting the construction of ensemble structure functions that describe typical amplitudes of variation as a function of time scale \citep[e.g.,][]{vwkabhirsyblns04}.  While some Seyfert AGN soften in \mbox{X-rays} as they brighten \citep[e.g.,][]{mev03}, the {\it optical} continua of SDSS quasars become bluer as they brighten \citep[e.g.,][]{wvkspbrb05}.  As for \mbox{X-ray} studies, the amplitude of fractional variability decreases with increasing luminosity, and a positive correlation of variability amplitude with redshift has also been observed \citep[e.g.,][]{csv00, vwkabhirsyblns04}.  The optical structure function is well-represented as a power law for time scales of $\lesssim 1$~yr, but appears to flatten at longer (multi-year) time scales \citep{iljahrrvtkgsss04}.  Some models invoking a combination of random, discrete emission events are disfavored because they do not reproduce optical structure functions \citep[e.g.,][]{kmut98, vwkabhirsyblns04}.  Damped random walk (DRW) models fit to individual AGN light curves generally indicate a damping time scale of $\sim$100~days or more \citep{kbs09, kkuwsskpsup10, mikkkbkswbgbd10}, corresponding to structure function flattening at longer time scales.

In the following sections, we describe the sample selection and data reduction process (\S\ref{obsRedSec}), explain the procedures we use to formally detect and characterize \mbox{X-ray} variability (\S\ref{analysisSec}), discuss some physical implications of our results (\S\ref{discussionSec}), and provide a concluding summary (\S\ref{concSec}).  Throughout, we use a cosmology in which $H_0 = 70$~km~s$^{-1}$~Mpc$^{-1}$, $\Omega_M = 0.3,$ and $\Omega_{\Lambda} = 0.7$.

\section{OBSERVATIONS AND DATA REDUCTION}\label{obsRedSec}

\subsection{The {\it Chandra} Sample\label{chandraSampleSec}}

In order to obtain a list of quasars observed by {\it Chandra}, we searched the {\it Chandra} archive to find all ACIS-S or ACIS-I observations of SDSS Data Release 5 (DR5) quasars \citep{s+07} that used no gratings and were public as of 13~Jan 2009.  We use this sample to design and calibrate our complex data-analysis pipeline for an initial study.  In a follow-up study, we will expand the sample to include recently-observed quasars and additional AGN selected using new metrics such as optical/UV variability.  For each spectroscopic quasar, we identified candidate {\it Chandra} observations that had telescope aimpoints within 15~arcmin of the given quasar.  Then, we investigated these candidate observations to determine whether a quasar fell on an active detector chip.  In cases where a source fell within 32~pixels of a chip boundary, we discarded the candidate observation because of increased uncertainty of the instrument response in those regions.  Where necessary, we corrected aspect offsets using the prescription available online.\footnote{http://cxc.harvard.edu/cal/ASPECT/fix\_offset/fix\_offset.cgi}  Our final sample of multiply-observed sources includes 763 {\it Chandra} observations of 264 SDSS quasars.

We reduced each observation of an SDSS quasar using CIAO~4.1.2\footnote{http://cxc.harvard.edu/ciao/} following the procedures listed online for the tool {\tt psextract}.\footnote{http://cxc.harvard.edu/ciao/threads/psextract/}  This tool does not handle cases where zero counts are present in the source extraction region.  We flagged these cases for special handling.  For each quasar observation, we generated instrument response files (ARFs and RMFs) either directly (using the {\tt mkarf} and {\tt mkrmf} tools for zero-count sources) or indirectly (through {\tt psextract}).  These response files include a correction for the buildup of contaminant on the ACIS chips.  The instrument response accounts for spatial and temporal variation as a function of detector position and time.

For point sources brighter than $r=20$ mag, SDSS astrometry is accurate to $\sim$50 milli-arcseconds\footnote{http://www.sdss.org/dr7/products/general/astrometry.html}, or about 10\% of a {\it Chandra} pixel.  We can therefore use SDSS astrometry to localize sources on {\it Chandra} CCDs.  We performed ``forced photometry'' on the \mbox{X-ray} images, extracting source counts from a circular region with radius equal to the 90\% encircled energy fraction at 1.5~keV for a given off-axis angle.  This was done even in cases where the source is not detected in an image.  Extraction radii were computed using data tabulated on the {\it Chandra} \mbox{X-ray} Center web site.\footnote{http://cxc.harvard.edu/cal/Hrma/psf/ECF/hrmaD1996-12-20hrci\_ecf\_N0002.fits}

\subsection{Background Estimation\label{bGEstSec}}

Backgrounds were generally extracted from an annular region centered on the source position.  We selected the inner and outer radii of the annulus to be $15+r_s$ and $45+r_s$ pixels, respectively, where $r_s$ is the source extraction radius.  In our experience, this prescription provides sufficient area to estimate backgrounds reliably at small off-axis angles without becoming large enough to be contaminated by numerous sources at large off-axis angles.  In cases where background regions fell partly off-chip, we visually selected a new, circular region that was near to the source and appeared not to be contaminated by other sources.
In addition to visual inspection for source contamination, we also checked each background programmatically, searching for sources detected by the {\it Chandra} tool {\tt wavdetect} that fell inside or near our background regions.  We generated a list of {\tt wavdetect} sources using a conservative {\tt sigthresh} threshhold of $10^{-5}$, which roughly corresponds to 10~false detections per chip.  In cases where sources were found to contaminate our background regions, we selected a nearby background region that was free of contamination.

\subsection{Sample Properties\label{samplePropertiesSec}}

\begin{figure} [ht]
  \begin{center}
      \includegraphics[width=5in, angle=-90]{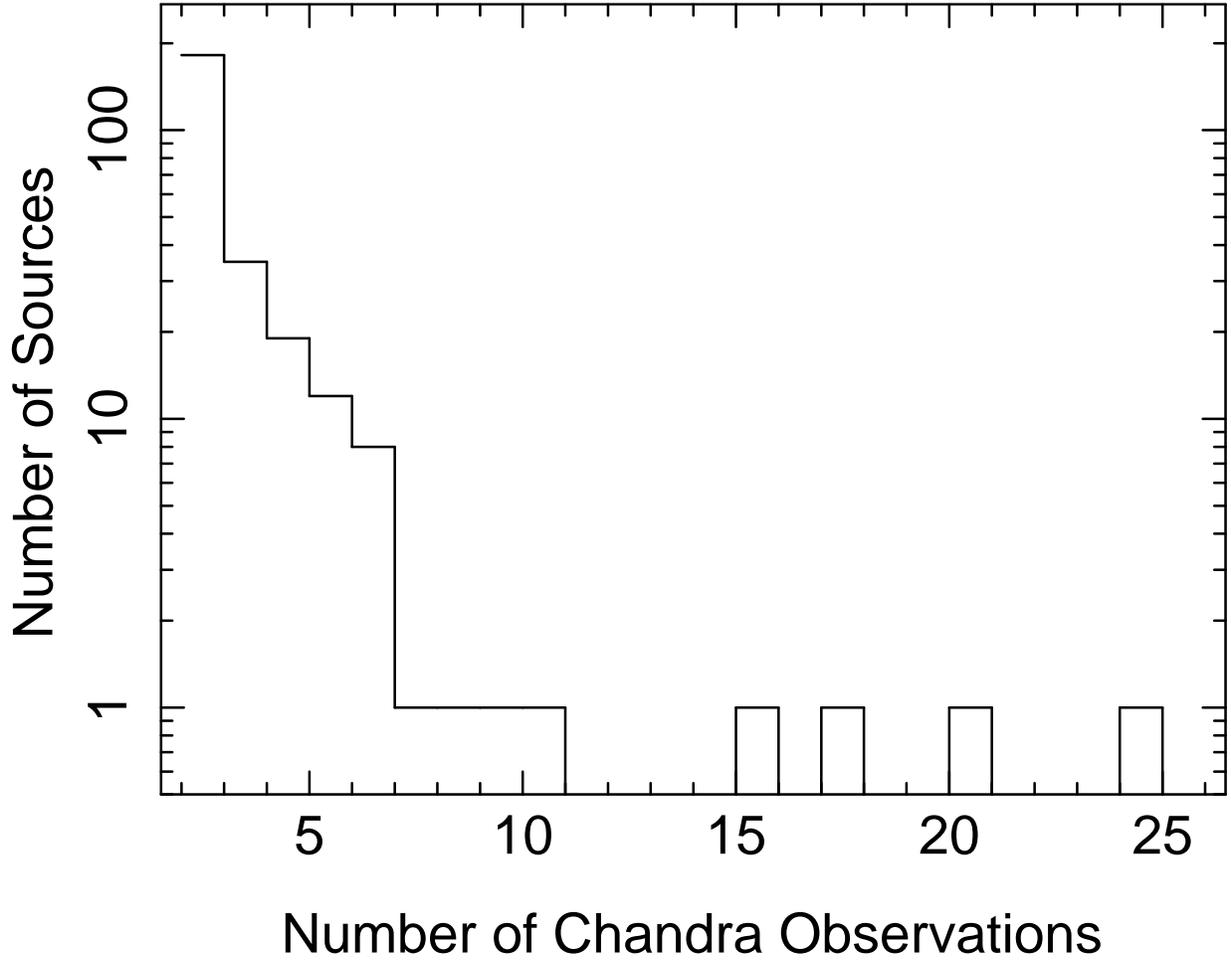}
      \caption{\label{numObsHistFig}Histogram showing the distribution of the number of {\it Chandra} observations per source, for our sample.  The $y-$axis is logarithmic to clearly illustrate the number of sources with more than 2~observations.}
   \end{center}
\end{figure}

\begin{figure} [ht]
  \begin{center}
      \includegraphics[width=5in, angle=-90]{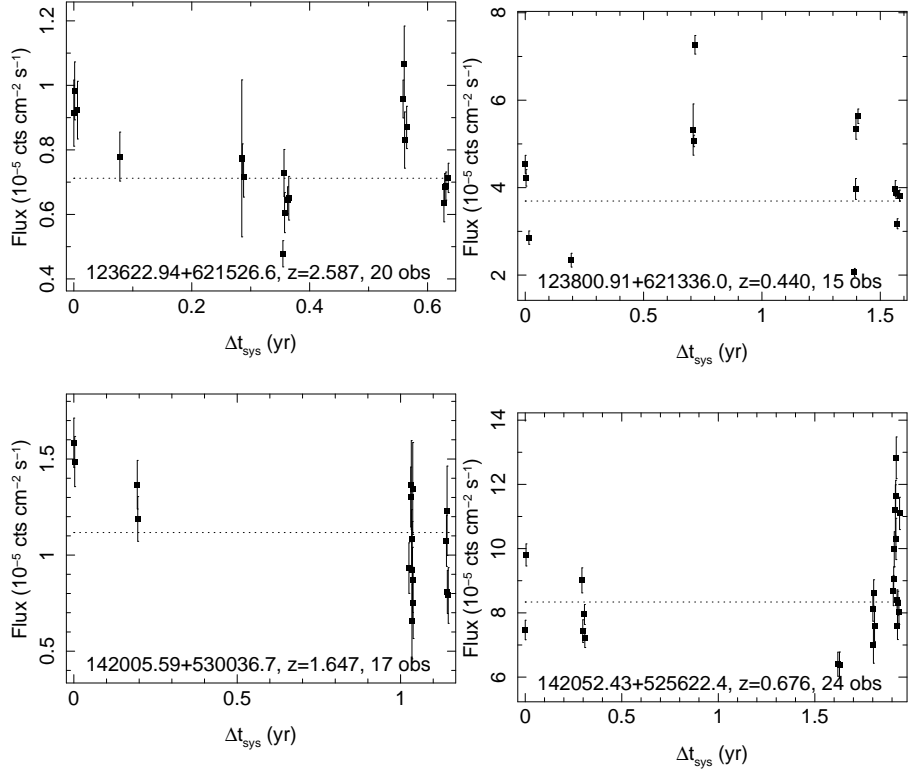}
      \caption{\label{highEpochsLCsFig}Light curves for sources having 15 or more observations in our sample.}
   \end{center}
\end{figure}

Figure~\ref{numObsHistFig} shows the distribution of the number of times a source was observed by {\it Chandra} in our sample.  Of 264~sources, 182~were observed 2~times, while the remaining 82~were observed 3~or more times.  Four sources were observed 15 or more times.  Figure~\ref{highEpochsLCsFig} shows example light curves for these sources.

\begin{figure} [ht]
  \begin{center}
      \includegraphics[width=5in, angle=-90]{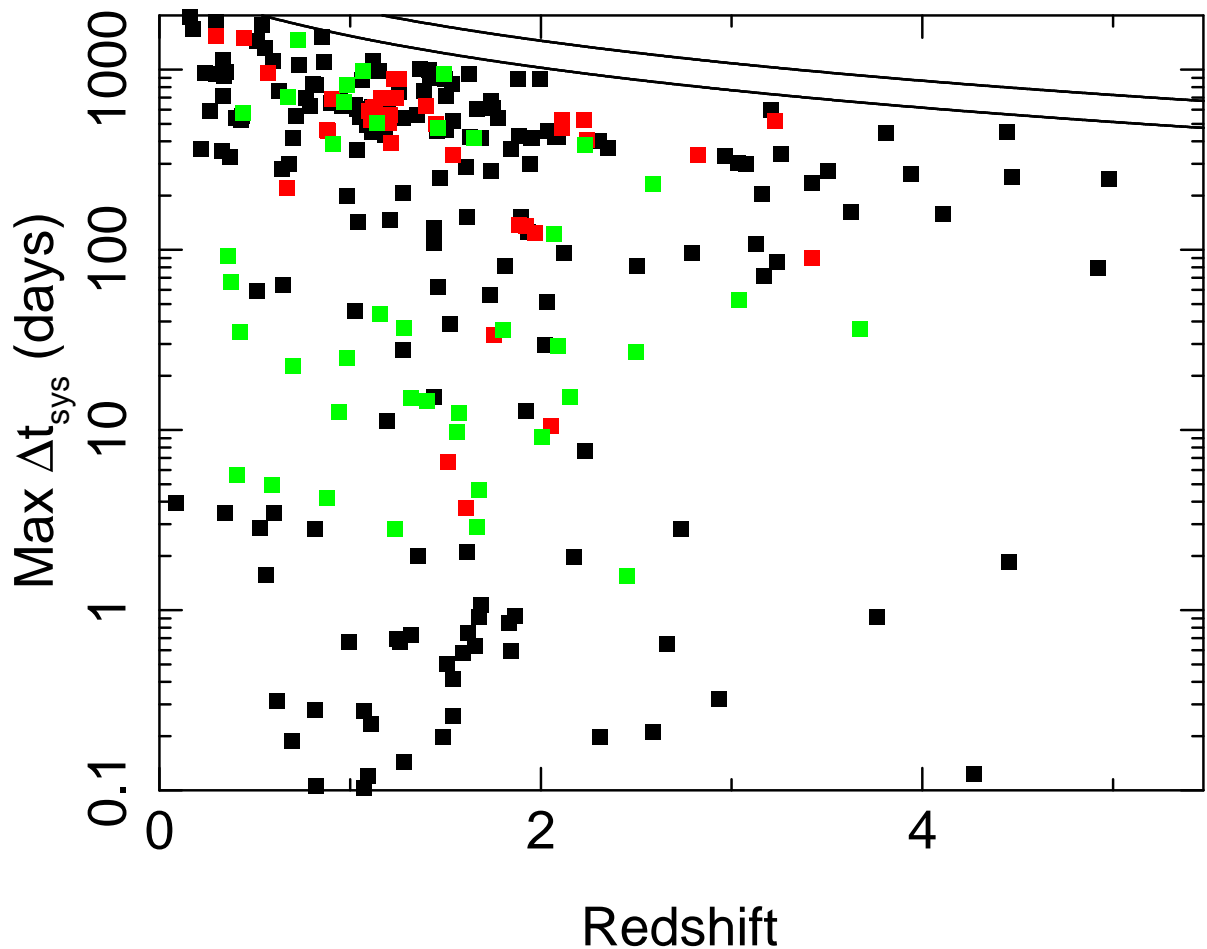}
      \caption{\label{dtSysVszFig}Redshift and maximum rest-frame time between observations (across all observing epochs) for our full sample of {\it Chandra} sources.  Sources observed in 2~epochs are shown as black squares.  Those with 3~epochs are plotted in red, and those with 4 or more epochs are plotted in green.  The solid line at the top represents the rest-frame time from the start of {\it Chandra} observations (14~Aug 1999 to 1~Jul 2011), which is an upper limit on time scales available in the {\it Chandra} archive.  The second solid line, slightly lower, represents cut-off time for an observation to be publicly available for our study.}
   \end{center}
\end{figure}

Some analyses depend on the time-domain and redshift coverage of the archive data.  To illustrate this coverage, Figure~\ref{dtSysVszFig} shows the maximal rest-frame time span (max $\Delta t_{sys}$) between observing epochs for each source in our sample.  Time spans between archived observations range from hours to $\gtrsim$5~yr for lower-redshift sources.  For sources with $z>2$, time spans of up to $\approx$1~yr are covered.  Of course, we can also examine time scales shorter than the maximum value of $\Delta t_{sys}$ for sources observed more than two times.

In an earlier study, \citet{gbsg08} found that {\it Chandra} detects nearly all ($\approx$100\%) non-BAL SDSS spectroscopic quasars in observations having exposure times $>$2.5~ks or off-axis angles $<$10~arcmin.  [The {\it Chandra} detection rate of SDSS spectroscopic quasars is lower, but still very high, for sources with shorter exposures and/or larger off-axis angles up to 15~arcmin \citep{gbsg08}.]  Motivated by these criteria, we define a high-quality sample, which we call ``Sample HQ'', consisting of 167~sources.  To construct Sample HQ, we culled all observations of any source that had exposures $<$2.5~ks and off-axis angles $>$10~arcmin.  We also required that sample-HQ observations be performed at an off-axis angle $>$1~arcmin, in order to eliminate bias that could be introduced by the target-selection criteria of the {\it Chandra} observatory.  A typical source has about 5~more counts per epoch (64 vs. 59 counts, on average) in sample HQ than in the full sample.

\begin{figure} [ht]
  \begin{center}
      \includegraphics[width=5in, angle=-90]{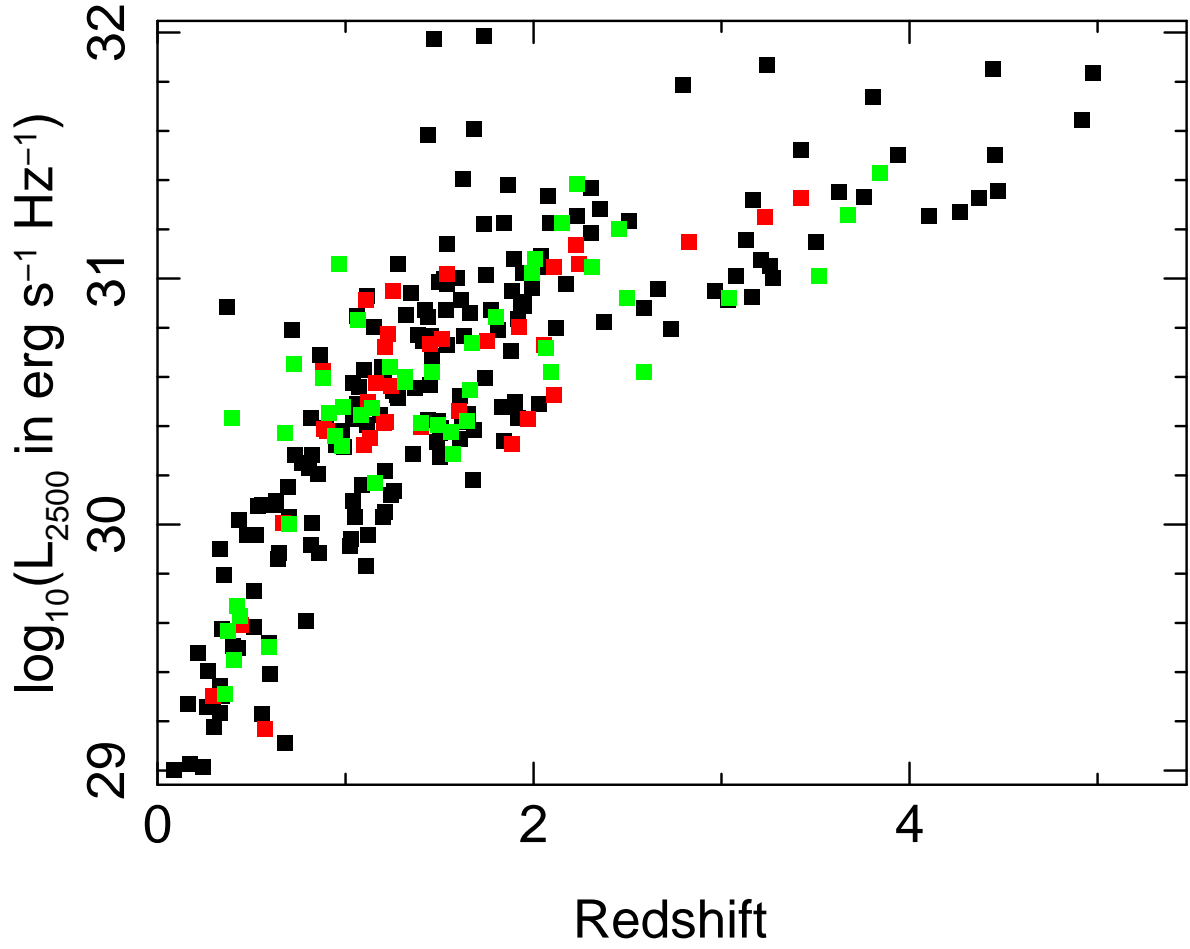}
      \caption{\label{l2500VszFig}$L_{2500}$ as a function of redshift for our full sample of {\it Chandra} sources.  The symbols are the same as in Figure~\ref{dtSysVszFig}.}
   \end{center}
\end{figure}

For each quasar, we calculate (or estimate) the monochromatic luminosity $L_{\nu}({\rm 2500~\mathring{A}})$ in units of ergs~s$^{-1}$~Hz$^{-1}$, which we denote $L_{2500}$.  This value is either calculated directly from our fits to the quasar continuum at (rest-frame) 2500~\AA, or is extrapolated by normalizing the composite quasar spectrum of \citet{v+01} to match the observed spectrum in the SDSS bandpass.  In the former case, continuum fits were performed using the method of \citet{gjbhswasvgfy09}, in which a reddened power law or a polynomial was fit to each spectrum, excluding regions with broad emission lines, BALs, or ionized iron emission.  A plot of $L_{2500}$ as a function of redshift is shown in Figure~\ref{l2500VszFig} for our full sample.  The bulk of the sample spans a factor of about 40 in luminosity, from $10^{29.8} < L_{2500} < 10^{31.4}$~erg~s$^{-1}$~Hz$^{-1}$.  Because the SDSS survey is flux-limited, $L_{2500}$ is strongly correlated with redshift.  Throughout this work, we use quasar redshifts reported in the SDSS quasar catalog.  These software-generated redshifts were verified by visual inspection.  Repeat observations of SDSS quasars have shown an rms difference in redshift of 0.006 \citep{wvkspbrb05}; this level of precision is sufficient for our study.

We also make use of central black hole masses ($M_{BH}$) determined from broad emission line fits to H$_{\beta}$, \ion{Mg}{2}, and \ion{C}{4} emission lines for SDSS spectroscopic quasars \citep{sgsrs08}.  We adopt these values with the understanding that accurate determination of $M_{BH}$ is an area of active research, with known discrepancies among existing methods.  \citet{sgsrs08} note that measurement errors for the quasar, continua, and emission lines are generally dominated by statistical uncertainties ($\gtrsim$0.3--0.4~dex) in the calibration of $M_{BH}$-estimation methods as well as unknown systematic effects in the use of, e.g., \ion{C}{4} emission as an estimator.  Figure~\ref{varArchPlotVsSeyfertsFig} shows estimated black hole masses and redshifts for quasars in our sample.  For comparison, we have also plotted black hole masses for Seyfert AGN and quasars determined through reverberation mapping and other methods \citep{bz03, bcrd04, pfgkmmmnopvw04, onpt05}.  (We note that in some cases, these masses may be controversial; this plot is simply intended to be illustrative.)  Our SDSS quasar sample extends well beyond the Seyfert~AGN regime to higher black hole masses ($M_{BH} > 10^9 M_{\astrosun}$) and, of course, higher redshifts.

We identify subsamples of quasars that are known to be radio-loud, following the method of \citet{gbsg08}.  Radio fluxes from FIRST \citep{bwh95} or NVSS \citep{ccgyptb98} were used to determine the ratio of flux densities at 5~GHz and 2500~\AA.  Sources having a ratio of $R^* >$10 ($\log(R^*) > 1$) were flagged as radio-loud.  We classify 25~quasars as radio-loud by this criterion.  For 97~quasars, we are not able to constrain $\log(R^*) < 1$ given the approximate 1~mJy (2.5~mJy) limit of the FIRST (NVSS) surveys.  Almost all (93) of these quasars are {\it at most} radio-intermediate (RIQs; $1 < \log(R^*) < 2$), which generally have \mbox{X-ray} properties similar to radio-quiet quasars \citep[e.g.,][]{mbsgsw11}, at least in single-epoch analyses.  Based on the fraction of radio-loud quasars in the subset of sources that we can classify unambiguously (i.e., radio sensitivity extends to $\log(R^*) < 1$), we roughly anticipate $\approx$7 RIQ contaminants in the radio-quiet quasar set.

We used the SDSS DR5 catalog of Broad Absorption Line quasars \citep{gjbhswasvgfy09} to identify any sources known to host absorbing BAL outflows along the line of sight.  BAL quasars typically have strong \mbox{X-ray} absorption, so we treat them separately.  We expect some contamination from unidentified BAL quasars at lower redshift ($z \lesssim 1.7$) because the \ion{C}{4} BAL-absorption region for these sources does not lie in the SDSS spectral bandpass; BALs are identified in $\sim$15\% of higher redshift AGN \citep[e.g.,][]{hf03, thrrsvkafbkn06, gjbhswasvgfy09}.  In the full sample of 264 sources, we unambiguously identify 18~cases of BAL quasars.

\subsection{Count Rates\label{chandraRedSec}}

For each ACIS chip that observed an SDSS quasar, we construct a weighted exposure map using the prescription given at the {\it Chandra} \mbox{X-ray} Center web site.\footnote{http://cxc.harvard.edu/ciao/threads/spectral\_weights/index.html}  This exposure map can be used to estimate the source count flux from the count rate observed in a given epoch.  We used a Galactic-absorbed power law for the exposure map weights.  In order to roughly estimate the shape of the ACIS spectrum obtained for each quasar observation, we fit each background-subtracted spectrum with a power law.  This model was fit to counts in the observed frame 0.5--8~keV energy band using the Cash statistic \citep{c79}.  Our goal was not to model features in each spectrum, but simply to describe the overall spectral shape in terms of a photon index ($\Gamma$) that would be used to construct the weighted exposure maps.  We constructed exposure maps for photon indices of 1.6, 1.8, 2.0, and 2.2, and selected the exposure map corresponding to the value of $\Gamma$ that most closely matched our fit value for each observation.

We extracted total counts from a circular aperture and estimated background counts from an annular (or offset circular) aperture as described in \S\ref{chandraSampleSec} and \S\ref{bGEstSec}.  Both total and background counts were multiplied by a factor of $1/c_A$ as an ``aperture correction.''  Because the fraction of encircled counts depends on photon energy, we used a different correction factor for our soft ($0.5-2.0$~keV), hard ($2.0-8.0$~keV), and full ($0.5-8.0$~keV) bands.  We calculate the correction factor for each observation by determining the median energy of a count in the appropriate band.  The correction factor $c_A$ does not differ greatly from the aperture correction $c_A = 0.90$ corresponding to the 90\% encircled fraction for 1.5~keV photons for our extraction radii.  For the soft, hard, and full bands, typical values of $c_A$ are $0.92$, $0.87$, and $0.91$, respectively.

Source counts were estimated by subtracting the background from the total number of counts.  We estimated the $1\sigma$ upper and lower limits on the number of source counts by applying Equations~7 and~11 of \citet{g86} to the numbers of total and background counts, then propagated the errors to determine errors in source count rates.  We applied a small correction factor to account for the influence of Galactic absorption in each band; the correction factor was determined using the Galactic column with our power law model fits to each spectrum.  We then divided the number of counts by the average value of the weighted exposure map in the aperture to obtain count rates (in counts~cm$^{-2}$~s$^{-1}$) in the soft, hard, and full bands for each observation of an SDSS quasar.  Properties for individual sources are given in Table~\ref{sourceTable}, while properties measured for individual epochs are given in Table~\ref{epochTable0} and Table~\ref{epochTable1}.\footnote{In these tables, we list some measurements, such as count rates and their errors up to three digits after the decimal.  Of course, the measurements should not be considered significant to this number of digits.  This approach is adopted to avoid accumulating round-off errors and assist machine interpretation of the data.}  The count rates used in this work are measured in the observed frame unless otherwise specified.  Our statistics are often dimensionless, so if we apply a factor of $(1+z)$ to the time dimension when we calculate count rates or derived quantities such as count rate errors, this factor would cancel out of the final statistic.  See, e.g., equations \ref{fracVarDefEqn}, \ref{excessVarianceEqn}, and \ref{sijDefEqn}.  For many of our analyses, we do not classify sources as ``detected'' or ``not-detected'' (at some confidence level), but instead work directly with the background-subtracted count rates.  For faint sources, these rates can be zero or even negative.  In any case, there is a high detection rate in our sample.  For example, full-band detections were obtained at $>$95\% confidence for 690 of 763 observations in the full sample and 490 of 507 observations meeting Sample HQ requirements.  Furthermore, for some analyses, we restrict the sample to sources with higher numbers of counts in order to maximize signal-to-noise.

We calculate ``count-rate luminosities'' from 0.5--8 keV count rates as:
\begin{eqnarray}
L_i &\equiv& P_i (1+z)^x 4\pi D^2_L\label{photonLumEqn},
\end{eqnarray}
where $P_i$ is the flux in counts~cm$^{-2}$~s$^{-1}$ (calculated by dividing net source counts by the exposure map), $D_L$ is the luminosity distance, and the exponent $x$ incorporates both a bandpass correction and a $K$-correction to account for the fact that the {\it Chandra} bandpass is sampling different spectral regions as a function of redshift.  We have assumed a power law with photon index of $\Gamma = 1.8$ for this correction, so that $x = 0.8$ \citep{sg86}.  We use a fiducial photon index of $1.8$ instead of our best-fit photon indices because we do not want to amplify any effects that are purely due to measurement error or modeling (e.g., absorption features that change dramatically between observations).  We do not convert count-rate luminosities into traditional luminosities (in units of erg~s$^{-1}$) because doing so would require assumptions that decrease the accuracy of our results.  Calculating the typical photon energy used as a conversion factor in a given bandpass involves making an assumption about the spectral shape or performing a model fit; both of these can introduce additional errors, especially for atypical and/or faint sources.

Figure~\ref{sampleVsOldFig} compares the luminosity and redshift distribution of our sample to that of A00, Pao04, and Pap08.  (Data for this figure were kindly provided by O.~Almaini and M.~Paolillo; the data for the M02 sample were not available.)  For illustrative purposes, we estimate $L_{0.5-8}$, the luminosity in the 0.5--8.0~keV band, from the count-rate luminosities in our sample assuming that the energy of each photon is $\approx$1~keV.  The Pao04 sample drawn from the CDFS and the Pap08 Lockman Hole sample cover lower luminosities (and shorter time scales), while our sample approximately covers the A00 sample and extends it to luminosities up to $\sim$10 times higher.  Our sample includes more quasars at higher redshifts (with 63~quasars at $z > 2$), as well.  Figure~\ref{sampledtVsOldFig} shows the maximum rest-frame time scales probed by each study, estimated for sources in A00, Pao04, Pap08 by dividing the maximum observed-frame time scale by $(1+z)$.  Using the {\it Chandra} archive, we can probe time scales up to 10~times longer than in the lower-luminosity sample of Pao04, and $\sim$100 times longer than in A00 and Pap08.  Figure~\ref{sampledtVsOldFig} shows (as a gray box) power spectrum break time scales estimated for some typical sources in our sample using the relation given in \citet{mkkuf06}.  The \mbox{X-ray} archives enable us to explore the regime beyond this break time scale.

\section{ANALYSIS}\label{analysisSec}

\subsection{Identifying Variation\label{identifyVarSec}}

For each quasar, we performed a one-parameter fit to the photon fluxes using our estimated errors to determine the best constant flux that would describe our data.  As expected, we find that the observed distribution of $\chi^2_{\nu}$ is skewed to higher values than would be expected if the constant model were a good fit to the data.  Of course, several factors affect the distribution of $\chi^2_{\nu}$ in our analysis, including non-normality of the count rate distribution and the estimation of source count rate errors.

For reasons such as these, we adopt a different method to determine whether sources can be robustly classified as ``variable.''  For each epoch of a given source, we calculate the expected total (source $+$ background) count rate in our extraction aperture from the measured background and best-fit constant flux.  We flag an epoch as ``variable'' if the observed count rate is higher or lower than the number of counts corresponding to a deviation from the best-fit value at $>$99\% confidence, according to a Poisson statistic.  Any source with at least one variable epoch is considered a ``variable source.''  For the full sample, 74 of 264 sources are classified as variable; 54 of 167 sources are variable in Sample HQ.  We classify the remaining sources as ``non-variable,'' with the caution that this term should be understood to mean that we did not detect variability {\it within the limits of our data}.

\subsection{Sensitivity to Variation\label{sensitivityToVarSec}}

\begin{figure} [ht]
  \begin{center}
      \includegraphics[width=5in, angle=-90]{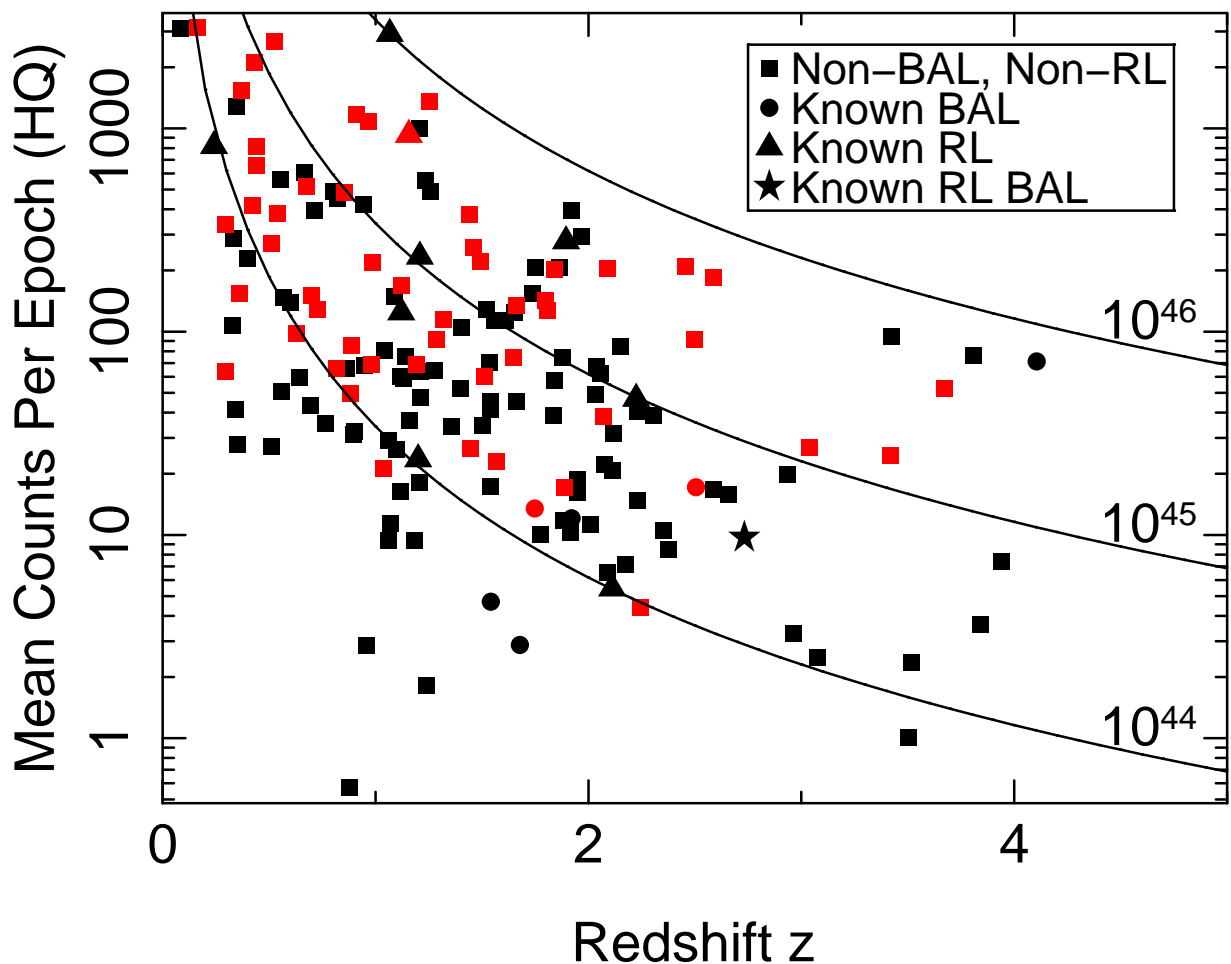}
      \caption{\label{meanCtsHQVsZFig}The mean number of $0.5-8$~keV counts for sources in Sample HQ as a function of redshift.  Sources that have been flagged as variable are plotted using red points; non-variable sources are plotted as black points.  Known BAL quasars are plotted as circles, known radio-loud quasars are plotted as triangles, and radio-loud BAL quasars are plotted as stars.  All other sources are plotted as squares.  The solid black lines show typical numbers of counts from a hypothetical quasar having an unabsorbed power-law spectrum with $\Gamma=2$ in 18~ks of on-axis exposure on an ACIS-I chip, for 0.5--8~keV luminosities $L_X = 10^{44}$, $10^{45}$, and $10^{46}$~erg~s$^{-1}$.}
   \end{center}
\end{figure}

Variability tests are more sensitive for sources with larger numbers of counts.  In order to examine this dependency in our data, we plot in Figure~\ref{meanCtsHQVsZFig} the mean number of source counts as a function of redshift.  The ``mean counts'' are calculated for each source using only epochs from Sample HQ that have high exposures and low off-axis angles.  Of course, the mean number of counts depends on variability characteristics, so this figure should be considered to be only a rough exploration of our data set.  Sources flagged as variable are plotted in red, while non-variable sources are plotted with black points.  Sources known to host BALs are plotted as circles, while known radio-loud sources are triangles and known radio-loud BAL quasars are stars.  The remaining sources, including those for which radio-intermediate status and BAL absorption could not be ruled out (\S\ref{samplePropertiesSec}), are plotted as squares.  As a rough guide, we have also plotted solid curves indicating typical numbers of counts for quasars with 0.5--8~keV luminosities of $10^{44}$, $10^{45}$, and $10^{46}$~erg~s$^{-1}$.  The curves were constructed assuming a hypothetical source with an unabsorbed, $\Gamma=2$ power-law spectrum observed on-axis on an ACIS-I chip for 18~ks (which is a typical, median exposure time in our sample).  Individual sources may differ from this hypothetical source in various respects.

Variability is primarily detected in quasars having $\gtrsim$50 counts per observation (on average), with about half of those sources flagged as variable.  Our sensitivity to variability is diminished at higher redshifts ($z \gtrsim 2$), where the majority of sources have $\lesssim$50 counts.  On the other hand, we detect variability in 9 of 14 sources with $\ge$700 average counts and 7 of 10 sources with $\ge$1000~average counts at redshifts $z \lesssim 1.2$.

\subsection{Fractional Variation\label{fracVarSec}}

\begin{figure} [ht]
  \begin{center}
      \includegraphics[width=5in, angle=-90]{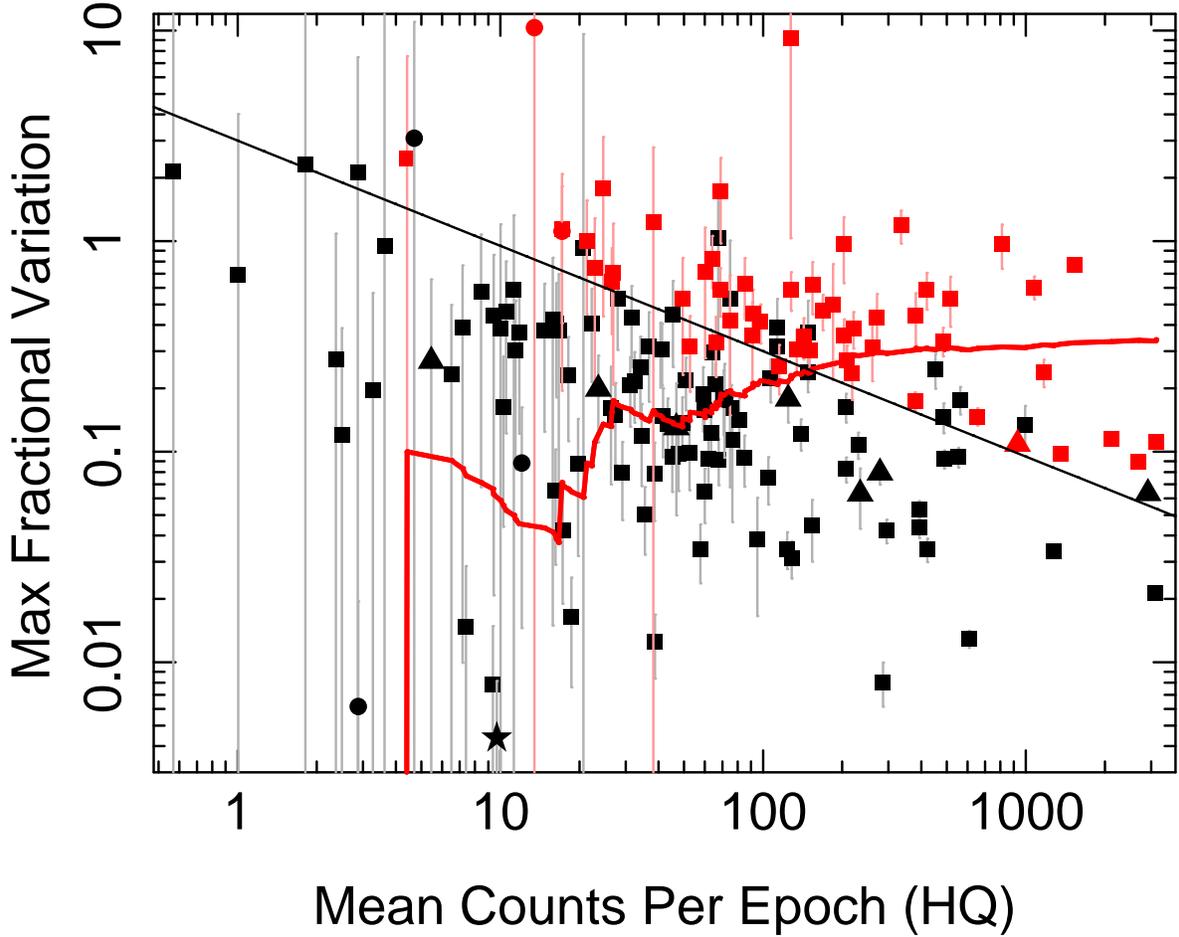}
      \caption{\label{fracVarAndDetectVsMeanCtsFig}The solid points show variability levels as a function of the number of counts for quasars in Sample~HQ, using the same symbols as in Figure~\ref{meanCtsHQVsZFig}.  Sources for which variability was detected (not detected) are red (black).  The $y$-axis is the maximum value of $|r/r_0 - 1|$ over all epochs for each source, where $r$ is the observed flux and $r_0$ is the best-fit flux.  The solid black line roughly indicates the $3\sigma$ variablity level, which approximates our variability-detection criterion.  The thick red line indicates the fraction of sources with mean counts $\le x$ that are identified as variable.}
   \end{center}
\end{figure}

In Figure~\ref{fracVarAndDetectVsMeanCtsFig}, we show the maximum variation from the best-fit constant count rate for the sources in Sample~HQ as a function of the mean number of counts per epoch.  The $y$-axis for the plotted points is the maximum of $|r/r_0 - 1|$ for all HQ observations of a source, where $r$ is the measured flux and $r_0$ is the best-fit constant flux for that source.  The solid black line shows the ``$3\sigma$ limit'' relation ($y = 3\sqrt{x}/x$ for a mean number of counts $x$), which is a rough approximation of our variability-selection criterion.  The thick, red line indicates the relation $y = f(x)$, where $f(x)$ is the fraction of sources that are identified as variable in the set of sources having mean counts $\le x$.  For the entire sample including sources with small numbers of counts, $\approx$30\% of sources are classified as variable at high significance.

In the following discussion, we calculate fractional variation using all available measurements for each source.  We define the fractional variation, $F$, using the formula:
\begin{eqnarray}
F &\equiv& (c_i - c_j) / (c_i + c_j),\label{fracVarDefEqn}
\end{eqnarray}
where $c_i$ is the count flux in the later epoch and $c_j$ is the flux in the earlier epoch.  Mathematically, $F$ represents half of the full distance between two measurements, $(c_i - c_j) / 2$, as a fraction of the average value of those two measurements, $(c_i + c_j) / 2$.  We have chosen this functional form to be symmetric (up to a sign) in $c_i$ and $c_j$ and to roughly represent the fractional deviation from some ``average'' flux.  Each measurement of $F$ between two epochs is associated with a rest-frame time between measurements, $\Delta t_{sys}$.

\subsubsection{Fractional Variation Over Time\label{fracVarTimeSec}}

Given the complex nature of measurement errors in our data set, we choose to assume that the intrinsic distribution of $F$ is Gaussian.  We use that assumption to constrain the distribution of fractional variation given the observed values of $F$ and errors in $F$.  The standard deviation of the Gaussian distribution, $\sigma(F)$, is calculated using the likelihood method of \citet[][hereafter M88]{mgwzs88} in bins of 100 pairs of epochs.  The errors on $\sigma(F)$ that are shown in the plot were calculated in the same way.  The epoch pairs were constructed using all pairs of measurements for each quasar.  A quasar that was observed $N$ times would therefore contribute $N(N-1)/2$ pairs.

\begin{figure} [ht]
  \begin{center}
      \includegraphics[width=5in, angle=-90]{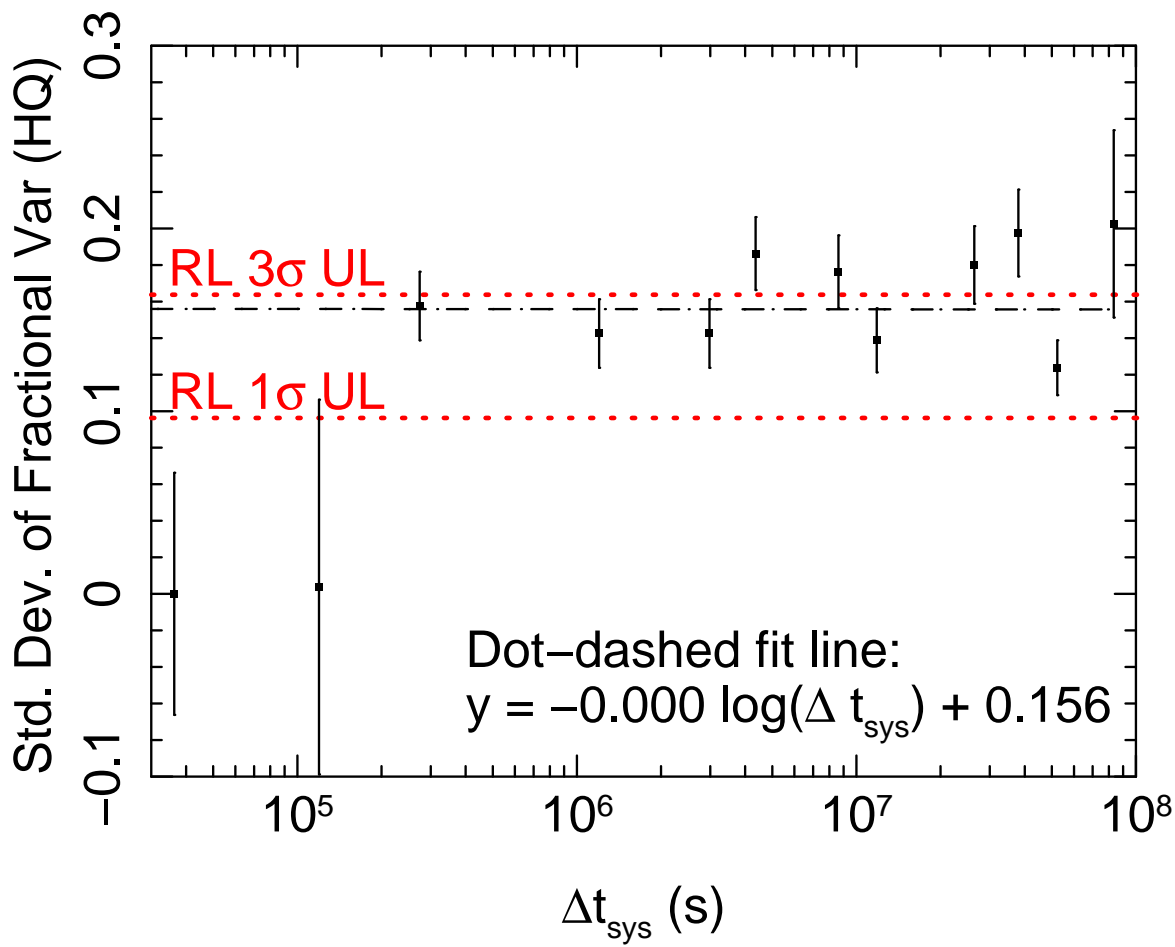}
      \caption{\label{fracVarVsdTFig}Each point shows the standard deviation of fractional variation (after accounting for the scatter due to measurement errors) in a bin of 100 epoch pairs in Sample~HQ as a function of the median time between epochs in that bin.  The black dot-dashed line shows a fit to values at rest-frame times $\Delta t_{sys} > 5\times 10^5$~s.  The red dotted lines indicate 1$\sigma$ and 3$\sigma$ upper limits for 15~pairs of epochs for {\it radio-loud, non-BAL} quasars.}
   \end{center}
\end{figure}

Figure~\ref{fracVarVsdTFig} shows our estimated values of $\sigma(F)$ as a function of $\Delta t_{sys}$.  Each point in the figure is placed at an $x$-coordinate corresponding to the median $\Delta t_{sys}$ in the bin for which $\sigma(F)$ was calculated.  The dot-dashed line represents a linear fit of $\sigma(F)$ at time scales $>5 \times 10^5$~s and is parameterized by:
\begin{eqnarray}
\sigma(F) &=& (0.000 \pm 0.012) \log(\Delta t_{sys}) + (0.156 \pm 0.083).\label{fitLongdtFracVarEqn}
\end{eqnarray}
At $\Delta t_{sys} \gtrsim 1$~day, the fractional variation is about 15.6\%.  At short time scales ($\Delta t_{sys} \lesssim 5\times 10^5$~s), there is no significant variation detected above our measurement errors.  If much more data were available to constrain $\sigma(F)$ as a function of time, we could map out the gap in the current plot where the fractional variation ``jumps'' from insignificant (on the shortest time scales) and breaks to a flatter shape at longer time scales.

We note that one source, J$142052.43+525622.4$, has so many observations that it dominates one of the long time-scale bins.  It varies somewhat less than the typical source, and if it is removed, the fit model shows a mild, but insignificant increase with longer $\Delta t_{sys}$.  Interestingly, J$142052.43+525622.4$ has weaker [\ion{O}{3}] emission, similar to the trend observed for non-variable sources (\S\ref{optSpecSec}), although it is technically classified as ``variable'' in our sample.

The red dotted lines in Figure~\ref{fracVarVsdTFig} indicate the 1$\sigma$ and 3$\sigma$ upper limits on $\sigma(F)$ for 15~epochs of radio-loud, non-BAL quasars with $\Delta t_{sys} > 5\times 10^5$~s.  Because the 3$\sigma$ upper limit for radio-loud quasars is only a little larger than the standard deviation of fractional variation for radio-quiet quasars, radio-loud quasars appear to be less variable than radio-quiet quasars in our sample.  If real, this effect could be due to additional, relatively-constant jet emission that dilutes the variable \mbox{X-ray} spectrum of the disk corona.

One shortcoming of this ``ensemble approach'' to measuring fractional variation is that sources with more observations have more influence on $\sigma(F)$, because they contribute more epochs to the ensemble average.  We can also measure quasar variation using a single epoch from each source and comparing it to the best-fit photon flux for that quasar.  This approach has the advantage of placing each quasar on an equal footing, so that a small number of quasars do not dominate the results.  However, we do not have enough epochs to map out variation in detail as a function of $\Delta t_{sys}$, and of course there is no particular ``time'' associated with the best-fit constant flux.  If we simply compare one epoch per source to the best fit count rate for that quasar, we find a time-independent standard deviation of fractional variation ($c_i/c_0 - 1$, where $c_0$ is the model flux) of about 16.7\% $\pm$ 2\%.  This result is consistent with values of $\sigma(F)$ found above for time scales $\gtrsim$1 week, which is reasonable given that such time scales represent the large majority of our data set.

\subsubsection{Symmetry of Variation\label{varSymmSec}}

Asymmetry in optical light curves can distinguish among different models of AGN emission \citep[e.g.][]{kmut98}.  Here we test whether any asymmetry is evident in the \mbox{X-ray} light curves of quasars in Sample~HQ.  We considered the fractional variation, $F$, between the earliest and latest HQ epochs for each source.  We find 87~cases with $F > 0$ and 80~cases with $F < 0$, indicating no strong tendency for quasars to get brighter or fainter over the time scale of our sample.

We also tested subsamples on longer and shorter time scales.  For times between epochs $\Delta t_{sys} > 10^7$~s, the split is 51~with $F > 0$ and 48~with $F < 0$.  For $\Delta t_{sys} < 10^7$~s, the split is 36 and 32, respectively.  Of course, there may be asymmetries at a level below what we can detect with the current data; the measurement errors do add some scatter to the distribution of fractional variation.  With a larger sample, it would also be possible to test for asymmetry on a smaller range of timescales.

\subsubsection{Extremes of Variation\label{extremesOfFracVarSec}}

\begin{figure} [ht]
  \begin{center}
      \includegraphics[width=5in, angle=-90]{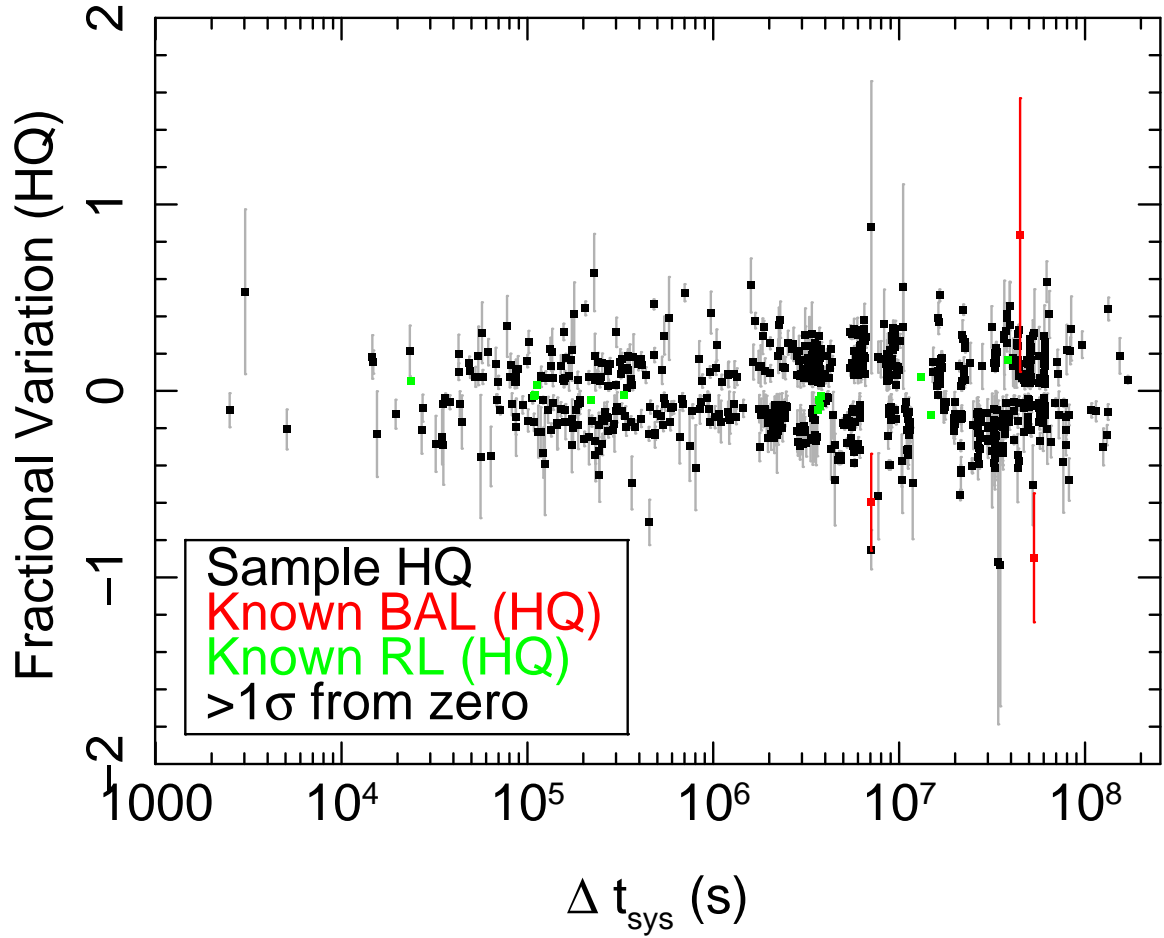}
      \caption{\label{fracVargt1SFromZeroFig}The fractional variation of Sample~HQ quasars as a function of time.  Red and green points indicate known BAL and radio-loud quasars, respectively.  {\it We only plot points that are $> 1\sigma$ from zero}, in order to clearly demonstrate the amplitudes of fractional variation.}
   \end{center}
\end{figure}

While our assumption of a Gaussian distribution of fractional variation permits us to characterize the typical extent of variation, it does not describe the behavior of outliers or extreme variation events.  Figure~\ref{fracVargt1SFromZeroFig} shows the fractional variation, $F$, between pairs of epochs in Sample~HQ.  For this plot, we have omitted any cases where the fractional variation is $< 1\sigma$ from zero, in order to clearly visualize any extreme values of $F$.  Black points show typical quasars, while red or green points signify known BAL or radio-loud quasars, respectively.  As the plot shows, $F$ is nearly always $< 50$\%, and $F > 100$\% is apparently quite rare.

\begin{figure} [ht]
  \begin{center}
      \includegraphics[width=5in, angle=-90]{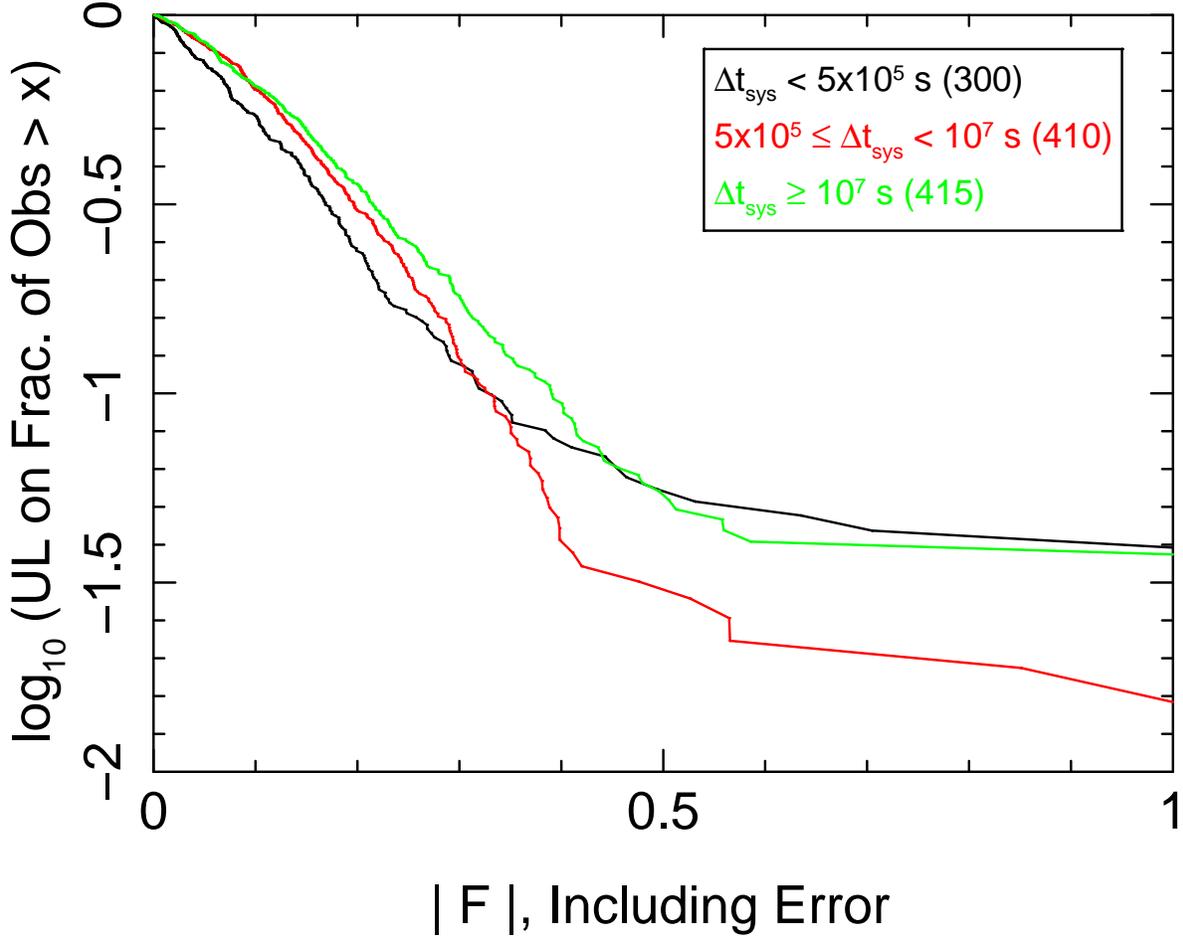}
      \caption{\label{uLOnRateOfFracVarFig}The $x$-axis represents the magnitude of fractional variation, $F$, {\it including} measurement error, observed over time scales $\Delta t_{sys} < 5\times 10^5$~s (black), $5\times 10^5 \le \Delta t_{sys} < 10^7$~s (red), and $\Delta t_{sys} \ge 10^7$~s (green).  We only consider radio-quiet, non-BAL quasars for this plot.  The $y$-axis indicates the upper limit (at 95\% confidence) on the true fraction of observations in our sample that have $|F|$ greater than the value on the $x$-axis.  The numbers in the legend in parentheses indicate subsample sizes.  The upper limits represented by these curves depend on several issues, including subsample sizes and measurement errors.}
   \end{center}
\end{figure}

Extreme variation has been observed in rare cases such as the narrow-line type 1 quasar PHL~1092, which decreased in flux by a factor of $\sim$200 over $\approx 3.2$~yr in the rest frame \citep{mfbgb09}.  We can use our data set to place limits on the frequency of such events.  Figure~\ref{uLOnRateOfFracVarFig} shows upper limits on the rate at which new observations sharing our Sample HQ properties would be expected to show a given magnitude of fractional variation.  The limits are constructed by assembling all values of fractional variation, $F$, over a given time frame.  The plot considers three time frames, $\Delta t_{sys} < 5\times 10^5$~s, $5\times 10^5 \le \Delta t_{sys} \le 10^7$~s, and $\Delta t_{sys} \ge 10^7$~s.  The upper limits for these time frames are plotted as a function of $|F|$ with black, red, and green curves, respectively.  For a given value of $|F_0|$, we calculated the upper limit on the $y-$axis by determining the maximum intrinsic rate of occurrences of values $|F| > |F_0|$ given the number of observed cases with fractional variation $< |F_0|$, according to a binomial statistic and using 95\% confidence limits.\footnote{The upper limits we calculate represent limits on the levels of variability that an observer might expect to measure, assuming that our data sample is representative of their observations.    This empirical approach may also include a contribution from exceptional outliers such as flares or flux drops that would not be modeled in an ensemble power spectrum.}  In cases where a source was not detected at $>$95\% confidence in an epoch, we conservatively forced the variation to be 100\%.  This was done for 19 of 1157 epoch pairs.

We caution that these upper limits depend on subsample sizes and {\it include} unmodeled variation due to measurement errors.  They are therefore indended only to constrain the rates at which variation greater than a given rate $F$ occurs.  (We could limit the impact of measurement error by dropping sources that have low average count rates, but that could introduce new biases.)  For example, fractional variation $F \ge $100\% should occur in $<10^{-1.4} \approx 4$\% of observations.  Fractional variation $F \ge 25$\% should occur in $<10^{-0.6} \approx 25$\% of observations.

\subsection{Excess Variance\label{excessVarianceSec}}

We calculate the excess variance of measured count rates \citep[e.g.,][]{ngmty97, tgnt99}, as:
\begin{eqnarray}
\sigma_{EV}^2 \equiv \frac{1}{N\mu^2} \sum^{N}_{i=1} \biggl((n_i - \mu)^2 - \sigma_i^2\biggr)\label{excessVarianceEqn},
\end{eqnarray}
where $n_i$ is the count rate in epoch $i$, $\mu$ is the mean of the values $n_i$, and $\sigma_i$ is the average statistical error in the measurement of $n_i$.  In order to avoid any bias introduced by having different numbers of epochs per source, we calculate $\sigma_{EV}$ using only the earliest and latest observation of each quasar.  $N$ is therefore always 2 in this calculation.  The excess variance statistic is not ideal for our data set, which consists of sources observed a small number of times over diverse time intervals.  However, we consider it for comparison to other published analyses.

\begin{figure} [ht]
  \begin{center}
      \includegraphics[width=5in, angle=-90]{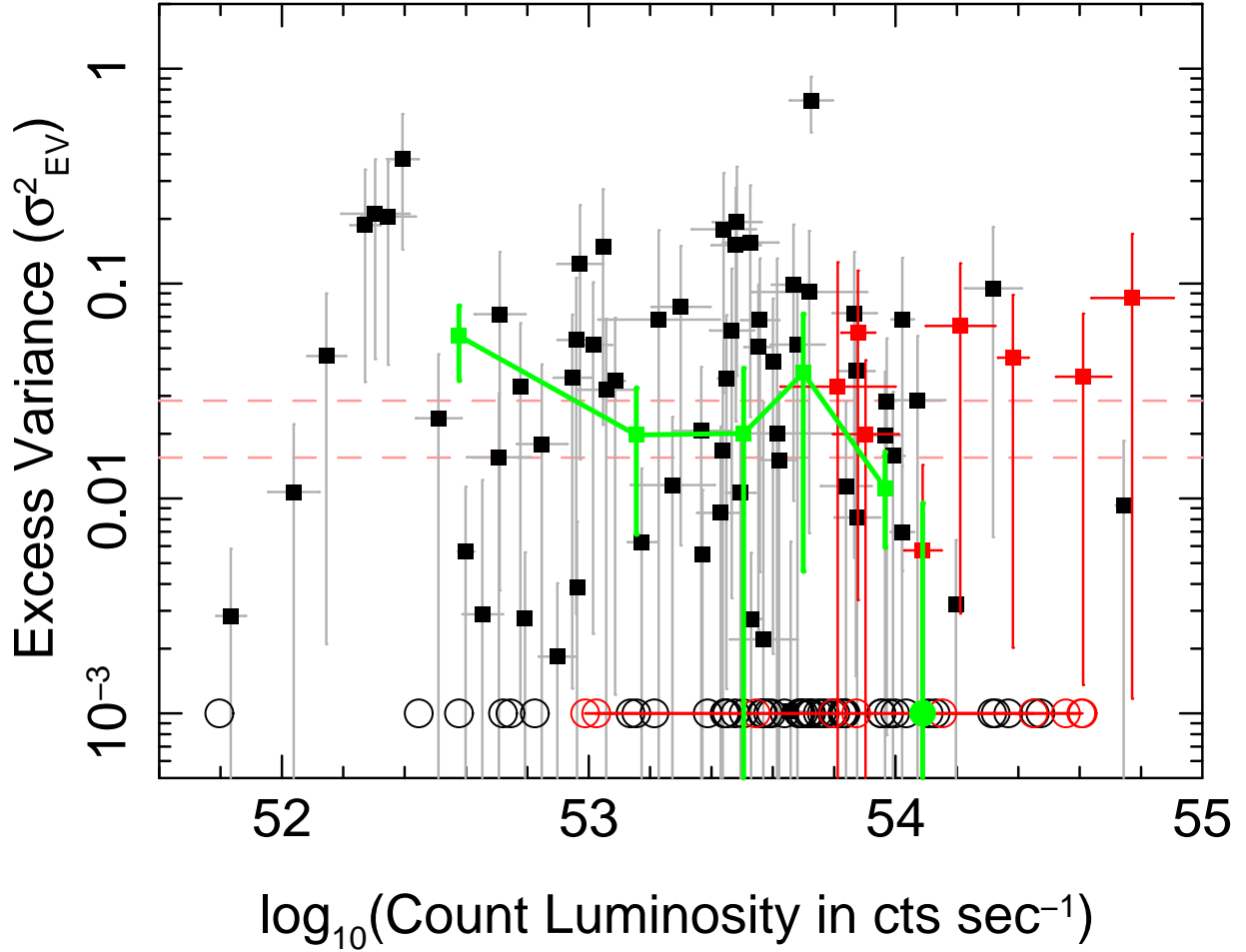}
      \caption{\label{plotRMSEVFig}Excess variance as a function of count-rate luminosity for radio-quiet, non-BAL quasars.  Sources with excess variance $< 0.001$ are plotted as open circles $y = 0.001$.  Red points are sources at redshift $z > 2$, while black points are sources at lower redshift.  The thick green line represents the mean excess variance calculated for bins of 21 sources at redshift $z < 2$.  The filled green circle represents the mean excess variance of sources at $z \ge 2$; the value is less than $0.001$.  The dashed red lines indicate 2$\sigma$ and 3$\sigma$ upper limits on the mean excess variance for quasars at $z > 2$.}
   \end{center}
\end{figure}

We limit our analysis to 131~radio-quiet, non-BAL sources that have $\ge$50 counts per epoch, on average, in order to screen out a large number of sources with small or negative $\sigma_{EV}^2$.  We estimate the error on $\sigma_{EV}$ using the formula given in \citet{tgnt99}.  Figure~\ref{plotRMSEVFig} shows $\sigma_{EV}$ as a function of count-rate luminosity.  Sources at $z > 2$ are plotted in red.  At lower redshifts ($z < 2$), 49 of 113~sources have $\sigma_{EV}^2 < 0.001$; these are plotted as black open circles at $y = 0.001$.  Similarly, 10 of 18 higher-redshift ($z > 2$) AGN have $\sigma_{EV}^2 < 0.001$ and are plotted with red open circles along the bottom of the plot.  No clear patterns are visible in the plots except a tendency for higher-redshift sources (plotted in red) to have higher luminosities, due to the flux-limited nature of the SDSS quasar survey.

In order to test for additional structure in the data, we calculated the mean of the excess variance values in bins of 21 sources each.  To calculate this mean, we used only lower-redshift sources at $z < 2$.  This mean $\sigma_{EV}^2$ is plotted in Figure~\ref{plotRMSEVFig} as a thick green line.  Error bars on the data points (placed at the median count-rate luminosity for each bin) represent the estimated error on the mean.  The green line shows a general trend of decreasing $\sigma_{EV}^2$ with luminosity, as is commonly observed (e.g., A00, M02, Pao04, Pap08, and references therein).  A filled green circle represents the mean $\sigma_{EV}^2$ for 18~sources at higher redshifts $z \ge 2$; its value is less than $0.001$.  Dashed red lines in Figure~\ref{plotRMSEVFig} indicate 2$\sigma$ and 3$\sigma$ upper limits on the excess variance for quasars at $z > 2$.  While we cannot completely rule out the possibility of an increase at high redshifts, the upper limits indicate that any increase would be at most very small.

The extent to which a quasar is measured to vary can depend on the time scales over which it is observed.  Local Seyfert AGN show increasing variations with time up to at least a break time scale.  As we discuss in detail in \S\ref{varHigherzSec}, it would be difficult to account for this effect.  We do not know the shape of the power spectrum for quasars, and many of our observations are likely beyond the break time scale (Figure~\ref{sampledtVsOldFig}).  Instead, we explore the possibility that the increase in variability could be timescale-dependent by calculating $\sigma_{EV}$ using only observations that are separated by a certain range of time scales.  For example, we constructed a subsample of observations with rest-frame time scales $\lesssim$1~yr.  We tried a variety of system-frame and observed-frame time scales, and in no case did we find evidence for increased variation at $z > 2$.

Using similar methods, we find no significant correlation between $\sigma_{EV}^2$ and the average hardness ratio (defined in \S\ref{hRSec}) of a source.  This agrees with the result of Pap08, who found no correlation between variability amplitude and spectral slope $\Gamma$ in their light curves.  Pap08 note that this result disagrees with earlier results \citep[e.g.,][]{gml93} indicating that nearby AGN with steeper spectra showed larger amplitude variations.  While this may be an indication that luminous quasars vary differently than local AGN, we caution that many factors affect these results, including the time scales probed, the baseline of spectral shapes spanned by a sample, and sensitivity to variation in fainter sources.  In following sections (\S\ref{hRSec} and \S\ref{specFitsBrightSourcesSec}), we examine the related issue of how spectral shape changes for a single source as it varies.

\subsection{Variability Dependence on Physical Properties\label{varDepOnPhysPropSec}}

The {\it Chandra} archive provides repeat observations of quasars having a wide range of luminosities, redshifts, and black hole masses.  While individual sources may not have sufficient observations to permit sensitive tests of variability properties, we can characterize the physical dependence of variability more effectively in an ensemble of observations.  We place each observation of a radio-quiet, non-BAL quasar in Sample~HQ on an equal footing in this ensemble by considering the quantity
\begin{eqnarray}
S_{ij} &\equiv& (f_{ij} - c_j) / \sigma_{ij}, \label{sijDefEqn}
\end{eqnarray}
where $f_{ij}$ is the $0.5-8$~keV count rate observed in epoch $i$ for source $j$, $c_j$ is the best-fit constant count rate for source $j$ over all epochs, and $\sigma_{ij}$ is the error on $f_{ij}$.  In the absence of variability, we would expect the distribution of $S_{ij}$ values to be roughly Gaussian with an rms of~1.  The mean value of $S_{ij}$ would be $\approx 0$ (by construction) in ideal conditions where variation is symmetric and not influenced by outliers.  This is generally the case except at the lowest redshift, which is influenced by a few ($\approx 5$) outliers that have higher $S_{ij}$ values.  We use $S_{ij}$ to test the hypothesis that intrinsic variability can be detected in the data at a certain redshift, luminosity level, or black hole mass.  Of course, determining the {\it amplitude} or detailed {\it pattern} of any intrinsic variability would require simulations to determine what pattern of $S_{ij}$ values could be expected from a given variability model.

\begin{figure} [ht]
  \begin{center}
      \includegraphics[width=5in, angle=-90]{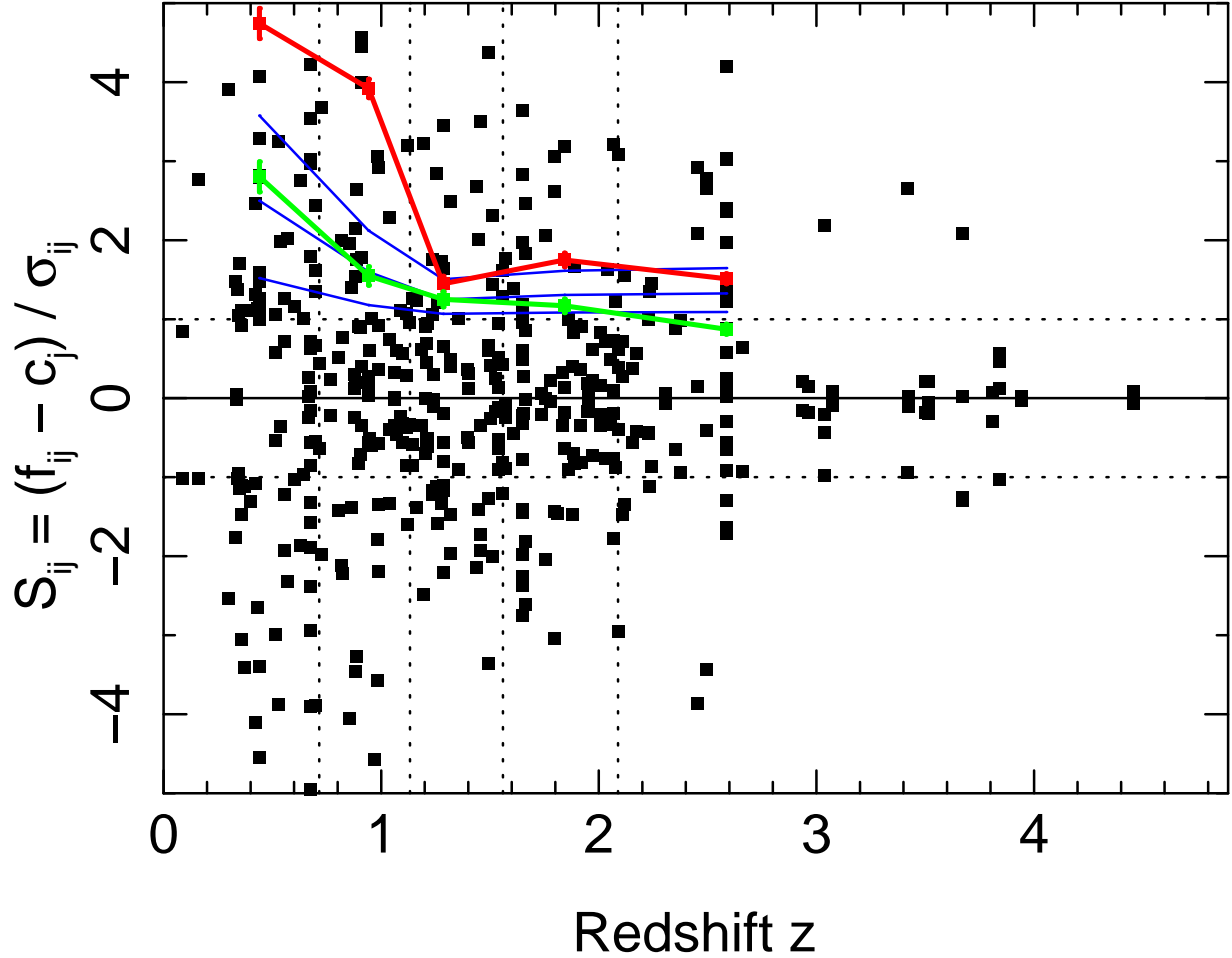}
      \caption{\label{vATVByzFig}The distribution of $S_{ij} = (f_{ij} - c_j) / \sigma_{ij}$ for all epochs $i$ of each source $j$.  The sources are ordered and binned by redshift.  Some points have been clipped off the plot boundary for visual clarity.  A horizontal dotted (solid) line indicates $y = \pm 1$ ($y = 0$), while vertical dotted lines indicate bin boundaries.  Red points (connected by red lines) indicate the standard deviation of points in each bin, while green points (connected by green lines) indicate the rms estimated from the median absolute deviation (MAD).  The red points can be driven by dramatically variable outliers in a given bin, while the green (MAD) points provide a representation of sample variability that is more robust (\S\ref{varDepOnPhysPropSec}).  Blue lines correspond to hypothetical sources that have intrinsic variation of 10\%, 20\%, or 30\% in each epoch.  At $|y| \le 1$, measurement error dominates any variability signal in the data.}
   \end{center}
\end{figure}

In Figure~\ref{vATVByzFig}, we plot $S_{ij}$ as a function of redshift.  We have binned the set of $S_{ij}$ values into 5~bins and calculated the sample standard deviation in each bin.  Red data points (placed at the median $x$-value of each bin) indicate the square root of the unbiased sample variance, with error bars estimated as $\Delta \sigma \equiv \sigma / \sqrt{2(N-1)}$ for a bin with $N$ data points.  (However,we note that scatter in $S_{ij}$ can be driven by non-Gaussian effects such as outliers.)  The unbiased sample variance reflects the influence of outliers, some of which have been clipped out of the plot range for visibility purposes.  A better estimate of the overall sample variability properties can be obtained using the median absolute deviation \citep[MAD; e.g.,][]{mmy06}, which is much less sensitive to strongly-variable outliers.\footnote{The MAD for a set of values is defined to be MAD$(x) \equiv median_i(|x_i - median(x)|)$.  I.e., it is the median of the absolute value of residuals from the sample median.  It is less sensitive to outliers than the sample variance is, and can be used to estimate the standard deviation $\sigma$ according to $\sigma \approx 1.483 \times$MAD.}  The green line shows this value, $\sigma_{MAD}$ calculated for each bin, with errors roughly estimated using the same formula as for the red data points.

As redshift increases, the amount of variability that we detect decreases.  $\sigma_{MAD}$ for $S_{ij}$ decreases to $\approx$1 as redshift increases, indicating that any intrinsic variability is dominated by measurement errors in this analysis at $z \gtrsim 2$.  As Figure~\ref{meanCtsHQVsZFig} shows, the average number of counts per epoch for a given quasar also decreases at higher redshift, so that variability becomes harder to detect at $z > 2.$  However, we find the same result --- $S_{ij}$ decreasing to the noise level ($S_{ij} = 1$) --- even when we limit our analysis to Sample~HQ sources that have 10 to 50 average counts per epoch.  We do not, therefore, find any indication that variability increases at higher redshifts in this data set.

Of course, these statements simply express the extent to which we can measure variability in the data.  Even bright AGN at high redshifts are presumably variable at some level, and their variability patterns may be complex.  In order to demonstrate how a simple pattern of variation would appear in the data, we have created a set of simulated data in which we fix a constant amount of variation (10\%, 20\%, or 30\%) in each observation of a quasar.  $\sigma_{MAD}$ for the simulated $S_{ij}$ values are plotted as blue lines in Figure~\ref{vATVByzFig}.  These decrease with redshift because measurement errors are generally increasing with redshift, but the simulated lines do not fall all the way to $S_{ij} = 1$, because they do possess intrinsic variability.  The green line corresponding to real quasars declines more steeply with redshift than the blue lines, again indicating that we see no significant increase in variability with redshift.

\begin{figure} [ht]
  \begin{center}
      \includegraphics[width=5in, angle=-90]{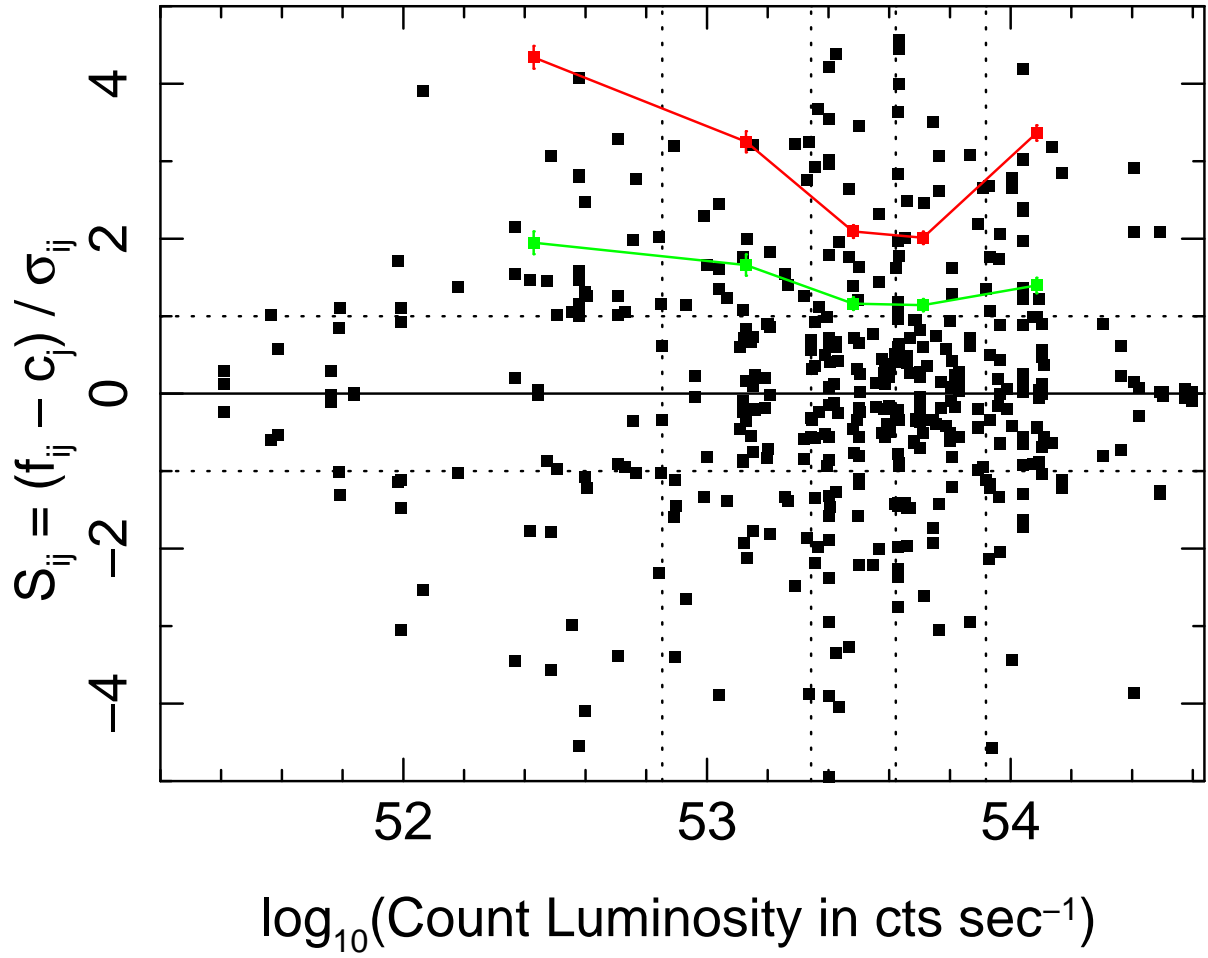}
      \caption{\label{vATVByLumFig}Similar to Figure~\ref{vATVByzFig}, but showing $S_{ij}$ as a function of count luminosity.}
   \end{center}
\end{figure}

We find similar results when plotting $S_{ij}$ as a function of luminosity or $M_{BH}$.  $\sigma_{MAD}$ decreases to the noise level in the highest $M_{BH}$ bins.  $\sigma_{MAD}$ decreases with luminosity as well, although there is a small (but insignificant) upturn at the highest luminosity values (Figure~\ref{vATVByLumFig}).

Finally, we have attempted to distentangle the degeneracy between redshift and luminosity by binning our data on both quantities.  Even with only four redshift bins (0--0.5, 0.5--1, 1--2, $>2$), this pushes the limits of available data, leaving generally 10--35 $S_{ij}$ points per $(L, z)$ bin.  In any case, we still find a general trend for $\sigma_{MAD}$ to decrease with both luminosity and redshift, and no evidence of an increase at $z > 2$.

Throughout this analysis, we have worked with a variability signal that is measured in the {\it observed}-frame $0.5-8$~keV band where {\it Chandra} is most sensitive.  To the extent that AGN variation differs from one part of the spectrum to another, our ability to detect variation is also redshift-dependent.  (See further discussion of this point in \S\ref{varHigherzSec}.)  The instrumental response varies strongly over the bandpass, further complicating any attempt to separate energy-dependent variation from redshift dependence.  For the current work, when we state that ``We do not detect variability'' in some instance, that statement should be understood in the context of the data that we have available, using the {\it Chandra} observed-frame bandpass and instrumental response.

\subsection{Variation in Spectral Shape\label{hRSec}}

Hardness ratios (HRs) --- here defined as the ratio $HR \equiv (H-S)/(H+S)$, where $H$ and $S$ are the observed-frame count rates in the hard (2.0--8.0~keV) and soft (0.5--2.0~keV) bands, respectively --- may be used to describe the overall spectral shape for sources that do not have sufficient counts for detailed spectral fitting.  We have computed observed-frame hardness ratios for all observations in Sample HQ, using the method of \citet{pksvzhw06} to determine $1\sigma$ HR errors.  We calculated separate weighted exposure maps for the soft and hard bands using the spectral slope obtained from the best fit to the full-band spectrum of each source.  (Narrower bandpasses and smaller count rates do not permit separately fitting the soft and hard band spectra.)

\begin{figure} [ht]
  \begin{center}
      \includegraphics[width=5in, angle=-90]{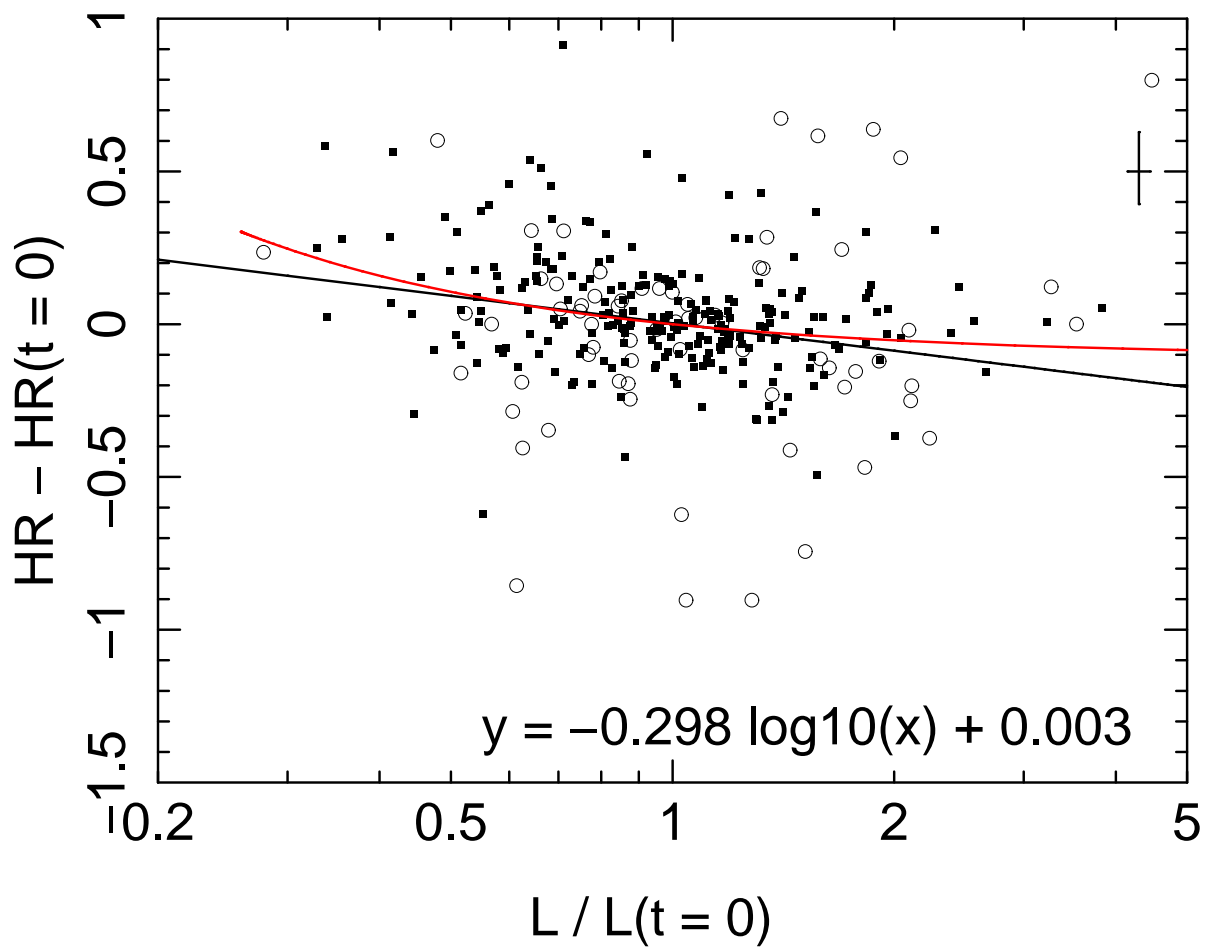}
      \caption{\label{vAPHRdHRVsFracLumFig}Change in $HR$ compared to change in luminosity for all epochs of non-BAL, radio-quiet quasars in Sample~HQ.  Squares represent quasars at lower redshifts ($z < 2$), while open circles represent quasars at higher redshifts ($z \ge 2$).  Typical (median) errors are shown in the upper right corner.  The solid black line represents a linear fit to the full set of points, while the red curve represents the two-component toy model described in \S\ref{hRSec}.}
   \end{center}
\end{figure}

We examine all epochs in Sample~HQ (excluding those from BAL and radio-loud quasars) to determine how $HR$ varies with luminosity.  Here we put all epochs of all sources on an equal footing, scaled by their $HR$ and luminosity in the earliest available epoch.  For each epoch, we calculate the change in $HR$ and luminosity from the first observation of that source.  These quantities [$HR - HR(t=0)$ and $L / L(t=0)$] are plotted in Figure~\ref{vAPHRdHRVsFracLumFig} for lower-redshift ($z < 2$, black squares) and higher-redshift ($z \ge 2$, open circles) quasars.  There is a significant anticorrelation (at $>$99.99\% confidence, according to a Spearman rank correlation test) between the change in $HR$ and the $L / L(t=0)$ for quasars at $z < 2$.  The correlation is not significant at $z > 2$, although the sample sensitivity is weakened by larger measurement errors for fainter sources as well as the fact that softer photons are shifted out of the {\it Chandra} bandpass at higher redshifts.  Defining $\Delta HR \equiv HR - HR(t=0)$ and $F \equiv L_{ph} / L_{ph}(t=0)$, we obtain the fit
\begin{eqnarray}
\Delta HR &=& (-0.298\pm 0.030)\times \log(F) + (0.003\pm 0.004)\label{dHRVsdLEqn}
\end{eqnarray}
for the full sample including all redshifts.

We constructed toy physical models of the trend between $\Delta HR$ and $L/L(t=0)$.  The change in spectral hardness could, of course, be associated with a single power law model that changes both shape and normalization, following the relation described in Equation~\ref{dHRVsdLEqn}.  The goal of our toy models is to determine whether we can reproduce Equation~\ref{dHRVsdLEqn} with only a single variable parameter.

We modeled several scenarios, including:  1) a double power law with a constant hard component ($\Gamma = 1.3$) added to a variable softer component ($\Gamma = 2$), 2) a constant power law with variable neutral absorption, and 3) a constant power law with an absorber that varies in ionization level.  For each toy model, we attempted to determine a set of fit parameters that caused the $HR$ and luminosity to vary together in a way that reproduced the trends observed in Figure~\ref{vAPHRdHRVsFracLumFig}.  The absorption-dominated models (cases 2 and 3) generally produced a much steeper trend of $HR$ with luminosity than we observed.  In contrast, the double power-law model (case 1) reproduced the observed trend reasonably well for a hard, constant component having $\Gamma = 1.3$ and a normalization at 1~keV of about 20\% of the variable, soft power-law component.  This model is shown as a red curve in Figure~\ref{vAPHRdHRVsFracLumFig}.  The impact of a two-component model on observations of high-redshift quasars is discussed further in \S\ref{specVarSec}.

\subsection{Spectral Fits for Bright Sources\label{specFitsBrightSourcesSec}}

For sources with large numbers of counts, we can fit spectra to determine how spectral shapes and features vary as a function of source brightness.  We have selected 16~radio-quiet, non-BAL quasars having at least 500 counts per epoch, on average, and fit them with a spectral model consisting of a Galactic-absorbed power law absorbed by one absorption edge.  The edge is constrained to have an energy threshhold of 0.739~keV in the rest frame, corresponding to absorption from the \ion{O}{7} ion.  The optical depth of the edge, $\tau$, is allowed to vary.  We have chosen the edge model to roughly characterize a power law spectrum with ionized absorption using a single parameter.  In fact, we constrain $\tau > 0$ at the 1$\sigma$ level in only 7 of 87 epochs.  We obtain similar results using a neutral, rest-frame absorption model, so the choice of this model is not strongly biasing our results.  While the single-edge model certainly does not reproduce the complex spectral features that an absorber may imprint on an emission spectrum, it does provide a basic model of variable absorption when spectra do not permit careful fitting of absorption models that have more degrees of freedom.  The results of the fits are given in Table~\ref{brightFitTable}.

\begin{figure} [ht]
  \begin{center}
      \includegraphics[width=5in, angle=-90]{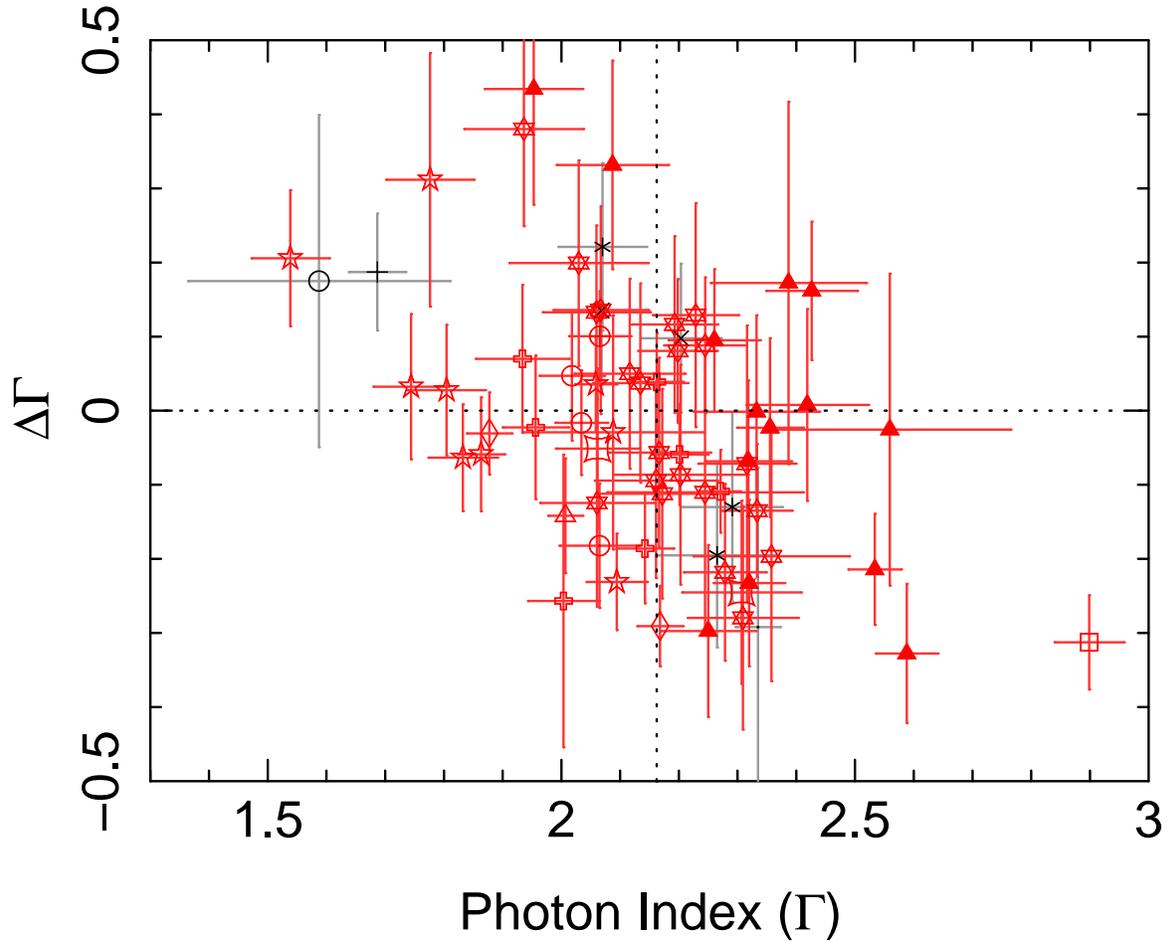}
      \caption{\label{vAEVPdGammaVsGammaTruncatedFig}The change in photon index $\Gamma$ between two observations of the same source as a function of $\Gamma$ in the earlier observation.  Sources classified as ``variable'' are plotted in red points, while ``non-variable'' sources are plotted in black.  Two outlier points have been clipped for visibility.  The horizontal dotted line indicates $y = 0$ and the vertical dotted line indicates the median value of $\Gamma$ for the data set.  All epochs of a single source are plotted with the same symbol.}
   \end{center}
\end{figure}

\begin{figure} [ht]
  \begin{center}
      \includegraphics[width=5in, angle=-90]{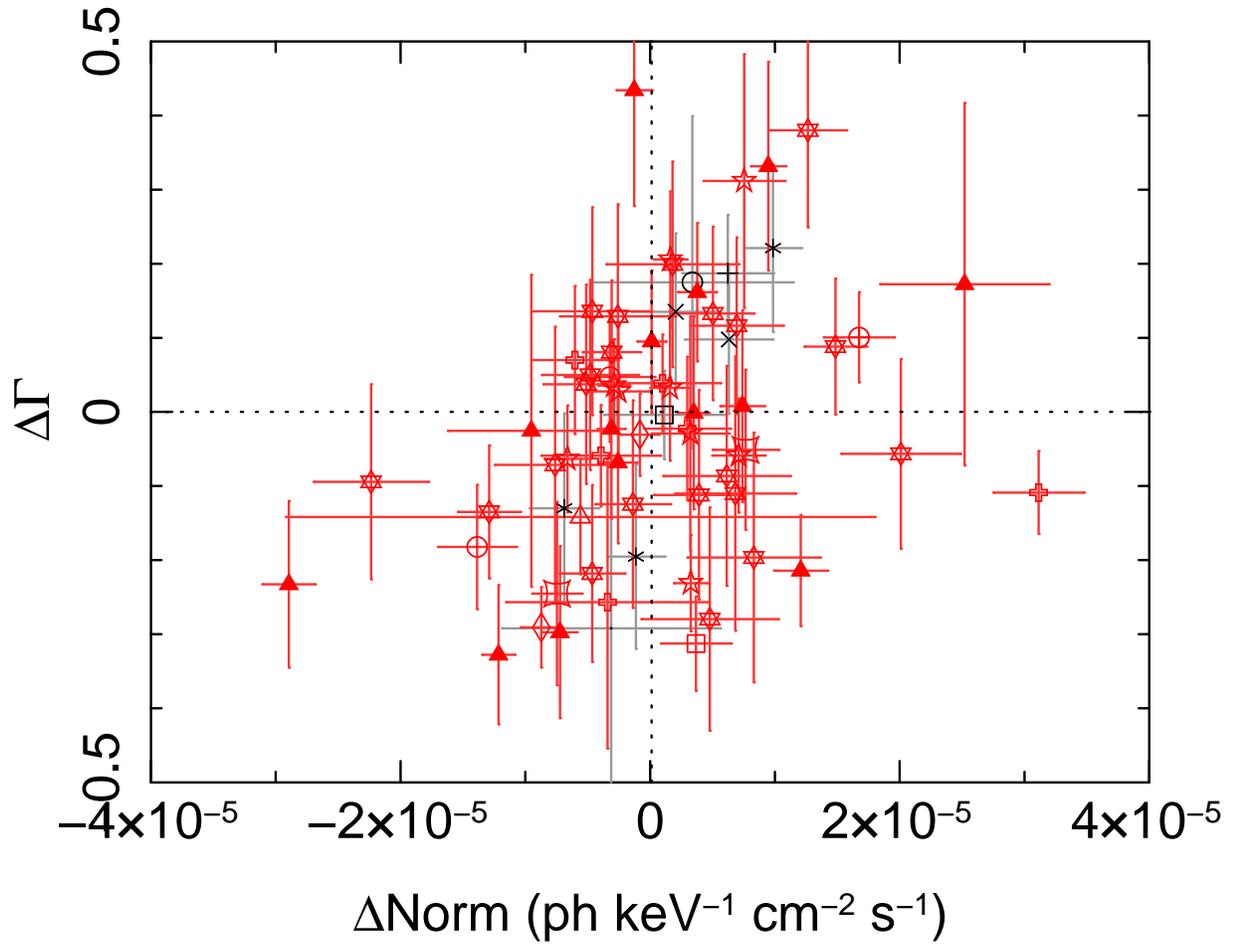}
      \caption{\label{vAEVPdGammaVsdNormTruncatedFig}Same as Figure~\ref{vAEVPdGammaVsdNormTruncatedFig}, but showing $\Delta\Gamma$ plotted against the change in power law normalization, $\Delta$Norm.}
   \end{center}
\end{figure}

Figure~\ref{vAEVPdGammaVsGammaTruncatedFig} shows the distribution of ($\Delta\Gamma_{ij}$, $\Gamma_{ij}$) pairs from our fit models.  We ordered the Sample HQ observations of each source chronologically, then calculated $\Delta\Gamma_{ij}$ as the difference in photon indices between epoch $i+1$ and epoch $i$ for source $j$.  $\Gamma_{ij}$ is the photon index from the first epoch (i.e., epoch $i$) in the pair.  Two outlier points have been clipped in this figure for visual clarity.  The plot shows a strong anti-correlation between $\Delta\Gamma$ and $\Gamma$ at the $>$99.99\% confidence level, according to a Spearman rank correlation test.  The median value of $\Gamma$ in this sample is $\Gamma_M \approx 2.16$.  (The value $\Gamma = 2.16$ is not necessarily representative of the full population; for example, the sources chosen for spectral fitting were selected to have larger numbers of \mbox{X-ray} counts.)  Spectra tend to steepen when they are flatter than this value, and tend to flatten when they are steeper.  Figure~\ref{vAEVPdGammaVsdNormTruncatedFig} shows how the best-fit photon index changes with the power law normalization.  There is a strong correlation (99.97\% confidence, according to a Spearman rank correlation test) between $\Delta\Gamma_{ij}$ and $\Delta$Norm$_{ij}$, where $\Delta$Norm$_{ij}$ is the difference in power law normalization values between epochs $i+1$ and $i$.  Overall, spectra steepen as they brighten (and/or flatten when getting fainter), consistent with what we have already observed using hardness ratios for a larger sample of sources (see Figure~\ref{vAPHRdHRVsFracLumFig} and \S\ref{hRSec}).

\subsection{Hard Flux and Soft Flux\label{fluxFluxSec}}

Some local Seyfert AGN show a linear relationship between hard- and soft-band \mbox{X-ray} count rates \citep[e.g.,][]{tum03}.  When the linear relationship is extrapolated to a zero soft-band count rate, a significant non-zero hard-band flux remains.  This suggests a two-component model of Seyfert AGN \mbox{X-ray} spectra, in which a constant hard spectral component is augmented by a softer, variable power-law.  The non-zero offset in the hard band comes from the underlying hard spectral component that remains when the variable power law component is absent.  The shape of the hard spectral component, determined from flux-flux plots in different \mbox{X-ray} bands, can resemble the expected emission from cold disk reflection \citep[e.g.,][]{tum03}.

For each radio-quiet, non-BAL source having two or more observations in Sample~HQ, we use the IDL {\tt FITEXY} routine to generate a linear fit to the observed frame hard-band (2--8~keV) count rates as a function of soft-band (0.5--2~keV) count rates.  That is, we fit for $a$ and $b$ in the equation:
\begin{eqnarray}
r_H &=& a r_S + b\label{linearHardSoftEqn},
\end{eqnarray}
where $r_H$ and $r_S$ are the hard- and soft-band count rates, respectively.  The FITEXY routine accounts for errors in both the hard and soft band count rates.  Our current data do not permit a precise test of whether the flux-flux relation is best modeled as linear, although visual inspection indicates that a linear model works well in most cases with more than 5~observations.  The linear model is not formally rejected by the fit statistic in 12 of 15 quasars with $\ge 5$ observations.  Because many sources have only 2 or 3 observations, our goal is not to test the linear model, but instead to determine a typical value for $y_0$ in the ensemble that would correspond to a constant hard-band offset.

Under the approximation that the $y$-intercept values are Gaussian distributed, we used the method of M88 to determine $\overline{y_0}$, the mean value of the $y$-intercept in these fits.  The mean $y$-intercepts, corresponding to a constant hard-band flux component, were greater than zero at $>$99\% confidence both for lower-redshift Sample HQ sources (51 at $z < 1$) and also for higher-redshift sources (116 at $z \ge 1$).  Then, to estimate the contribution of this hard, constant component to the hard \mbox{X-ray} spectrum, we re-ran the fits, but this time normalized the hard-band count rates by the mean hard-band count rate for that source.  The normalized fractional offset has a value of $0.428 \pm 0.140$ for sources at $z < 1$ and $0.753 \pm 0.165$ for sources at $z \ge 1$.  The increase in the fractional offset (although with a large error bar) suggests that the the shape of a putative constant component would be flatter than the shape of the variable component, so that it contributes an increasing fraction of flux at higher energies.  Local Seyfert AGN show hard-spectrum fractional contributions of 20--40\% at energies $\ge$2~keV, possibly due to Compton reflection from the accretion disk; these fractions increase at higher energies \citep[e.g.,][]{tum03, vf04}.  If a similar model holds in the quasar luminosity regime, we might expect the fractional contribution of the constant component to increase as higher energies are shifted into the {\it Chandra} bandpass.  (See \S\ref{specVarSec} for further discussion about the effects on observations of high-redshift quasars.)

\subsection{Intra-Observation Variability\label{intraObsVarSec}}

The counts in {\it Chandra} event lists are tagged with arrival times, permitting a search for variability in the light curves of each observation.  We use a Kolmogorov-Smirnov (KS) test to identify any light curve for which the count arrival times are significantly non-uniform.  (Here we define ``light curve'' to mean the time sequence of all counts that fall inside the source extraction region.)  At a 99\% KS-test confidence level, we identify 17~{\it potentially} variable light curves.  We have removed five additional observations from consideration because visual inspection of their light curves indicates that instrumental factors such as background flaring are dominating the light curves.  A time-dependent analysis of background variation is beyond the scope of this study, but we note that in our bright sample (with total count rates $> 0.01$~s$^{-1}$, described below), the estimated background count rate is $<$10\% of the total rate in $\approx$86\% of the light curve exposure.  Background effects should not strongly affect our overall results.

Six of the 17~light curves that the KS test flags as variable belong to a single source, J$123800.91+621336.0$ (hereafter ``J1238'') at $z=0.44$ with absolute $i-$magnitude $M_i = -23.04$.  This source was observed many times as it falls in the {\it Chandra} Deep Field-North \citep[e.g.,][]{abbghsv03}.  Given that we have searched $\approx$800~light curves for a result at 99\% confidence, we cannot consider the variability criterion to be highly significant for the remaining 11~light curves.  This result is consistent with our previous observation that the level of intrinsic variation {\it between} observations falls below our detection limit for time scales shorter than $\sim 10^5$~s (\S\ref{fracVarSec}).

\begin{figure} [ht]
  \begin{center}
      \includegraphics[width=5in, angle=-90]{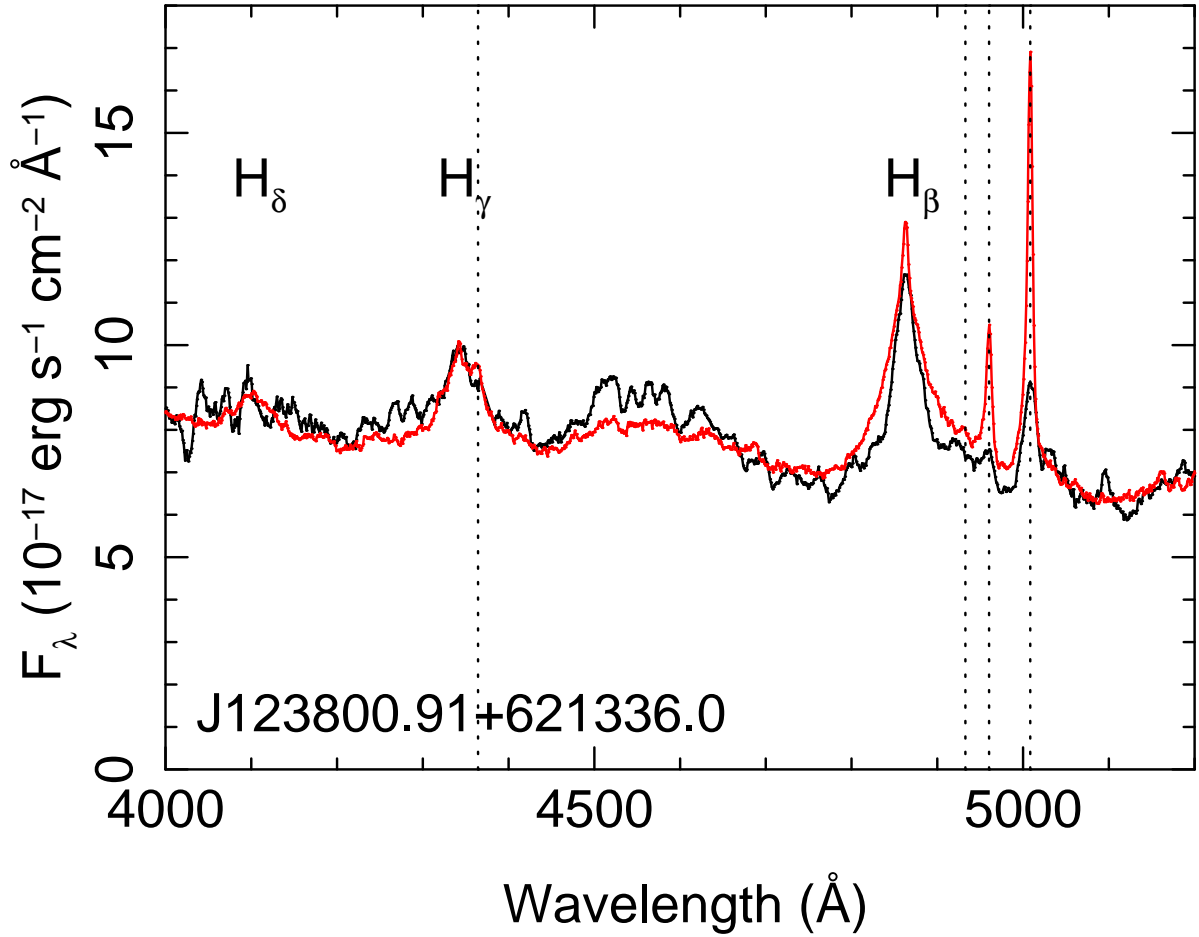}
      \caption{\label{j1238Fig}SDSS spectrum of J$123800.91+621336.0$, plotted in black.  The spectrum has been smoothed with a boxcar window 11~bins wide.  The red spectrum shows the quasar composite, normalized to overlap at 5100\AA.  Vertical dotted lines indicate the wavelengths of [\ion{O}{3}] lines, while Balmer lines are labeled with text.  We interpolated over the spectrum at 4372--4378\AA\ due to contamination from sky lines.}
   \end{center}
\end{figure}

\begin{figure} [ht]
  \begin{center}
      \includegraphics[width=5in, angle=-90]{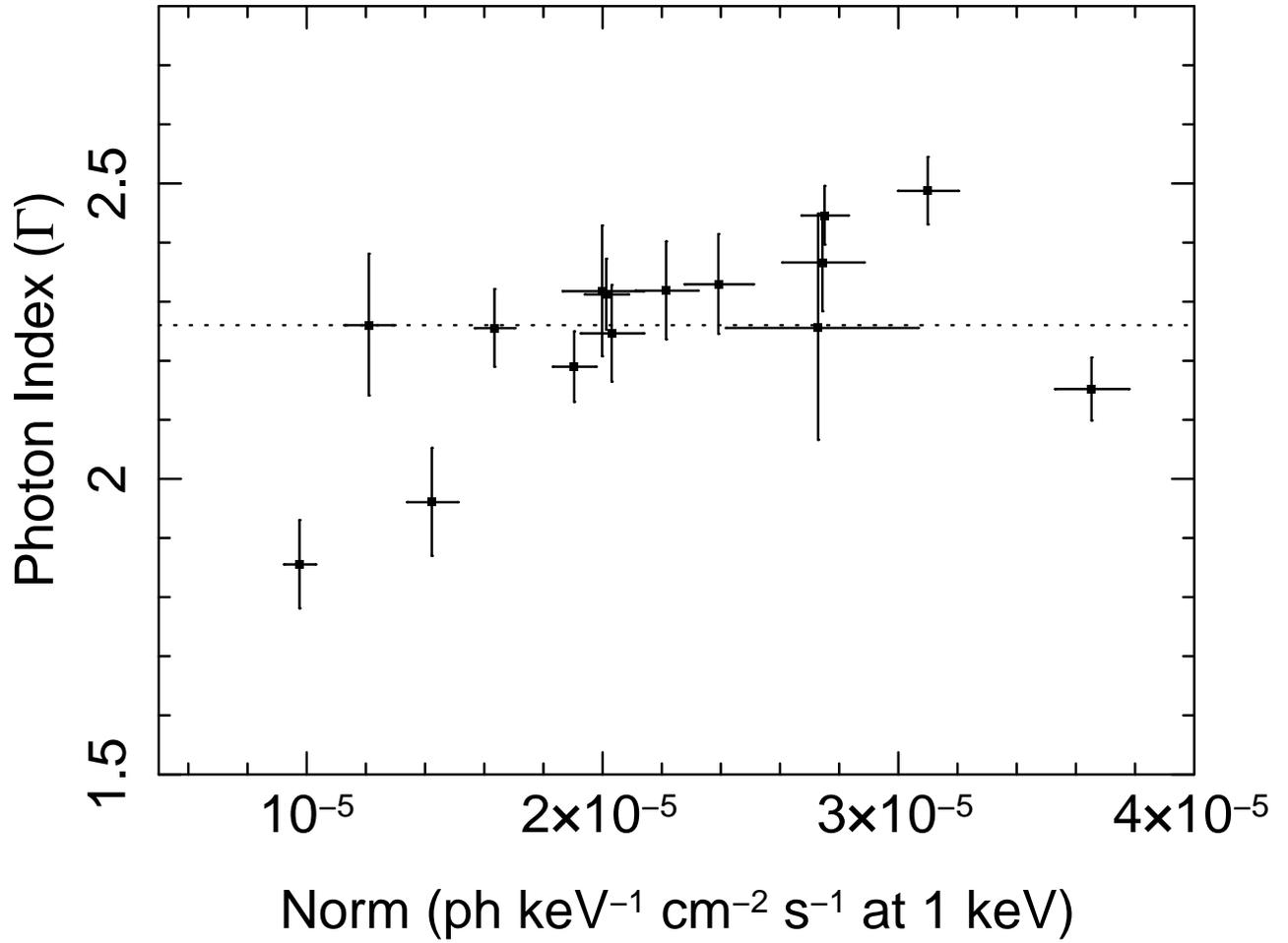}
      \caption{\label{gammaJ1238Fig}Results of fitting a Galactic-absorbed power law to each epoch of the observed-frame (0.5--8~keV) spectra of J1238.  Error bars represent 1$\sigma$ confidence regions.  The median value of $\Gamma$ is shown as a dotted horizontal line.}
   \end{center}
\end{figure}

Figure~\ref{j1238Fig} shows the 4000--5200\AA\ region of the SDSS spectrum of J1238 (in black) with the quasar composite of \citet{v+01} overplotted in red for comparison.  The FWHM of the H$_{\beta}$ and \ion{Mg}{2} $\lambda$2800 (not shown) lines is $\approx$1860~km~s$^{-1}$, while the [\ion{O}{3}] $\lambda$5007 line is quite weak in comparison to the composite value.  J1238 also shows an excess of ionized iron emission in the 4500--4600\AA\ region.  Fitting a Galactic-absorbed power law to the observed-frame 0.5--8~keV region, we find that the spectrum is steeper than typical quasars, with $2.2 < \Gamma < 2.4$ in most epochs (Figure~\ref{gammaJ1238Fig}).  In several epochs, $\Gamma$ deviates from the median fit value of $\Gamma = 2.6$ by 3.3--5.4$\sigma$, including two epochs with flatter spectra ($\Gamma < 2$) at lower flux levels.  All of these traits indicate that J1238 is a quasar analog of the Narrow Line Seyfert~1 (NLS1) population \citep[see, e.g., reviews in][]{pogge00, komossa07}.  NLS1 AGN are often observed to be highly variable in \mbox{X-rays}, and are thought to have smaller black hole masses and correspondingly high accretion rates \citep[e.g.,][and references therein]{bbf96, leighly99, komossa07}.

J1238 was not detected in the VLA FIRST survey, and is classified as radio-quiet in our study.  It was detected as an unresolved source in the deep radio survey of \citet{richards2000}, with a 1.4~GHz flux of $190\pm 13$~$\mu$Jy, corresponding to $\log(R^*) \approx -0.004$.  Although J1238 does show radio emission at faint levels, it is far below the radio-loud limit for our sample.

As a second test for short-term variability, we implemented an algorithm to search for flares or dramatic, short-term absorption (or dimming) events similar to events that have been previously reported at quasar luminosities \citep[e.g.,][]{rgboh91}, or from the Galactic center at much lower luminosities \citep[e.g.,][]{bbbcfgmmrtw01}.  The algorithm breaks each light curve into segments 1000~s wide.  This segment size was chosen to provide a reasonable balance between the number of segments per light curve and the number of counts per segment.  The first stage of our algorithm {\it iteratively} determines the baseline count rate of the light curve, excluding any candidate flare regions.  In each light curve, we identify the segment that shows the most significant deviation from the mean count rate (according to a Poisson statistic), where the latter is calculated using segments that have not already been flagged as potentially variable and are not the current segment under test.  If the deviation is significant at $>$99\% confidence according to a Poisson statistic, we flag that segment as {\it potentially} variable.  Then we iterate the process to find the next-most-variable segment, and so on.  When no more segments can be flagged as potentially variable, we have determined a ``clean'' estimate of the baseline count rate of the source.  Next, the second stage of our algorithm uses this new baseline count rate to determine which segments should be classified as variable at $>$99\% confidence.

Our algorithm identifies hundreds of candidate variable segments out of about 21652 segments, or 0.69~yr of combined light curves.  To evaluate the significance of this result, we re-run our algorithm on {\it simulated} light curves that have the same mean count rates and no intrinsic variability.  For each segment flagged as variable in either the real or simulated light curves, we record the baseline (expected) count rate and the count rate ``fractional deviation'' $F_c$, defined as:
\begin{eqnarray}
F_c &\equiv& (C_{obs} - \langle C\rangle) / \langle C\rangle,
\end{eqnarray}
where $C_{obs}$ is the observed count rate in that segment and $\langle C\rangle$ is the mean count rate for the light curve.

\begin{figure} [ht]
  \begin{center}
      \includegraphics[width=5in, angle=-90]{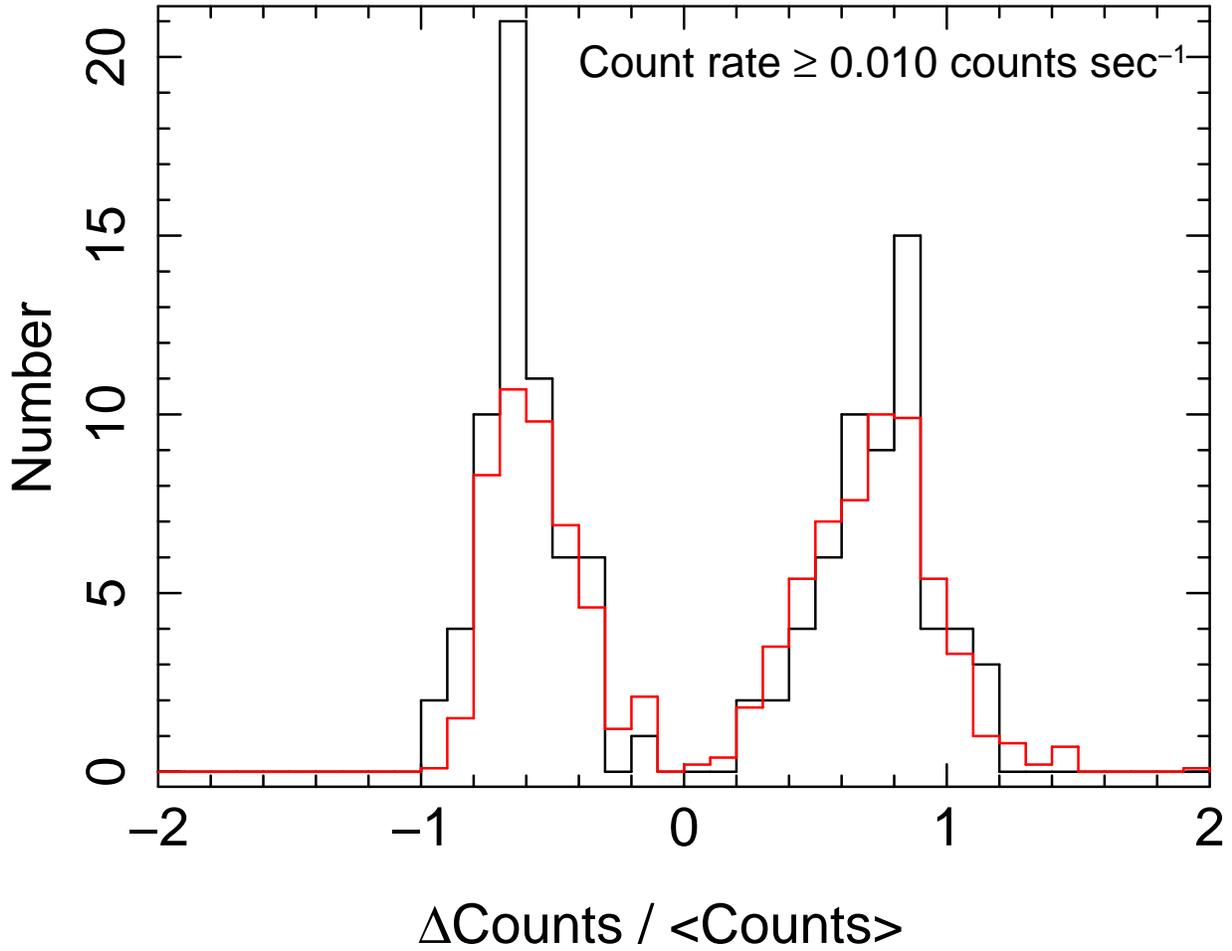}
      \caption{\label{vASELCFlareCountsBrightHistWSimFig}The black histogram shows the distribution of fractional change in counts, $F_c$ for segments flagged as (potentially) variable in light curves having mean count rates $\ge$0.01~counts~s$^{-1}$.  The red curve shows the same distribution for our simulated sources that have no intrinsic variability.  Positive values of $F_c$ correspond to emission flares, while negative values correspond to absorption or dimming events.}
   \end{center}
\end{figure}

The distribution of $F_c$ depends on $\langle C \rangle$, because weaker variability can be detected at higher count rates.  In Figure~\ref{vASELCFlareCountsBrightHistWSimFig}, we show the distribution of $F_c$ for count rates higher than 0.01~counts~s$^{-1}$.  We also consider the case of lower count rates separately (not shown).  For a baseline count rate of 0.01~counts~s$^{-1}$, a significant ``event'' would have a flux increase of at least about 80\%, or a decrease of about 70\%.  For higher baseline count rates, the amplitude would be smaller.  Events in the high-count-rate subsample could therefore be relatively similar to the flare reported by \citet{rgboh91} in the quasar PKS~0558--504, which increased by up to 67\% over 3~minutes and lasted for 10-20 minutes.  Additional rapid flux changes have been observed for this source \citep[e.g.,][]{wmkmn01, bagf04}.  A more extreme example would be PHL~1092, which was observed to brighten by a factor of $\approx$3.8 over a rest-frame time $<$3.6~ks \citep{bbfr99}.  Rapid, strong flares have also been observed in PDS~456 \citep{rovlwspe00}; NLS1-type AGN such as RX~J1702.5+3247 \citep{gblmw01}, IRAS~13224--3809 \citep{bbff97, dbsl02}; and other AGN.

A KS test finds no evidence in most cases of a significant difference in the distribution of $F_c$ between the real and simulated data sets either for the distribution as a whole or for the separate cases of absorption/dimming ($F_c < 0$) and flaring ($F_c > 0$) events.  The difference in distributions is significant (at $>$99.9\% confidence) for the absorption/dimming ($F_c < 0$) events in the low-counts case; presumably this is because our simulations have not modeled the background and therefore do not reach a ``floor'' where the counts in a time segment are dominated by the background.  Based on these results from the KS test, we conclude that the distributions of $F_c$ do not differ remarkably from distributions in the null case of no intrinsic flaring or dimming.

Of course, it may still be the case that we are detecting variable segments of light curves at a higher {\it rate} than predicted from the simulations, even if the {\it distribution} of $F_c$ is not significantly different.  Of the four cases under test (low-count-rate absorption/dimming, low-count-rate flaring, high-count-rate absorption/dimming, high-count-rate flaring), two show event rates that are marginally higher than expected from the simulations.  According to a binomial statistic, the probability of seeing more absorption/dimming events in the high-counts case is 1.4\%, while the probability of observing more flaring events in the low-counts case is 3.5\%.  In each of the four cases, the flare rate is consistent with zero at 99\% confidence.  Upper limits on the rates of 1~ks events (after subtracting off the simulated rates, in numbers per observed-frame year) are $<6.7$~yr$^{-1}$ (low-count-rate absorption/dimming), $<29.3$~yr$^{-1}$ (low-count-rate flaring), $<40.5$~yr$^{-1}$ (high-count-rate absorption/dimming), $<9.0$~yr$^{-1}$ (high-count-rate flaring).  The limits quoted are for 99\% (single-sided) confidence limits.  The upper limits are higher in cases that showed marginal evidence for significant event rates above the simulated values because a higher (but not highly-significant) number of events was detected in these cases.  For the full sample (combining both low and high count rates), we have the following rates:  $<37.3$~yr$^{-1}$ (absorption/dimming) and $<51.5$~yr$^{-1}$ (flaring).  At most a small fraction of time ($\lesssim 0.2$\%) is spent by quasars in the short (ks-time scale) flaring or absorption/dimming states that our algorithm is designed to detect.

The algorithm presented here is only a first empirical attempt to characterize any possible flaring activity in the large sample of archived light curves.  One possible enhancement would be to search for flares that cover longer time scales.  As a preliminary test, we applied our algorithm to search for flares in light curve segments 5~ks long (rather than 1~ks).  (At 0.01~counts~s$^{-1}$, a significant event would require an increase by 36\% or a decrease by 32\%.)  Our results are qualitatively similar, but the upper limits we can place are much weaker due to the smaller numbers of light curve segments.

\subsection{Optical Spectral Properties\label{optSpecSec}}

The SDSS spectra of our sources have a wide range of spectral shapes and signal-to-noise values.  In order to conduct a basic comparison of the optical spectral properties of variable and non-variable sources in the spectral region that contains the H$\beta$ and [\ion{O}{3}]$\lambda$5007 emission lines, we construct composite spectra of all sources with redshifts $0 < z < 0.8$ and $\ge$50 source counts per epoch, on average.  We have selected this region of spectrum because the H$\beta$ and [\ion{O}{3}] emission lines are commonly used to estimate black hole masses and accretion rates, and to identify subclasses of AGN such as Narrow Line Seyfert~1s.  The minimum count rate requirement ensures that we can detect variability at the $\gtrsim$50\% level at $\approx 3\sigma$ confidence (\S\ref{sensitivityToVarSec}).

The composite spectra are constructed by calculating the geometric mean of the spectra in a given (rest-frame) wavelength bin.  The input spectra are normalized to 1 at 5100~\AA\ in each case; note that the output spectrum depends somewhat on the choice of normalization wavelength.  We chose to normalize at 5100~\AA\ because it is a commonly-used spectral region that is less affected by complex emission or absorption features.  We omitted from the calculation any radio-loud quasars or known hosts of BAL outflows.  The variable composite was derived from SDSS spectra of 22~sources, while the non-variable composite represents 15~sources.

\begin{figure} [ht]
  \begin{center}
      \includegraphics[width=5in, angle=-90]{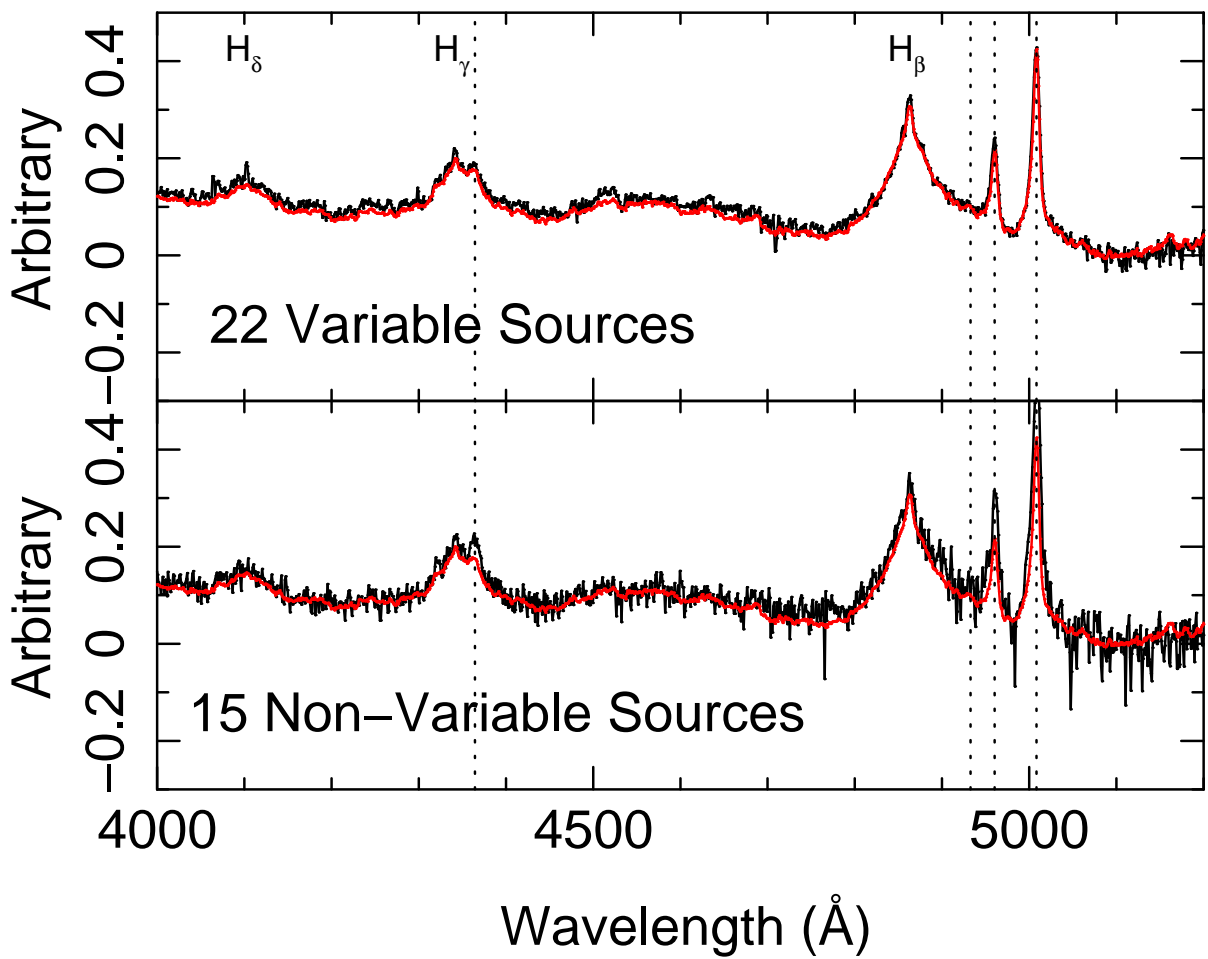}
      \caption{\label{vAMGMSVNV5000Fig}Composite SDSS spectra (plotted in black) for radio-quiet, non-BAL sources with redshifts $0 < z < 0.8$.  The $y$-axis corresponds to the logarithm of the composite spectrum, normalized to 1 at 5100~\AA.  The SDSS quasar composite from \citet{v+01} is overplotted in red for comparison.  The top panel shows the composite (in black) for sources identified as having at least one variable \mbox{X-ray} epoch.  The bottom panel shows the composite (in black) for non-variable sources.  Vertical dotted lines indicate the wavelengths of [\ion{O}{3}] lines, while prominent Balmer lines are labeled in the top panel.}
   \end{center}
\end{figure}

In Figure~\ref{vAMGMSVNV5000Fig}, we plot the composite spectra in black for variable (top panel) and non-variable (bottom panel) sources.  As before, ``variable'' sources are those for which the \mbox{X-ray} count rate is inconsistent with a constant value for at least one epoch.  For comparison, we have plotted a renormalized SDSS quasar composite spectrum from \citet{v+01} in red in each panel.

We find at most mild differences in the composite spectra for the H$\beta$~$\lambda$4861 line or the broad pattern of ionized Fe emission in the plotted region.  However, [\ion{O}{3}] emission for non-variable sources is noticeably stronger than typical in three lines (4364, 4960, and 5007~\AA) where [\ion{O}{3}] is clearly present in emission.  Visual inspection of the individual spectra supports this observation, although there is much diversity.

\begin{figure} [ht]
  \begin{center}
      \includegraphics[width=5in, angle=-90]{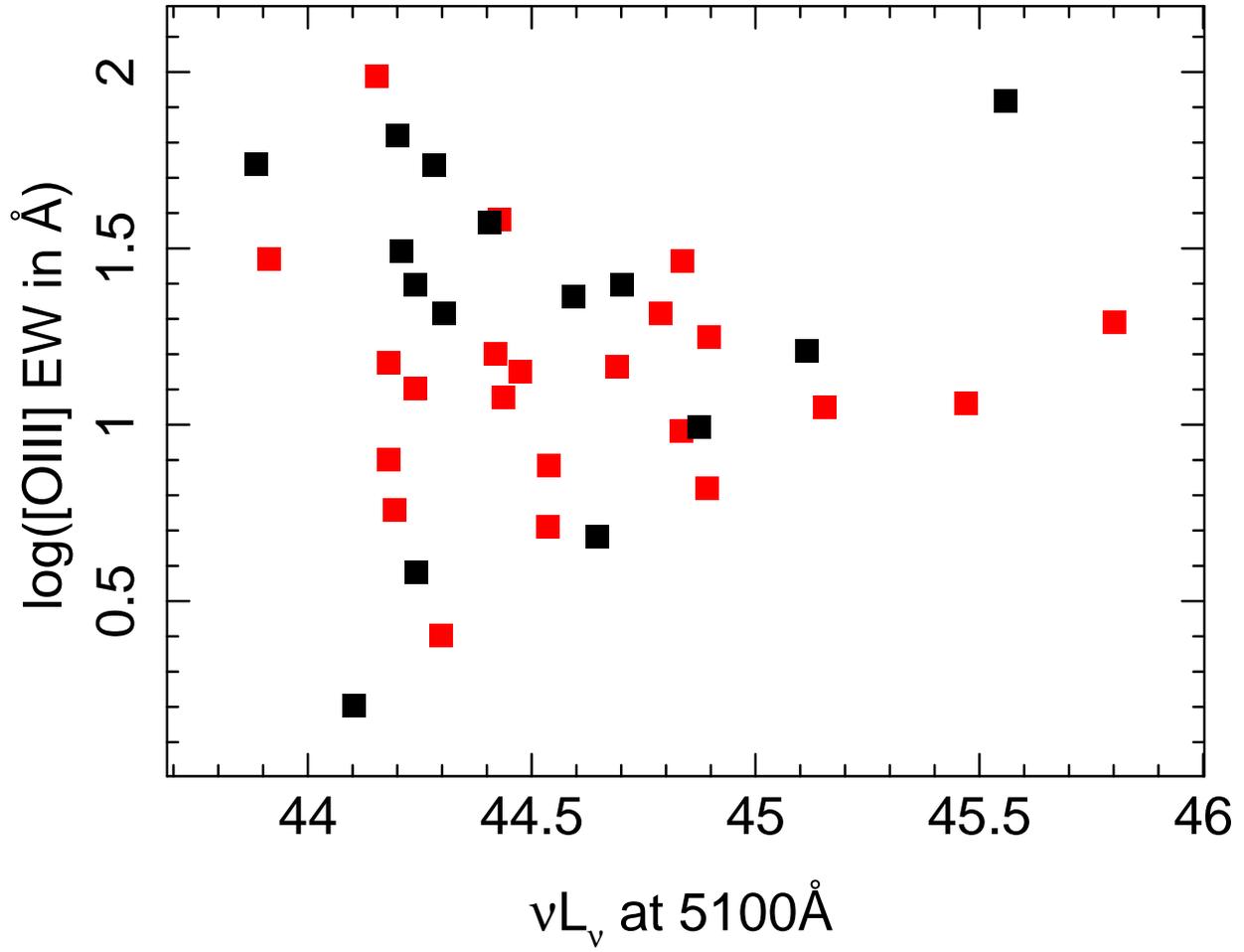}
      \caption{\label{vAMGMSOIIIEWVsL5100Fig}Estimated [\ion{O}{3}] equivalent widths as a function of monochromatic luminosity at 5100~\AA\ for sources classified as variable (red) and non-variable (black).}
   \end{center}
\end{figure}

To investigate this trend further, we estimated the monochromatic luminosity $\nu L_{\nu}$ at 5100~\AA\ and also the equivalent width (EW) of the [\ion{O}{3}]~$\lambda$5007 emission line for each spectrum.  We used a simple linear continuum fit across the H$\beta$ and [\ion{O}{3}] emission line region to estimate the EW.  No attempt was made to deblend ionized Fe emission; although we note that the composite spectra do not indicate a strong difference in Fe emission between the two subsamples, in general.  Figure~\ref{vAMGMSOIIIEWVsL5100Fig} shows the results of these calculations.  While both subsamples (variable and non-variable) cover a similar range of luminosities and EWs, the non-variable sources are more likely to occupy the high-EW portion of the plot, with $\log(EW) > 1.3$.  A KS test finds no evidence of a difference in the luminosity distributions, but a suggestive probability (97\% confidence) that the EW distributions differ.

Because we have limited this study to sources with at least 50 counts per epoch on average, our classifications of ``variable'' and ``non-variable'' are very similar to placing a cut on the amplitude of variability.  Still, the individual observations do have different sensitivities, so the two concepts are not exactly similar.  In order to investigate [\ion{O}{3}] strength as a function of variability amplitude, we define
\begin{eqnarray}
S^{\prime}_j &\equiv& \max_i(|S_{ij}|)\label{sPrimeEqn},
\end{eqnarray}
where $S_{ij}$ is defined in Equation~\ref{sijDefEqn} and the maximum is taken over all epochs $i$ for source $j$.  $S^\prime_j$ represents the maximum amplitude of variation (in units of $\sigma$) for each of the 37 sources we are investigating.  In fact, we find that sources with larger values of $S^{\prime}_j$ are generally the sources we flagged as variable; the two criteria are nearly equivalent.  If we cut the sample at the median value $S^{\prime}_j = 2.8$, then a KS test indicates that the [\ion{O}{3}] distributions of the two subsamples are incompatible at 96\% confidence.  If we place the cut closer to the division between the ``variable'' and ``non-variable'' sources, at $S^{\prime}_j = 1.5$ to $1.9$, then the KS test indicates a sample difference at $\gtrsim$99.5\% confidence.  (We find similar results, though less significant by a few percent, using a definition in which variability amplitude is defined as $\equiv \max_i(|(f_{ij} - c_j) / c_j|)$, with the best-fit count rate in the denominator instead of the measurement error.)  As before, we find that there are some indications of a real effect in the data, but larger samples are needed to test this result more sensitively.

It is also possible that some other effects in the data are influencing our ``variable'' and ``non-variable'' subsamples.  If we had enough sources, it would be ideal to control for factors such as the number of observations of a given source and the time scales over which the sources were observed.  With our small sample, this is difficult to do.  Although our variability criterion is not associated with a specific time scale (but is simply compared to the best-fit constant rate), we can investigate what happens if we remove ten cases of ``variable'' sources that were observed more than two times, under the extreme assumption that they would not have been flagged as variable from their first two observations alone.  In this case, the split in [\ion{O}{3}] EWs becomes less significant (with a 19\% probability of greater difference in EW distributions), but this is expected due to the fact that the remaining sample is small.  Effectively, this cut has removed many ``variable'' sources without changing the cluster of non-variable sources having stronger-than-average [\ion{O}{3}] emission.

[\ion{O}{3}] emission and \mbox{X-ray} variation are two effects that are associated with the well-known ``Eigenvector~1'' description of AGN.  Our current study has been partially motivated by a desire to determine if and how such factors are actually related, and what the physical significance of such a relation would be.  We discuss this issue further in \S\ref{discOpticalSec}.

\section{DISCUSSION}\label{discussionSec}

\subsection{Chandra Sensitivity and Quasar Variation\label{chandraSensSec}}

{\it Chandra}'s ability to detect significant \mbox{X-ray} variation depends strongly on the number of counts obtained for a source.  Figure~\ref{fracVarAndDetectVsMeanCtsFig} demonstrates the sensitivity of {\it Chandra} to variation in archived observations of SDSS spectroscopic quasars.  A minimum of $\sim$10 counts per epoch are needed before variability can be detected at the 100\% level (corresponding to a source doubling or halving its count rate compared to the best-fit constant rate).  For sources with fewer than 20--30 counts per epoch, we see minimal evidence of variability.

Overall, variability is detected in $\approx$30\% of the sample, including sources with low numbers of counts.  For sources with larger numbers of counts, variability can be measured with greater precision.  Eleven of 19 sources having $>$500 mean counts are detected as variable at the $\gtrsim$13\% level, and 7~of 10~sources having $>$1000 mean counts are detected as variable at the $\gtrsim$10\% level.  Based on this small sample, we expect that \mbox{X-ray} variability at the $> 10$\% level would be observed in most, if not all, quasars with sufficient numbers of counts.  \citet{psgkg04} likewise detected variablity in over half the AGN in their analysis of the {\it Chandra} Deep Field--South, and estimated that intrinsic variability could be present in a fraction approaching 90\% of their sources.

The level of fractional variation ($\sigma(F) \approx$ 16\%) we observe certainly contributes to the scatter in observed \mbox{X-ray-to-optical} flux ratios for non-BAL SDSS quasars.  This amount of variation is smaller than that required to explain the scatter in \mbox{X-ray-to-optical} spectral slopes as due to variability alone \citep{sbsvv05, ssbaklsv06, gbs08}, at least over time scales of $\lesssim$2~yr.  Using simultaneous UV/\mbox{X-ray} observations from {\it XMM-Newton}, \citet{vtta10} have also found that the observed scatter in \mbox{X-ray-to-optical} flux ratios cannot be explained by variability alone.  Some physical differences in quasars must also contribute to an intrinsic scatter in \mbox{X-ray-to-optical} flux ratios.

On time scales shorter than $\Delta t_{sys} \sim 2\times 10^5$~s, we do not detect quasar variability.  At longer time scales, the amount of intrinsic variability is roughly constant or perhaps slowly increasing.  If we assume that $5\times 10^5$~s represents a ``break'' in the quasar luminosity structure function, an assumed linear relationship purely between the break time scale ($T_B$) and $M_{BH}$ would underestimate $M_{BH}$ in our sources by an order of magnitude or more \citep[e.g.,][]{mgug05}.  For luminous quasars, an additional term dependent on bolometric luminosity \citep[e.g.,][]{mkkuf06} may be important to increase the predicted black hole mass for high-$M_{BH}$, high-luminosity quasars.

\subsection{Variability at Higher Redshifts\label{varHigherzSec}}

Although earlier studies have suggested that variability increases at higher redshifts \citep{alsebggsg00, mal02, psgkg04}, we have not been able to confirm that result in our data (\S\ref{excessVarianceSec}, \S\ref{varDepOnPhysPropSec}).  In fact, any increase in excess variance at $z > 2$ would be above the $\approx 2\sigma$ upper limit on the excess variance we actually measure (\S\ref{excessVarianceSec}).  We also do not detect any increase in variation at high black hole masses.  Increased variability at higher redshifts was not detected even when we cut the sample to remove sources with lower numbers of counts, thus maximizing the sensitivity to any variability.  Looking forward, dedicated \mbox{X-ray} observations targeting higher-redshift quasars that are matched to lower-redshift quasars in other sample properties would be ideal for a definitive test of redshift-dependence.

We stress that our approach has been to look for an empirical signature of variation using the data and observational bandpass that are available to us.  Some earlier studies (A00, M02) have generated correction factors for their measured variance that model the effects of sampling sources on different time scales.  These corrections rely on an assumed shape of a quasar power spectrum to boost the observed variation of sources observed over shorter times, as quasars tend to vary more over longer time scales.  However, several factors argue against attempting such corrections to our study.  It is not clear what the shape of the power spectrum should be over the wide range of time scales in our data, especially since many of our data points sample time scales beyond the power spectrum break (see Figure~\ref{sampledtVsOldFig}).  We know even less about this region of the power spectrum for quasars than we do about the slope of the quasar power spectrum on shorter time scales.  Furthermore, some statistics in our study (such as $S_{ij}$) do not have a well-defined time scale associated with them.  For the case of the $\sigma_{EV}$ statistic, which we carefully constructed to depend on only one time scale per source, we did not find evidence for increased variability at higher redshifts even when we limited the range of system-frame or observed-frame time scales (\S\ref{excessVarianceSec}).

Applying a time dilation correction factor of $(1+z)^{1/4}$ (A00) would have only a small (40\%) effect even at $z=3$, and would still not result in a detection of ``increased variability'' at higher redshifts in our data.  Similarly, the study of P04 notes that ``no correction for time dilation has been applied to this data since it would require us to make a somewhat arbitrary assumption about the power density spectrum (PDS) of our sources....''  That assumption becomes even more arbitrary for the longer time scales in our data.  There is, of course, some danger that if we mis-characterise the correction for our sample, we will be {\it creating} a redshift-dependent trend in the data.  For these reasons, we prefer to leave our results in terms of observational, rather than modeled, constraints.  Using the data we provide, readers can straightforwardly test the effects of modeling different correction factors.

Another topic that has received somewhat less attention is the possibility that the variability signal could decrease with redshift because different spectral regions are being shifted across the {\it Chandra} bandpass.  We have performed a basic study of how spectral variation may change as a function of photon energy.  In our case (\S\ref{fluxFluxSec} and \S\ref{specVarSec}), we found some evidence that hard spectral regions may show increasing influence of a constant component that shifts at higher redshifts into an energy range where {\it Chandra} is most sensitive.  At our current stage of understanding, attempting to correct for this effect would introduce systematic uncertainty, and we prefer to stay with our empirical approach.

Increased variability at higher redshifts would have intriguing physical implications, although we note that it is not obviously expected in the currently-favored ``cosmic downsizing'' picture for SMBH growth \citep[e.g.,][and references therein]{cbbbg03, mrghms04, bh05}.  In this picture, we observe multiple largely-independent generations of growing SMBHs as we look back in cosmic time, with the more massive SMBHs generally growing earlier.  At a fixed high luminosity, we then do not expect to see SMBH masses dropping (and thus perhaps variability strength rising) toward higher redshifts.

\subsection{Spectral Variation\label{specVarSec}}

Quasar variation appears similar to that observed for lower-luminosity Seyfert AGN in several respects.  The \mbox{X-ray} spectra of quasars vary in shape as well as brightness.  In general, the spectrum of a given source is flatter when that source is in a fainter state, and the spectrum steepens as the source brightens.  The change in $HR$ is anti-correlated with the change in luminosity (Figure~\ref{vAPHRdHRVsFracLumFig}), demonstrating the steepening-brightening trend.  Direct spectral fits for multiple epochs of high-S/N sources (Figure~\ref{vAEVPdGammaVsGammaTruncatedFig}) demonstrate a strong anti-correlation between $\Delta\Gamma$ and $\Gamma$ that tends to move spectra toward $\Gamma \sim 2$.

The power law spectral index (derived from our simple fit model) may vary significantly for a given source.  Based on our fits to sources with large numbers of counts, intrinsic variation by $\Delta\Gamma \sim 0.1-0.2$ is common.  Some caution is therefore warranted when using the spectral slope as an indicator of quasar physical properties such as black hole mass or accretion rate, as the spectral slope (measured according to our simple model) can vary with \mbox{X-ray} luminosity over time.  Of course, the variation mechanism may contribute to a relationship between $\Gamma$ and accretion rate ($L / L_{Edd}$, where $L_{Edd}$ is the Eddington rate) that has been found in single-epoch observations of quasar ensembles \citep[e.g.,][]{sbnmk08, rye09}.  For the epochs in Figure~\ref{vAEVPdGammaVsdNormTruncatedFig}, we measure a best-fit relationship:
\begin{eqnarray}
\Delta\Gamma &=& (0.534 \pm 0.104) \log(N_1 / N_0) - (0.039 \pm 0.012)\label{dGammaVsNormRatioEqn},
\end{eqnarray}
where $N_1 / N_0$ is the ratio of power law normalizations between the two epochs, and $\Delta\Gamma \equiv \Gamma_1 - \Gamma_0$ is the change in $\Gamma$.  (We note that this fit may not adequately describe the properties of individual sources, as we are simply assuming that one model describes the variation in all epochs.  We do not have sufficient data to fit each source individually.)  The slope in Equation~\ref{dGammaVsNormRatioEqn} is a bit steeper than the slope ($0.31 \pm 0.01$) in Equation~(1) of \citet{sbnmk08} that related $\Gamma$ to $L / L_{Edd}$ for individual AGN.  The overall effect may be similar to the ``intrinsic'' and ``global'' Baldwin effects \citep[e.g.,][]{b77, krk90, pp92}, in that the variation in an individual object may be more extreme than the overall trend observed across single-epoch measurements of a large number of sources.  At the least, intrinsic variation contributes to scatter observed in the sample of sources.

Simple absorption models did not adequately describe the observed anti-correlation between spectral hardness and luminosity, although we cannot rule out the possibility of more complex, multi-parameter absorption models.  However,  a two component emission model did reproduce the trend reasonably well without fine-tuning.  The emission model consists of a variable power law with $\Gamma = 2$ and a hard ($\Gamma = 1.3$), constant component normalized to $\sim 20$\% of the variable power law flux at 1~keV.  Previously, Pao04 suggested that the increasing contribution of a flatter, reflected component may result in lower variability levels for sources with ``absorbed,'' or harder, spectra.  In our data, the relationship between hard- and soft-band \mbox{X-ray} fluxes, extrapolated to zero soft-band flux, also suggests that a constant, hard spectral component is present in our ensemble of sources.  Similar models have been used to describe flux variation in Seyfert AGN such as MCG~--6--30--15 \citep[e.g.,][]{tum03, vf04}.

If a typical quasar spectrum includes a non-negligible constant component that is flatter than the dominant component with power law index $\Gamma = 2$, we might expect that this component changes the observed spectral shape for higher-redshift quasars.  There has been disagreement over whether spectral shapes evolve with redshift, but so far strong evidence for evolution has not been discovered.  \citet{sbvsfrs05} find that the spectral slope for a joint fit to 10 of the most luminous quasars at $4 < z < 6.3$ is $\Gamma \approx 2$, similar to that of lower-redshift quasars.  However, this result applies to the {\it rest}-frame 2--10~keV range, and not necessarily to a harder component that becomes more evident at higher energies.  \citet{vbsk05} report a slightly harder ($\Gamma = 1.9^{+0.10}_{-0.09}$) {\it observed}-frame photon index for a joint fit of 48 quasars detected in \mbox{X-rays} at $4.0 < z < 6.3$.  \citet{garbchkkkvamkmswt09} find no evidence for spectral evolution in a large sample of optically-selected quasars out to $z < 5.4$.  However, \citet{bsscjhaed03} previously observed spectral flattening in a sample of 17 optically-selected quasars at $3.7 < z < 6.3$, and \citet{kbs09} observed marginally-significant spectral flattening with redshift out to $z \sim 4.7$.

To determine how a two-component model would affect the observed spectrum at high redshift, we randomly generated spectra for 9 {\it Chandra} sources at $z = 4.2$.  We simulated each observation from a model with two power laws having $\Gamma = 2$ and $\Gamma = 1.3$.  The flatter power law was normalized to 20\% of the steeper component at rest-frame 1~keV; it starts to rise above the steeper component at rest-frame $10 / (1+z)$~keV.  Then we fit each spectrum with a power law model to determine typical measured photon indices, finding an average value of $\Gamma \approx 1.72 \pm 0.17$.

For comparison, we fit Galactic-absorbed power law models to higher-energy regions of spectra for quasars at $z > 4$.  The quasars were described in \citet{sbvsfrs05}, with spectra kindly provided to us by O.~Shemmer.  We fit the {\it observed}-frame energy range $E_0$--$E_1$~keV, where $E_0 \equiv 10/(1+z)$~keV and $E_1$ varied, in order to characterize the region where a hard component may dominate (in our simple model).  We fit the spectra of all three EPIC cameras simultaneously, requiring them to have the same photon index, but allowing slightly different normalizations to account for cross-calibration differences between instruments.  In all cases, the photon indices for the observed frame spectra were flatter than for the rest-frame 2--10~keV region.  For a fit range extending up to $E_1 = 5$~keV in the observed frame, the spectrum flattened by $\Delta \Gamma = 0.1$ to $0.9$.  For higher values of $E_1$, the spectrum appeared to flatten more, although background contamination becomes higher at these energies as well.  This result suggests that a hard spectral component is becoming more evident at high energies.  Although inspection indicates that backgrounds are generally well below count rates in our fit region, it would be useful to obtain larger, high-quality samples to examine the shape of the high-energy quasar spectrum more reliably.

In addition to the hypothesis of a constant harder component in quasar spectra, the assumption (based on studies of local Seyfert AGN) that the relationship between soft and hard \mbox{X-ray} count rates can be extrapolated {\it linearly} should also be tested with larger, high-quality data sets.  The two-power-law model we use here is the simplest attempt to explain the putative constant component, but more sophisticated models may be required that cut off at higher energies if the flatter component is not evident in high-redshift spectra.  Our results also suggest that studies of high-redshift populations should test multi-component models against individual sources in their data to constrain spectral complexity beyond a single power law shape.  The spectral shape measured from a stack or joint fit can be biased toward the brighter sources in a sample, so care should ideally be taken to account for selection effects and brightening-steepening behavior as sources vary.

\subsection{[\ion{O}{3}] Emission Strength\label{discOpticalSec}}

In our sample, sources classified as \mbox{X-ray} non-variable have [\ion{O}{3}]~$\lambda$5007\AA\ emission lines that are about 12\AA\ stronger (on average) than those of variable sources (\S\ref{optSpecSec}).  This was the most prominent effect we found when comparing composite spectra of variable and non-variable AGN.  In the {\it optical} bandpass, variability has been observed to be related to both [\ion{O}{3}] and H$_{\beta}$ emission.  Quasars with stronger [\ion{O}{3}] emission have {\it larger} optical variability amplitudes \citep{gmkns99, mww09}, although with signficant scatter.  The cause of this relation is not known; it has been suggested that the effect may be due to an enhanced ionization rate from a variable ionizing continuum \citep{gmkns99}, or perhaps associated with accretion instabilities or star formation \citep{mww09}.  However, in the \mbox{X-rays}, we see the opposite effect:  stronger [\ion{O}{3}] emission is associated with lower levels of \mbox{X-ray} variability.

The effect may be anomalous.  The set of sources for which we can sensitively measure [\ion{O}{3}] emission and \mbox{X-ray} variability is relatively small, and the effect is not highly significant.  However, we do not see a significant difference (according to a KS test) between the distributions of monochromatic luminosity ($\nu L_{\nu}$ at $5100$\AA), \mbox{X-ray} count-rate luminosity, or mean numbers of \mbox{X-ray} counts for the variable and non-variable sources.  There may be other factors differentiating the two subsamples, or the effect may be a random deviation in our small sample.

In our search for short term variability on kilosecond time scales (\S\ref{intraObsVarSec}), we identified an unusual quasar, SDSS J$1238$, that was flagged as variable in multiple epochs.  This quasar is NLS1-like, with atypically weak [\ion{O}{3}] $\lambda$5007\AA\ emission, a H$_{\beta} \lambda$4862 FWHM $\approx$1860~km~s$^{-1}$, and stronger ionized Fe emission compared to the average SDSS quasar.  A tendency for rapid, high-amplitude \mbox{X-ray} variation is a well-known property of NLS1 AGN \citep[e.g.,][]{bbf96, pogge00, komossa07}.

With EWs from a few to $\sim$100\AA, the [\ion{O}{3}]~$\lambda$5007\AA\ lines in our study span the range of EWs originally used by \citet{bg92} to define ``Eigenvector~1.''  This empirically-based collection of properties includes an anticorrelation between \ion{Fe}{2} and [\ion{O}{3}] emission strengths; it also includes trends with radio-loudness and some H$_{\beta}$ line features such as FWHM and asymmetry.  NLS1 AGN fall at one end of this eigenvector, generally having narrow H$_{\beta}$ lines, weak [\ion{O}{3}] emission, and strong ionized Fe emission.  The quasar J1238 also shows similar properties when compared to typical SDSS quasars.  In our sample, \mbox{X-ray} ``non-variable'' quasars may exemplify the (radio-quiet) population at the opposite (strong-[\ion{O}{3}]) end of Eigenvector~1, while quasars showing moderate \mbox{X-ray} variability have [\ion{O}{3}] emission levels more typical of ordinary SDSS quasars.\footnote{On the long timescales we generally sample, it is not immediately clear how \mbox{X-ray} variability should relate to Eigenvector~1.  For example, \citet{maupsbp07} find a break in the power spectral density (PSD) function at $\approx 10^{-6}$~Hz for the NLS1 Ark~564, below which its PSD drops significantly.  However, the black hole masses for our sample are generally much higher than those for which PSD studies have been performed (Figure~\ref{varArchPlotVsSeyfertsFig}).  We therefore expect any PSD breaks for quasar-luminosity NLS1 analogs in our sample to be at lower frequencies of perhaps $\sim 10^{-8}$~Hz, which corresponds to longer timescales than we adequately sample in this work.}  The sample of strong-[\ion{O}{3}] emitters that have been observed multiple times in \mbox{X-rays} should be expanded in order to test the potential relation between \mbox{X-ray} variation and Eigenvector~1 more sensitively.  Some new sources can be added from the archives as \mbox{X-ray} and optical surveys progress, but targeted observations would be most effective to obtain a significant ensemble of the strongest-[\ion{O}{3}] emitters.

A link between \mbox{X-ray} temporal properties and [\ion{O}{3}] emission would be interesting because it would represent another physical connection between small-scale (disk corona) and large-scale (NLR) AGN physics \citep[e.g.,][]{bb98}.  (Of course, with larger data sets we could also test more sensitively whether additional parameters, such as Balmer line profiles, are related to \mbox{X-ray} variation.)  In one category of models, a third parameter modulates both \mbox{X-ray} and NLR emission, creating an {\it indirect} relation between the two.  For example, [\ion{O}{3}] emission strength is believed to be dominated by geometric factors \citep[e.g.,][]{bl05}.  The Eddington ratio ($L/L_{Edd}$) of bolometric to Eddington luminosity could control both the obscuration of \mbox{X-ray}/UV radiation essential for NLR ionization \citep[e.g.,][]{acn80, bg92, cqm07} and also \mbox{X-ray} emission properties.  Alternatively, some physical effect could create a more {\it direct} connection between the corona and NLR.  Very large NLRs that produce the strongest [\ion{O}{3}] emission lines may require replenishment \citep{nsmoccd04}, and small-scale jets have been observed to interact with the NLRs of {\it radio-quiet} Seyfert AGN \citep[e.g.][]{fws98, hp01, lfbh06, wfermkz11}.  Jet activity has been associated with other Eigenvector~1 properties, including radio loudness and \mbox{X-ray} spectral slope, and radio-loudness (indicative of jet activity) is associated with decreased variability in our subsample of {\it radio-loud} quasars (\S\ref{fracVarTimeSec}).

\section{SUMMARY AND CONCLUSIONS}\label{concSec}

In this study, we have examined archived {\it Chandra} \mbox{X-ray} observations of 264~SDSS spectroscopic quasars to search for \mbox{X-ray} variability, characterize it, and test whether this variability is related to other quasar properties.  Our findings include the following:

\begin{enumerate}
\item{We find strong evidence of \mbox{X-ray} variation in $\approx$30\% of the quasars in our sample overall.  Our sensitivity to variation increases with the number of source counts; 70\% of sources with $\ge$1000 counts per epoch are detected as variable.}
\item{Quasars in our sample typically vary with a standard deviation of fractional variation of $\approx$16\%.  This amount of variation is not large enough to explain the scatter in \mbox{X-ray-to-optical} ratios as being due to variation alone.  On time scales shorter than a few $\times 10^5$~s, the ensemble variability falls below our detection limit.  Coupled with the flatter trend of variability on longer time scales, this suggests a ``break'' in the trend at $\sim(2-5)\times 10^5$~s.}
\item{We find no evidence that higher-redshift quasars are more variable than lower-redshift quasars, as has been suggested in previous studies.  However, this analysis is complicated by the fact that we do not have many quasars at high redshift with large numbers of counts ($\gtrsim$100) to enable sensitive tests.}
\item{The \mbox{X-ray} spectra of quasars tend to be flatter when fainter and steeper when brighter, as is seen in the case of some local Seyfert AGN.  We were able to reproduce this trend with a simple, two-parameter power-law model that has been used to describe Seyfert variability.  Spectral fits to bright sources show an anti-correlation between $\Delta\Gamma$ and $\Gamma$; quasar spectra tend to flatten or steepen as necessary to bring them back to $\Gamma \sim 2$ as they vary.}
\item{As soft-band count rates are extrapolated to zero, a significant hard-band flux remains.  This suggests that quasar spectra have an underlying constant, hard spectral component, following the model proposed for some Seyfert AGN.  The constant fraction of the hard-band count rate (measured in observed-frame bands) likely increases with redshift as different segments of the constant and variable spectral components shift through the bandpass.}
\item{A search for intra-observation variation on time scales of 1~ks revealed one unusual source, J1238, with strong, short-term variability.  The optical spectrum of J1238 shows it is an NLS1-type object.  For the full sample, we constrain the rates of significant variation in 1~ks bins to be $<37.3$~yr$^{-1}$ (absorption/dimming) and $<51.5$~yr$^{-1}$ (emission), using observed-frame years.}
\item{We generated upper limits on the rate of observations showing at least a magnitude $F$ of fractional variation, where an ``observation'' is representative of the epochs in Sample~HQ.  As examples, $|F| > 100$\% is rare, occuring $\lesssim$4\% of the time.  $|F| \ge 25$\% occurs in fewer than 25\% of observations.}
\item{Median spectra suggest that sources with higher (detectable) levels of \mbox{X-ray} variability have weaker [\ion{O}{3}] emission.  Additional data are required to confirm the relation between Eigenvector~1 properties such as [\ion{O}{3}] strength and \mbox{X-ray} variability, and (more generally) to test how this phenomenon could connect small-scale (corona) and large-scale (NLR) AGN structures.}
\end{enumerate}

The sample of serendipitously-observed quasars continues to expand for variability studies.  The {\it Chandra} archive and spectroscopic quasar catalogs continue to grow over time.  Incorporating {\it ROSAT} PSPC observations would extend the observed-frame time baseline to $>$20~yr, while archived {\it XMM-Newton} data can greatly increase sample sizes and provide simultaneous optical/UV monitoring.  {\it eROSITA}, scheduled to launch within the next two years, will provide a sensitive new survey scanning the \mbox{X-ray} sky multiple times, with each scan taking about 6~months to complete \citep{cappelluti+2011}.  The SDSS-III Baryon Oscillation Spectroscopic Survey (BOSS) quasar survey \citep[e.g.,][]{ross+2011} is more than doubling the current SDSS spectroscopic quasar sample, with most of the new quasars having redshifts $z > 2.2$.  Although these sources will generally be fainter, they will provide much-needed leverage for studies of accretion evolution over cosmic time.  In addition to the sorts of analyses we have conducted in this paper, these expanded data sets may permit us to place detailed constraints on the quasar variability power spectrum by comparing the data to light curves that are simulated from different PSD models.

New AGN are increasingly being identified by temporal properties, permitting current and future surveys to go beyond the quasar realm to consider AGN that are blended with host galaxy emission.  However, there is a great need for new data-mining and statistical techniques that will appropriately characterize the properties of fainter sources in these new surveys, while accounting for instrumental cross-calibration and perhaps selection biases in the samples.



\acknowledgements

We gratefully acknowledge support from NASA {\it Chandra} grant AR9-0015X (RRG), NASA ADP grant NNX10AC99G (WNB), and NASA grant NNX09AP83G (WNB).  We thank I.~McHardy, O.~Shemmer, and P.~Uttley for helpful discussions during the preparation of this paper.  We thank O.~Almaini and M.~Paolillo for providing data used to construct Figure~\ref{sampleVsOldFig}, and O.~Shemmer for providing spectra of high-redshift quasars.  We also thank Meagan Albright and Jolene Tanner for their contributions to this study as part of the University of Washington Pre-MAP program.





\bibliographystyle{apj3}
\bibliography{apj-jour,bibliography}

\begin{deluxetable}{llcccrrcrc}
\tabletypesize{\scriptsize}
\tablecolumns{10}
\tablewidth{0pc}
\tablecaption{Properties Common to Each Source\label{sourceTable}}
\tablehead{\colhead{SDSS Name} & \colhead{$z$} & \colhead{Number} & \colhead{Known\tablenotemark{a}} & \colhead{Known} & \colhead{$F_{2500\mathring{A}}$\tablenotemark{b}} & \colhead{Const Rate} & \colhead{Is} & \colhead{Const Rate HQ} & \colhead{Is} \\ \colhead{(J2000)} & \colhead{} & \colhead{of Obs} & \colhead{RL} & \colhead{BAL} & \colhead{10$^{-27}$ erg} & \colhead{10$^{-6}$ cts} & \colhead{Var} & \colhead{10$^{-6}$ cts} & \colhead{Var} \\ \colhead{} & \colhead{} & \colhead{} & \colhead{} & \colhead{} & \colhead{cm$^{-2}$ s$^{-1}$ Hz$^{-1}$} & \colhead{cm$^{-2}$ s$^{-1}$} & \colhead{} & \colhead{cm$^{-2}$ s$^{-1}$} & \colhead{HQ}}
\startdata
$000622.60-000424.4$ & $1.038$ & $2$ & $1$ & 0 & $0.444$ & $219.874\pm16.620$ & 0 &  & \\
$004054.65-091526.7$ & $4.976$ & $2$ & $*$ & 0 & $1.584$ & $3.963\pm2.731$ & 0 &  & \\
$011513.16+002013.1$ & $2.119$ & $2$ & $*$ & 0 & $0.587$ & $3.226\pm1.189$ & 0 & $0.323\pm0.119$ & 0\\
$014219.01+132746.5$ & $0.267$ & $2$ & $1$ & 0 & $1.453$ & $2.618\pm10.106$ & 0 &  & \\
$014320.96+132429.7$ & $1.739$ & $2$ & $*$ & 1 & $0.525$ & $-1.975\pm8.862$ & 0 &  & \\
$015254.04+010434.6$ & $0.570$ & $3$ & $*$ & 0 & $0.213$ & $48.463\pm19.566$ & 0 & $4.808\pm0.654$ & 0\\
$015258.66+010507.4$ & $0.647$ & $2$ & $$ & 0 & $0.700$ & $53.334\pm17.655$ & 0 &  & \\
$015309.12+005250.1$ & $1.161$ & $3$ & $$ & 0 & $1.075$ & $12.372\pm13.836$ & 0 & $1.217\pm0.373$ & 0\\
$015313.28+005307.3$ & $1.399$ & $3$ & $*$ & 0 & $0.497$ & $17.971\pm10.860$ & 0 & $1.838\pm0.461$ & 0\\
$020039.15-084554.9$ & $0.432$ & $2$ & $$ & 0 & $2.207$ & $120.719\pm9.207$ & 1 &  & \\
$021013.64-001200.5$ & $1.505$ & $2$ & $*$ & 0 & $0.329$ & $10.423\pm8.147$ & 0 &  & \\
$021025.93-001624.0$ & $0.595$ & $2$ & $*$ & 0 & $0.358$ & $22.578\pm15.931$ & 1 &  & \\
$022408.28+000301.2$ & $1.608$ & $2$ & $*$ & 0 & $0.514$ & $8.646\pm2.552$ & 0 & $0.865\pm0.255$ & 0\\
$022430.60-000038.8$ & $0.431$ & $2$ & $$ & 0 & $0.667$ & $107.135\pm5.049$ & 1 & $10.713\pm0.505$ & 1\\
$022518.36-001332.3$ & $3.627$ & $2$ & $*$ & 0 & $0.848$ & $8.326\pm5.196$ & 0 &  & \\
$022644.03+003305.8$ & $2.374$ & $2$ & $*$ & 0 & $0.509$ & $4.442\pm3.556$ & 0 & $0.444\pm0.356$ & 0\\
$022726.10+004827.6$ & $1.111$ & $2$ & $*$ & 0 & $0.211$ & $14.511\pm5.484$ & 0 &  & \\
$022730.14+004733.6$ & $1.485$ & $2$ & $*$ & 0 & $0.388$ & $19.458\pm6.504$ & 0 &  & \\
$022908.57+003908.1$ & $1.209$ & $2$ & $*$ & 0 & $0.296$ & $26.365\pm6.700$ & 0 & $2.637\pm0.670$ & 0\\
$022934.06+004524.7$ & $1.900$ & $2$ & $*$ & 0 & $0.359$ & $9.585\pm5.499$ & 0 &  & \\
$022938.18+002716.6$ & $1.490$ & $2$ & $*$ & 0 & $0.470$ & $4.923\pm4.254$ & 0 &  & \\
$023044.91+003459.5$ & $1.678$ & $2$ & $*$ & 1 & $0.216$ & $1.420\pm2.502$ & 0 & $0.142\pm0.250$ & 0\\
$023137.42+003706.0$ & $0.558$ & $2$ & $*$ & 0 & $0.246$ & $28.371\pm6.838$ & 0 & $2.837\pm0.684$ & 0\\
$024103.25+002727.3$ & $1.457$ & $2$ & $$ & 0 & $1.084$ & $9.342\pm13.738$ & 0 &  & \\
$024110.02+002301.4$ & $0.790$ & $2$ & $*$ & 0 & $0.247$ & $15.505\pm13.262$ & 0 &  & \\
$024142.62+003910.1$ & $1.685$ & $2$ & $*$ & 1 & $0.343$ & $10.834\pm22.382$ & 0 &  & \\
$024145.19+003028.4$ & $1.051$ & $2$ & $1$ & 0 & $0.373$ & $50.500\pm15.929$ & 0 &  & \\
$024230.65-000029.6$ & $2.505$ & $2$ & $$ & 1 & $1.195$ & $0.597\pm0.352$ & 1 & $0.060\pm0.035$ & 1\\
$074408.41+375841.1$ & $0.881$ & $4$ & $$ & 0 & $1.933$ & $3.429\pm1.572$ & 1 & $0.343\pm0.157$ & 1\\
$074417.47+375317.2$ & $1.067$ & $5$ & $1$ & 0 & $2.541$ & $148.129\pm8.189$ & 1 & $14.988\pm0.738$ & 0\\
$074502.90+374947.0$ & $0.593$ & $4$ & $*$ & 0 & $0.349$ & $30.388\pm5.093$ & 0 &  & \\
$074524.97+375436.7$ & $0.406$ & $4$ & $*$ & 0 & $0.671$ & $19.774\pm4.516$ & 1 &  & \\
$074545.01+392700.8$ & $1.629$ & $2$ & $$ & 0 & $0.875$ & $22.872\pm3.357$ & 0 &  & \\
$075502.11+220346.8$ & $0.400$ & $2$ & $$ & 0 & $0.799$ & $9.354\pm1.006$ & 0 & $0.935\pm0.101$ & 0\\
$080731.76+211754.3$ & $1.194$ & $2$ & $$ & 0 & $1.175$ & $19.689\pm4.555$ & 1 & $1.969\pm0.455$ & 1\\
$080749.15+212122.3$ & $2.232$ & $2$ & $$ & 0 & $1.537$ & $10.819\pm3.168$ & 0 & $1.082\pm0.317$ & 0\\
$081426.45+364713.5$ & $2.732$ & $2$ & $1$ & 1 & $0.375$ & $4.104\pm2.895$ & 0 & $0.410\pm0.290$ & 0\\
$083454.89+553421.1$ & $0.241$ & $2$ & $1$ & 0 & $0.733$ & $37.978\pm3.691$ & 0 & $3.798\pm0.369$ & 0\\
$083633.54+553245.0$ & $1.614$ & $2$ & $$ & 1 & $1.421$ & $18.285\pm6.333$ & 0 &  & \\
$084207.58+322646.7$ & $1.685$ & $2$ & $$ & 0 & $6.344$ & $49.105\pm4.445$ & 0 &  & \\
$084308.18+362439.6$ & $1.502$ & $2$ & $*$ & 0 & $0.412$ & $9.442\pm3.151$ & 0 & $0.944\pm0.315$ & 0\\
$084905.07+445714.7$ & $1.259$ & $2$ & $*$ & 0 & $0.333$ & $19.417\pm1.425$ & 0 & $1.942\pm0.143$ & 0\\
$084943.70+450024.2$ & $1.592$ & $2$ & $$ & 0 & $1.820$ & $30.711\pm2.160$ & 0 &  & \\
$090900.43+105934.8$ & $0.162$ & $2$ & $$ & 0 & $2.984$ & $722.397\pm30.768$ & 1 & $72.240\pm3.077$ & 1\\
$090928.50+541925.9$ & $3.760$ & $2$ & $*$ & 1 & $0.768$ & $0.641\pm0.732$ & 0 &  & \\
$091029.03+542719.0$ & $0.526$ & $2$ & $$ & 0 & $1.274$ & $156.138\pm4.701$ & 1 & $15.614\pm0.470$ & 1\\
$091127.61+055054.0$ & $2.793$ & $2$ & $$ & 1 & $3.561$ & $21.430\pm2.804$ & 0 &  & \\
$091210.34+054742.0$ & $3.241$ & $2$ & $$ & 0 & $3.348$ & $5.948\pm3.310$ & 0 &  & \\
$091752.54+414530.5$ & $1.277$ & $2$ & $$ & 0 & $0.777$ & $8.723\pm3.373$ & 0 &  & \\
$092108.62+453857.3$ & $0.174$ & $2$ & $*$ & 0 & $1.476$ & $538.732\pm21.725$ & 1 &  & \\
$092314.48+510020.7$ & $1.388$ & $2$ & $$ & 0 & $1.323$ & $17.125\pm10.509$ & 0 &  & \\
$094745.14+072520.6$ & $0.086$ & $2$ & $$ & 0 & $5.939$ & $320.078\pm8.782$ & 0 & $32.008\pm0.878$ & 0\\
$095240.16+515249.9$ & $0.553$ & $2$ & $$ & 0 & $1.511$ & $92.083\pm9.037$ & 0 & $9.208\pm0.904$ & 0\\
$095243.04+515121.0$ & $0.862$ & $2$ & $$ & 0 & $2.522$ & $110.763\pm10.977$ & 0 &  & \\
$095542.12+411655.2$ & $3.420$ & $3$ & $*$ & 0 & $0.886$ & $3.037\pm7.793$ & 1 & $0.302\pm0.215$ & 1\\
$095544.91+410755.0$ & $1.921$ & $3$ & $*$ & 1 & $0.710$ & $1.604\pm5.929$ & 0 & $0.160\pm0.104$ & 0\\
$095548.13+410955.3$ & $2.308$ & $2$ & $$ & 0 & $1.232$ & $17.079\pm7.844$ & 0 &  & \\
$095640.38+411043.5$ & $1.887$ & $3$ & $*$ & 0 & $0.244$ & $2.368\pm6.652$ & 1 & $0.237\pm0.181$ & 1\\
$095820.44+020303.9$ & $1.356$ & $2$ & $*$ & 0 & $0.409$ & $2.876\pm0.963$ & 0 & $0.288\pm0.096$ & 0\\
$095835.98+015157.0$ & $2.934$ & $2$ & $*$ & 0 & $$ & $1.892\pm0.965$ & 0 & $0.189\pm0.097$ & 0\\
$095857.34+021314.5$ & $1.024$ & $2$ & $*$ & 0 & $0.299$ & $91.785\pm4.438$ & 1 &  & \\
$095858.68+020138.9$ & $2.454$ & $4$ & $$ & 0 & $1.154$ & $19.584\pm3.237$ & 1 & $1.958\pm0.324$ & 1\\
$095902.76+021906.3$ & $0.345$ & $2$ & $$ & 0 & $0.671$ & $108.981\pm4.718$ & 0 & $10.898\pm0.472$ & 0\\
$095918.70+020951.4$ & $1.157$ & $4$ & $1$ & 0 & $0.426$ & $83.168\pm6.029$ & 1 & $8.317\pm0.603$ & 1\\
$095924.46+015954.3$ & $1.236$ & $4$ & $$ & 0 & $1.100$ & $53.427\pm5.123$ & 0 & $5.343\pm0.512$ & 0\\
$095949.40+020140.9$ & $1.753$ & $3$ & $*$ & 0 & $0.730$ & $19.564\pm2.730$ & 0 & $1.956\pm0.273$ & 0\\
$095957.97+014327.3$ & $1.618$ & $2$ & $*$ & 0 & $0.399$ & $7.425\pm1.990$ & 0 &  & \\
$100012.91+023522.8$ & $0.699$ & $5$ & $$ & 0 & $0.788$ & $31.959\pm6.387$ & 1 & $2.948\pm0.595$ & 1\\
$100014.13+020054.4$ & $2.497$ & $5$ & $*$ & 0 & $0.588$ & $7.728\pm2.699$ & 1 & $0.795\pm0.223$ & 1\\
$100024.39+015053.9$ & $1.664$ & $6$ & $*$ & 0 & $0.510$ & $3.951\pm2.383$ & 1 & $0.400\pm0.161$ & 0\\
$100024.64+023149.0$ & $1.318$ & $5$ & $$ & 0 & $0.886$ & $21.383\pm5.368$ & 1 & $2.138\pm0.537$ & 1\\
$100025.24+015852.0$ & $0.373$ & $4$ & $$ & 0 & $1.060$ & $141.057\pm8.725$ & 1 & $14.106\pm0.873$ & 1\\
$100043.13+020637.2$ & $0.360$ & $6$ & $*$ & 0 & $0.632$ & $19.413\pm5.233$ & 1 & $1.941\pm0.523$ & 1\\
$100055.39+023441.3$ & $1.403$ & $4$ & $*$ & 0 & $0.514$ & $15.301\pm3.707$ & 0 & $1.559\pm0.322$ & 0\\
$100058.84+015400.2$ & $1.559$ & $5$ & $*$ & 0 & $0.386$ & $18.397\pm5.385$ & 0 & $1.840\pm0.538$ & 0\\
$100104.32+553521.6$ & $1.537$ & $2$ & $$ & 0 & $1.247$ & $16.386\pm3.820$ & 0 & $1.639\pm0.382$ & 0\\
$100114.29+022356.8$ & $1.799$ & $4$ & $$ & 0 & $0.876$ & $13.170\pm2.702$ & 1 & $1.317\pm0.270$ & 1\\
$100116.78+014053.5$ & $2.055$ & $3$ & $*$ & 0 & $0.529$ & $10.119\pm2.994$ & 0 & $1.182\pm0.268$ & 0\\
$100120.26+023341.3$ & $1.834$ & $2$ & $*$ & 0 & $0.365$ & $4.476\pm1.415$ & 0 & $0.448\pm0.142$ & 0\\
$100130.37+014304.3$ & $1.571$ & $5$ & $*$ & 0 & $0.312$ & $4.226\pm3.122$ & 0 & $0.407\pm0.277$ & 1\\
$100145.15+022456.9$ & $2.032$ & $2$ & $*$ & 0 & $0.311$ & $4.832\pm1.187$ & 0 & $0.483\pm0.119$ & 0\\
$100201.51+020329.4$ & $2.008$ & $6$ & $$ & 0 & $1.232$ & $1.998\pm4.013$ & 0 & $0.210\pm0.247$ & 0\\
$100205.36+554257.9$ & $1.151$ & $2$ & $$ & 0 & $1.834$ & $16.737\pm5.599$ & 0 &  & \\
$100219.49+015537.0$ & $1.509$ & $2$ & $*$ & 0 & $0.428$ & $17.506\pm3.740$ & 0 &  & \\
$102350.94+041542.0$ & $1.809$ & $2$ & $*$ & 0 & $0.766$ & $1.854\pm1.639$ & 1 & $0.185\pm0.164$ & 1\\
$103222.85+575551.1$ & $1.243$ & $2$ & $1$ & 0 & $0.330$ & $22.592\pm2.706$ & 0 &  & \\
$103227.93+573822.5$ & $1.968$ & $3$ & $*$ & 0 & $0.287$ & $34.075\pm3.835$ & 0 & $3.407\pm0.383$ & 0\\
$104829.95+123428.0$ & $0.442$ & $3$ & $$ & 0 & $0.790$ & $49.208\pm4.251$ & 1 & $4.921\pm0.425$ & 1\\
$105015.58+570255.7$ & $3.273$ & $2$ & $*$ & 0 & $0.450$ & $8.182\pm7.308$ & 0 &  & \\
$105039.54+572336.6$ & $1.447$ & $2$ & $1$ & 0 & $0.694$ & $23.350\pm8.775$ & 0 &  & \\
$105050.14+573820.0$ & $1.281$ & $2$ & $$ & 0 & $0.797$ & $26.146\pm8.842$ & 0 &  & \\
$105239.60+572431.4$ & $1.112$ & $3$ & $$ & 0 & $2.532$ & $51.458\pm12.947$ & 1 &  & \\
$105316.75+573550.8$ & $1.205$ & $3$ & $$ & 0 & $0.684$ & $81.026\pm15.179$ & 0 & $8.103\pm1.518$ & 0\\
$105518.08+570423.5$ & $0.696$ & $2$ & $$ & 0 & $1.123$ & $47.098\pm12.371$ & 0 & $4.710\pm1.237$ & 0\\
$111354.66+124439.0$ & $0.680$ & $2$ & $1$ & 0 & $0.107$ & $30.643\pm19.016$ & 1 &  & \\
$111422.47+531913.2$ & $0.885$ & $3$ & $$ & 0 & $1.187$ & $38.261\pm10.991$ & 1 & $3.826\pm1.099$ & 1\\
$111452.84+531531.7$ & $1.213$ & $3$ & $$ & 0 & $0.688$ & $43.474\pm11.329$ & 0 & $4.347\pm1.133$ & 0\\
$111518.58+531452.7$ & $1.540$ & $3$ & $$ & 0 & $1.735$ & $16.952\pm5.661$ & 0 & $1.695\pm0.566$ & 0\\
$111520.73+530922.1$ & $0.877$ & $3$ & $$ & 0 & $2.082$ & $0.183\pm2.450$ & 0 & $0.018\pm0.245$ & 0\\
$111816.95+074558.1$ & $1.736$ & $2$ & $$ & 0 & $13.013$ & $114.067\pm6.977$ & 1 &  & \\
$111840.56+075324.1$ & $1.463$ & $2$ & $$ & 0 & $0.888$ & $24.815\pm4.632$ & 0 &  & \\
$111946.94+133759.2$ & $2.023$ & $2$ & $$ & 0 & $1.127$ & $8.410\pm2.999$ & 0 &  & \\
$112026.20+134024.6$ & $0.982$ & $2$ & $$ & 0 & $0.950$ & $68.886\pm11.089$ & 1 &  & \\
$112048.99+133821.9$ & $0.513$ & $2$ & $$ & 0 & $0.793$ & $3.115\pm1.099$ & 0 & $0.312\pm0.110$ & 0\\
$112106.07+133824.9$ & $1.944$ & $2$ & $$ & 0 & $1.152$ & $45.943\pm14.019$ & 0 &  & \\
$112213.65+041548.7$ & $3.517$ & $4$ & $*$ & 0 & $0.406$ & $1.305\pm4.408$ & 0 & $0.142\pm0.321$ & 0\\
$112320.73+013747.4$ & $1.469$ & $2$ & $$ & 0 & $17.207$ & $108.419\pm10.153$ & 0 &  & \\
$112404.52+040418.1$ & $3.841$ & $5$ & $*$ & 0 & $0.929$ & $2.392\pm5.626$ & 0 & $0.228\pm0.459$ & 0\\
$114636.88+472313.3$ & $1.895$ & $2$ & $1$ & 0 & $1.366$ & $51.248\pm4.892$ & 0 & $5.125\pm0.489$ & 0\\
$114651.21+471732.5$ & $3.130$ & $2$ & $*$ & 1 & $0.692$ & $0.999\pm2.414$ & 0 &  & \\
$114656.73+472755.6$ & $0.668$ & $3$ & $$ & 0 & $0.872$ & $109.369\pm8.629$ & 0 & $10.937\pm0.863$ & 0\\
$115324.46+493108.7$ & $0.334$ & $2$ & $1$ & 0 & $2.854$ & $1481.928\pm22.813$ & 1 &  & \\
$115838.56+435505.8$ & $1.208$ & $2$ & $1$ & 0 & $0.434$ & $15.222\pm3.356$ & 0 & $1.522\pm0.336$ & 0\\
$115911.43+440818.3$ & $1.438$ & $2$ & $$ & 0 & $1.320$ & $22.118\pm6.177$ & 0 &  & \\
$120104.66+575846.9$ & $1.842$ & $2$ & $$ & 0 & $2.019$ & $26.017\pm8.028$ & 1 & $2.602\pm0.803$ & 1\\
$120106.14+580336.6$ & $1.087$ & $2$ & $*$ & 0 & $0.470$ & $20.459\pm6.911$ & 0 & $2.046\pm0.691$ & 0\\
$120233.39+580501.8$ & $3.424$ & $2$ & $*$ & 0 & $1.389$ & $11.738\pm4.683$ & 0 & $1.174\pm0.468$ & 0\\
$120924.07+103612.0$ & $0.395$ & $5$ & $$ & 1 & $6.871$ & $7.827\pm26.396$ & 0 &  & \\
$120924.80+102553.9$ & $0.263$ & $2$ & $*$ & 0 & $1.072$ & $110.776\pm38.543$ & 0 &  & \\
$120937.02+103756.9$ & $1.994$ & $5$ & $$ & 0 & $1.102$ & $6.221\pm23.135$ & 0 &  & \\
$120949.46+102146.8$ & $2.312$ & $6$ & $*$ & 0 & $0.898$ & $12.216\pm35.370$ & 0 &  & \\
$120959.05+104320.1$ & $1.314$ & $4$ & $$ & 0 & $0.855$ & $13.081\pm404.062$ & 0 &  & \\
$121013.57+104853.7$ & $1.080$ & $4$ & $1$ & 0 & $0.912$ & $52.349\pm35.155$ & 0 &  & \\
$121111.46+100826.8$ & $1.994$ & $2$ & $$ & 0 & $0.952$ & $15.664\pm9.156$ & 0 &  & \\
$121342.95+025248.9$ & $0.641$ & $2$ & $$ & 0 & $0.674$ & $14.385\pm3.101$ & 0 & $1.438\pm0.310$ & 0\\
$121440.27+142859.1$ & $1.625$ & $2$ & $$ & 1 & $3.824$ & $1.143\pm3.444$ & 0 &  & \\
$122418.00+070949.2$ & $0.981$ & $7$ & $$ & 0 & $0.829$ & $5.848\pm4.537$ & 1 & $0.497\pm0.211$ & 1\\
$122448.15+125413.3$ & $1.062$ & $2$ & $$ & 0 & $2.384$ & $0.639\pm0.454$ & 0 & $0.064\pm0.045$ & 0\\
$122511.91+125153.6$ & $1.255$ & $3$ & $$ & 0 & $2.188$ & $86.787\pm4.564$ & 1 & $8.679\pm0.456$ & 1\\
$122515.65+124441.0$ & $1.664$ & $2$ & $$ & 0 & $1.046$ & $18.494\pm2.974$ & 1 & $1.849\pm0.297$ & 1\\
$122722.12+075555.0$ & $3.168$ & $2$ & $1$ & 0 & $0.986$ & $8.822\pm8.352$ & 0 &  & \\
$122826.33+130106.2$ & $3.229$ & $3$ & $*$ & 0 & $0.816$ & $3.757\pm5.737$ & 0 &  & \\
$122923.73+075359.2$ & $0.854$ & $2$ & $$ & 0 & $0.838$ & $70.518\pm6.482$ & 1 & $7.052\pm0.648$ & 1\\
$123320.92+110702.4$ & $1.206$ & $2$ & $$ & 0 & $1.107$ & $15.059\pm7.321$ & 0 &  & \\
$123346.21+130905.7$ & $1.368$ & $2$ & $$ & 0 & $0.746$ & $17.454\pm6.481$ & 0 &  & \\
$123410.72+111732.6$ & $0.817$ & $2$ & $$ & 0 & $1.556$ & $39.691\pm8.514$ & 1 & $3.969\pm0.851$ & 1\\
$123527.75+121338.8$ & $0.726$ & $4$ & $$ & 0 & $3.252$ & $71.251\pm17.906$ & 1 & $7.565\pm1.375$ & 1\\
$123540.19+123620.7$ & $3.208$ & $2$ & $*$ & 0 & $0.552$ & $2.839\pm5.799$ & 0 &  & \\
$123618.94+121010.0$ & $0.993$ & $2$ & $$ & 0 & $0.803$ & $5.519\pm6.425$ & 0 &  & \\
$123622.94+621526.6$ & $2.587$ & $20$ & $*$ & 0 & $0.277$ & $7.122\pm3.914$ & 1 & $0.712\pm0.391$ & 1\\
$123715.99+620323.3$ & $2.068$ & $6$ & $*$ & 0 & $0.511$ & $1.963\pm2.437$ & 1 & $0.196\pm0.244$ & 1\\
$123759.56+621102.3$ & $0.909$ & $10$ & $$ & 0 & $1.309$ & $59.321\pm8.443$ & 1 & $5.932\pm0.844$ & 1\\
$123800.91+621336.0$ & $0.440$ & $15$ & $$ & 0 & $0.858$ & $36.941\pm8.695$ & 1 & $3.694\pm0.870$ & 1\\
$124107.11+113701.7$ & $1.413$ & $2$ & $$ & 0 & $1.093$ & $26.428\pm7.978$ & 0 &  & \\
$124210.41+115223.8$ & $0.298$ & $2$ & $$ & 0 & $0.689$ & $72.599\pm8.249$ & 1 &  & \\
$124255.31+024956.9$ & $1.459$ & $4$ & $$ & 0 & $0.770$ & $22.482\pm4.444$ & 1 & $2.248\pm0.444$ & 1\\
$125849.83-014303.3$ & $0.967$ & $8$ & $$ & 0 & $5.162$ & $136.037\pm16.400$ & 1 & $13.604\pm1.640$ & 1\\
$125919.26+124829.0$ & $0.701$ & $2$ & $$ & 0 & $0.839$ & $26.000\pm4.759$ & 1 &  & \\
$130216.13+003032.1$ & $4.468$ & $2$ & $*$ & 0 & $0.618$ & $0.498\pm0.912$ & 0 &  & \\
$132852.11+472218.3$ & $1.932$ & $2$ & $$ & 0 & $0.874$ & $8.923\pm3.570$ & 0 &  & \\
$132938.57+471854.6$ & $1.027$ & $2$ & $*$ & 0 & $0.316$ & $26.418\pm10.304$ & 0 &  & \\
$133004.72+472301.0$ & $2.825$ & $3$ & $1$ & 1 & $0.805$ & $1.095\pm5.995$ & 0 &  & \\
$133223.26+503431.3$ & $3.807$ & $2$ & $$ & 0 & $1.922$ & $6.185\pm2.478$ & 0 & $0.618\pm0.248$ & 0\\
$134425.94-000056.2$ & $1.096$ & $2$ & $$ & 0 & $1.362$ & $5.921\pm4.200$ & 0 &  & \\
$135854.44+623913.1$ & $1.228$ & $3$ & $$ & 0 & $1.523$ & $63.993\pm6.630$ & 1 &  & \\
$140041.11+622516.2$ & $1.878$ & $2$ & $*$ & 0 & $0.586$ & $9.450\pm3.582$ & 0 & $0.945\pm0.358$ & 0\\
$140146.53+024434.7$ & $4.441$ & $2$ & $*$ & 0 & $1.957$ & $9.610\pm2.363$ & 1 &  & \\
$140354.57+543246.8$ & $3.258$ & $2$ & $*$ & 0 & $0.508$ & $4.817\pm1.396$ & 1 &  & \\
$141500.38+520658.5$ & $0.424$ & $6$ & $$ & 0 & $1.016$ & $48.668\pm9.485$ & 1 & $4.867\pm0.949$ & 1\\
$141533.89+520558.0$ & $0.986$ & $6$ & $$ & 0 & $1.178$ & $28.762\pm6.984$ & 1 & $2.876\pm0.698$ & 1\\
$141551.27+522740.6$ & $2.585$ & $2$ & $*$ & 0 & $0.503$ & $3.131\pm2.255$ & 0 & $0.313\pm0.226$ & 0\\
$141551.59+520025.6$ & $1.514$ & $3$ & $$ & 0 & $0.980$ & $8.662\pm4.682$ & 1 & $0.866\pm0.468$ & 1\\
$141642.42+521812.7$ & $1.284$ & $9$ & $*$ & 0 & $$ & $18.385\pm7.574$ & 1 & $1.839\pm0.757$ & 1\\
$141647.20+521115.2$ & $2.153$ & $5$ & $$ & 0 & $1.530$ & $16.566\pm5.502$ & 0 & $1.635\pm0.327$ & 0\\
$141905.17+522527.7$ & $1.606$ & $3$ & $*$ & 0 & $0.447$ & $13.304\pm2.754$ & 0 &  & \\
$141908.18+062834.8$ & $1.437$ & $2$ & $1$ & 0 & $7.273$ & $1305.172\pm45.235$ & 1 &  & \\
$142005.59+530036.7$ & $1.647$ & $17$ & $*$ & 0 & $0.389$ & $11.174\pm7.145$ & 1 & $1.117\pm0.714$ & 1\\
$142015.64+523718.8$ & $1.674$ & $6$ & $*$ & 0 & $0.781$ & $14.832\pm6.621$ & 0 &  & \\
$142052.43+525622.4$ & $0.676$ & $24$ & $$ & 0 & $1.969$ & $83.371\pm22.501$ & 1 & $8.358\pm2.188$ & 1\\
$142147.09+532405.7$ & $3.039$ & $5$ & $*$ & 0 & $0.421$ & $3.556\pm2.422$ & 1 & $0.356\pm0.242$ & 1\\
$142301.08+533311.8$ & $1.863$ & $2$ & $$ & 0 & $2.837$ & $22.557\pm2.538$ & 0 & $2.256\pm0.254$ & 0\\
$142305.04+240507.8$ & $4.105$ & $2$ & $*$ & 1 & $0.561$ & $4.479\pm1.408$ & 0 & $0.448\pm0.141$ & 0\\
$142455.69+351356.6$ & $1.255$ & $2$ & $$ & 0 & $0.861$ & $5.769\pm3.806$ & 0 &  & \\
$142507.32+323137.4$ & $0.478$ & $2$ & $1$ & 0 & $1.544$ & $79.144\pm14.726$ & 0 &  & \\
$142530.09+335217.3$ & $1.185$ & $2$ & $$ & 0 & $0.768$ & $5.538\pm4.512$ & 0 & $0.554\pm0.451$ & 0\\
$142532.83+330124.9$ & $1.200$ & $2$ & $1$ & 0 & $0.286$ & $22.939\pm8.534$ & 0 & $2.294\pm0.853$ & 0\\
$142539.01+331009.4$ & $2.306$ & $2$ & $$ & 0 & $1.881$ & $35.656\pm9.731$ & 0 & $3.566\pm0.973$ & 0\\
$142543.30+335543.6$ & $1.133$ & $3$ & $$ & 0 & $0.675$ & $38.937\pm11.438$ & 0 & $3.894\pm1.144$ & 0\\
$142545.53+332603.3$ & $2.963$ & $2$ & $*$ & 0 & $0.469$ & $3.237\pm4.394$ & 0 & $0.324\pm0.439$ & 0\\
$142551.17+350113.0$ & $0.898$ & $2$ & $$ & 0 & $1.164$ & $28.423\pm8.816$ & 0 & $2.842\pm0.882$ & 0\\
$142557.63+334626.2$ & $0.351$ & $2$ & $$ & 0 & $2.016$ & $23.779\pm8.883$ & 0 & $2.378\pm0.888$ & 0\\
$142620.30+351712.1$ & $1.748$ & $2$ & $$ & 1 & $1.364$ & $2.221\pm6.837$ & 1 & $0.222\pm0.684$ & 1\\
$142622.66+334202.3$ & $1.349$ & $2$ & $$ & 0 & $2.069$ & $35.988\pm11.628$ & 0 &  & \\
$142623.15+351154.9$ & $3.503$ & $2$ & $*$ & 0 & $0.561$ & $0.879\pm3.865$ & 0 & $0.088\pm0.387$ & 0\\
$142640.83+332158.7$ & $1.542$ & $2$ & $$ & 1 & $0.891$ & $2.121\pm5.310$ & 0 & $0.212\pm0.531$ & 0\\
$142730.19+324106.4$ & $1.776$ & $2$ & $$ & 0 & $0.955$ & $1.899\pm3.277$ & 0 & $0.190\pm0.328$ & 0\\
$142734.80+352543.4$ & $0.340$ & $2$ & $$ & 0 & $1.303$ & $37.178\pm10.303$ & 0 & $3.718\pm1.030$ & 0\\
$142738.21+351132.1$ & $1.208$ & $3$ & $$ & 0 & $1.389$ & $15.525\pm9.494$ & 0 & $1.664\pm0.726$ & 0\\
$142738.36+325320.0$ & $0.822$ & $2$ & $*$ & 0 & $0.574$ & $89.634\pm10.516$ & 0 & $8.963\pm1.052$ & 0\\
$142810.31+353847.0$ & $0.804$ & $2$ & $$ & 0 & $1.006$ & $99.180\pm10.800$ & 0 & $9.918\pm1.080$ & 0\\
$142813.94+334759.6$ & $2.243$ & $3$ & $$ & 0 & $0.967$ & $1.590\pm5.479$ & 1 & $0.256\pm0.509$ & 1\\
$142817.81+354021.9$ & $0.335$ & $2$ & $$ & 0 & $0.791$ & $59.363\pm10.160$ & 0 & $5.936\pm1.016$ & 0\\
$142848.32+350315.5$ & $2.111$ & $3$ & $$ & 0 & $1.049$ & $5.417\pm5.954$ & 0 & $0.542\pm0.595$ & 0\\
$142858.01+344149.9$ & $3.076$ & $2$ & $*$ & 0 & $0.505$ & $2.851\pm4.813$ & 0 & $0.285\pm0.481$ & 0\\
$142910.22+352946.8$ & $2.224$ & $3$ & $1$ & 0 & $1.181$ & $9.653\pm4.223$ & 0 & $0.965\pm0.422$ & 0\\
$142911.17+330941.3$ & $1.116$ & $2$ & $$ & 0 & $0.779$ & $14.136\pm6.838$ & 0 & $1.414\pm0.684$ & 0\\
$142912.87+340959.0$ & $2.229$ & $4$ & $$ & 0 & $2.063$ & $14.479\pm10.076$ & 0 & $1.468\pm0.895$ & 0\\
$142915.19+343820.3$ & $2.351$ & $2$ & $$ & 0 & $1.495$ & $10.811\pm6.959$ & 0 & $1.081\pm0.696$ & 0\\
$142917.20+342130.3$ & $1.275$ & $2$ & $$ & 0 & $0.800$ & $60.240\pm13.005$ & 0 & $6.024\pm1.300$ & 0\\
$142942.64+335654.7$ & $1.121$ & $3$ & $1$ & 0 & $0.961$ & $84.645\pm15.627$ & 0 & $8.464\pm1.563$ & 0\\
$142949.65+324653.9$ & $2.175$ & $2$ & $*$ & 0 & $0.852$ & $8.675\pm7.314$ & 0 & $0.867\pm0.731$ & 0\\
$142954.70+330134.7$ & $2.076$ & $2$ & $$ & 0 & $2.103$ & $20.133\pm8.187$ & 0 & $2.013\pm0.819$ & 0\\
$143031.78+330042.5$ & $1.071$ & $2$ & $$ & 0 & $1.212$ & $13.942\pm8.611$ & 0 & $1.394\pm0.861$ & 0\\
$143034.83+335945.3$ & $1.115$ & $2$ & $$ & 0 & $2.628$ & $43.009\pm9.442$ & 0 & $4.301\pm0.944$ & 0\\
$143106.77+340910.8$ & $1.098$ & $3$ & $$ & 0 & $0.669$ & $20.320\pm9.116$ & 0 & $2.032\pm0.912$ & 0\\
$143107.51+342730.9$ & $4.270$ & $2$ & $*$ & 0 & $0.548$ & $2.269\pm4.713$ & 0 &  & \\
$143132.13+341417.3$ & $1.040$ & $2$ & $$ & 0 & $1.330$ & $57.982\pm12.491$ & 0 & $5.798\pm1.249$ & 0\\
$143157.94+341650.2$ & $0.715$ & $2$ & $$ & 0 & $4.634$ & $272.680\pm22.804$ & 0 & $27.268\pm2.280$ & 0\\
$143201.74+343526.2$ & $1.071$ & $2$ & $$ & 0 & $0.934$ & $13.472\pm8.661$ & 0 &  & \\
$143219.53+341728.9$ & $0.629$ & $2$ & $$ & 0 & $1.206$ & $106.076\pm19.235$ & 1 & $10.608\pm1.924$ & 1\\
$143243.92+330746.6$ & $2.088$ & $2$ & $$ & 0 & $1.621$ & $6.423\pm5.359$ & 0 & $0.642\pm0.536$ & 0\\
$143244.26+350100.4$ & $1.038$ & $2$ & $$ & 0 & $0.957$ & $14.640\pm7.434$ & 1 & $1.464\pm0.743$ & 1\\
$143307.88+342315.9$ & $1.950$ & $2$ & $*$ & 0 & $0.839$ & $19.433\pm8.290$ & 0 & $1.943\pm0.829$ & 0\\
$143331.79+341532.7$ & $0.957$ & $2$ & $$ & 0 & $0.887$ & $1.822\pm4.439$ & 0 & $0.182\pm0.444$ & 0\\
$143335.68+350133.1$ & $0.618$ & $2$ & $$ & 0 & $1.213$ & $3.466\pm5.645$ & 0 &  & \\
$143345.10+345939.9$ & $0.815$ & $2$ & $*$ & 0 & $0.472$ & $2.857\pm5.754$ & 0 &  & \\
$143506.45+335526.0$ & $3.940$ & $2$ & $*$ & 0 & $1.055$ & $6.056\pm4.582$ & 0 & $0.606\pm0.458$ & 0\\
$143547.62+335309.6$ & $2.111$ & $3$ & $1$ & 0 & $0.315$ & $4.438\pm6.344$ & 0 & $0.586\pm0.574$ & 0\\
$143559.19+334640.1$ & $0.948$ & $2$ & $$ & 0 & $0.899$ & $70.576\pm14.211$ & 0 & $7.058\pm1.421$ & 0\\
$143604.64+350428.5$ & $3.033$ & $2$ & $*$ & 1 & $0.414$ & $0.033\pm2.619$ & 0 & $0.003\pm0.262$ & 0\\
$143617.81+353726.1$ & $1.448$ & $3$ & $$ & 0 & $1.016$ & $27.628\pm12.945$ & 0 & $2.457\pm0.964$ & 1\\
$143624.30+353709.4$ & $0.767$ & $2$ & $$ & 0 & $1.151$ & $34.920\pm10.075$ & 0 & $3.492\pm1.007$ & 0\\
$143624.61+352537.2$ & $1.060$ & $2$ & $$ & 0 & $1.052$ & $31.300\pm9.896$ & 0 & $3.130\pm0.990$ & 0\\
$143626.63+350029.7$ & $1.242$ & $3$ & $$ & 0 & $0.915$ & $0.347\pm3.547$ & 0 & $0.035\pm0.355$ & 0\\
$143627.78+343416.8$ & $1.883$ & $2$ & $$ & 0 & $1.026$ & $11.659\pm6.829$ & 0 & $1.166\pm0.683$ & 0\\
$143628.08+335524.3$ & $0.903$ & $3$ & $$ & 0 & $1.128$ & $30.806\pm12.188$ & 1 & $2.700\pm0.834$ & 0\\
$143632.99+344253.4$ & $1.948$ & $2$ & $$ & 0 & $0.874$ & $14.846\pm6.737$ & 0 & $1.485\pm0.674$ & 0\\
$143651.51+343602.4$ & $0.296$ & $2$ & $$ & 0 & $0.877$ & $51.688\pm13.078$ & 1 & $5.169\pm1.308$ & 1\\
$143706.20+343659.2$ & $4.369$ & $2$ & $*$ & 0 & $0.600$ & $3.378\pm6.450$ & 0 &  & \\
$143841.95+034110.3$ & $1.737$ & $2$ & $$ & 0 & $2.346$ & $22.186\pm5.090$ & 0 & $2.219\pm0.509$ & 0\\
$143859.05+033547.4$ & $0.734$ & $2$ & $$ & 0 & $1.291$ & $35.976\pm6.427$ & 0 &  & \\
$144642.92+012552.4$ & $1.421$ & $2$ & $$ & 0 & $1.449$ & $14.527\pm6.972$ & 0 &  & \\
$145206.45+580625.9$ & $1.440$ & $2$ & $*$ & 0 & $0.504$ & $36.140\pm2.998$ & 1 & $3.614\pm0.300$ & 1\\
$145207.32+580454.7$ & $1.920$ & $2$ & $*$ & 0 & $0.301$ & $34.166\pm2.700$ & 0 & $3.417\pm0.270$ & 0\\
$145215.59+430448.7$ & $0.296$ & $3$ & $$ & 0 & $0.927$ & $44.665\pm7.644$ & 1 & $4.466\pm0.764$ & 1\\
$150407.51-024816.5$ & $0.217$ & $2$ & $1$ & 0 & $2.657$ & $92.683\pm5.856$ & 1 &  & \\
$150948.65+333626.7$ & $0.512$ & $2$ & $$ & 0 & $0.568$ & $44.471\pm6.827$ & 1 & $4.447\pm0.683$ & 1\\
$151413.58+553500.7$ & $1.319$ & $2$ & $$ & 0 & $1.584$ & $14.912\pm2.713$ & 1 &  & \\
$151451.28+552602.3$ & $1.842$ & $2$ & $*$ & 0 & $0.263$ & $6.818\pm1.632$ & 0 & $0.682\pm0.163$ & 0\\
$151545.08+553518.4$ & $1.652$ & $2$ & $*$ & 0 & $0.412$ & $9.899\pm1.551$ & 0 & $0.990\pm0.155$ & 0\\
$153308.65+301820.7$ & $4.455$ & $2$ & $*$ & 0 & $0.874$ & $-0.038\pm0.992$ & 0 & $-0.004\pm0.099$ & 0\\
$155633.77+351757.3$ & $1.495$ & $2$ & $1$ & 1 & $1.708$ & $21.838\pm3.413$ & 0 &  & \\
$160410.22+432614.7$ & $1.538$ & $2$ & $$ & 0 & $2.301$ & $8.437\pm2.099$ & 0 & $0.844\pm0.210$ & 0\\
$160630.60+542007.5$ & $0.820$ & $2$ & $$ & 0 & $1.085$ & $58.018\pm14.034$ & 0 &  & \\
$160856.78+540313.7$ & $1.915$ & $2$ & $*$ & 0 & $0.767$ & $10.137\pm6.277$ & 0 & $1.014\pm0.628$ & 0\\
$164025.02+464449.0$ & $0.537$ & $2$ & $$ & 0 & $1.604$ & $35.889\pm8.166$ & 1 & $3.589\pm0.817$ & 1\\
$164733.23+350541.5$ & $0.861$ & $2$ & $*$ & 0 & $0.395$ & $38.988\pm8.173$ & 0 & $3.899\pm0.817$ & 0\\
$165108.85+345633.7$ & $1.541$ & $2$ & $$ & 0 & $1.590$ & $8.492\pm3.861$ & 0 & $0.849\pm0.386$ & 0\\
$170224.52+340539.0$ & $2.038$ & $2$ & $$ & 0 & $1.232$ & $5.852\pm3.762$ & 0 & $0.585\pm0.376$ & 0\\
$170441.37+604430.5$ & $0.372$ & $2$ & $$ & 0 & $22.112$ & $402.461\pm12.583$ & 1 &  & \\
$171957.81+263416.6$ & $3.160$ & $2$ & $1$ & 0 & $0.399$ & $1.537\pm3.592$ & 0 &  & \\
$172026.47+263816.0$ & $1.141$ & $4$ & $$ & 0 & $0.880$ & $17.377\pm44.859$ & 0 & $1.739\pm0.468$ & 0\\
$172211.65+575652.0$ & $1.610$ & $2$ & $*$ & 0 & $0.342$ & $7.416\pm4.865$ & 0 &  & \\
$173744.88+582829.6$ & $4.918$ & $2$ & $*$ & 0 & $1.041$ & $-0.130\pm3.845$ & 0 &  & \\
$173801.22+583012.1$ & $0.330$ & $2$ & $*$ & 0 & $0.630$ & $73.887\pm12.107$ & 0 & $7.389\pm1.211$ & 0\\
$173836.16+583748.5$ & $1.279$ & $2$ & $$ & 0 & $2.726$ & $4.692\pm4.609$ & 0 &  & \\
$221453.84+140022.2$ & $1.523$ & $2$ & $*$ & 0 & $1.700$ & $21.435\pm3.355$ & 0 & $2.144\pm0.335$ & 0\\
$221458.45+135344.7$ & $3.673$ & $4$ & $*$ & 0 & $0.672$ & $8.918\pm3.335$ & 1 & $0.892\pm0.334$ & 1\\
$221738.41+001206.6$ & $1.121$ & $2$ & $*$ & 0 & $0.275$ & $6.937\pm1.132$ & 1 & $0.694\pm0.113$ & 1\\
$221751.29+001146.4$ & $1.491$ & $5$ & $*$ & 0 & $0.450$ & $8.723\pm1.626$ & 1 & $0.872\pm0.163$ & 1\\
$221755.20+001512.3$ & $2.092$ & $4$ & $*$ & 0 & $0.442$ & $8.789\pm1.548$ & 1 & $0.879\pm0.155$ & 1\\
$232007.52+002944.3$ & $0.942$ & $4$ & $$ & 0 & $0.987$ & $87.608\pm9.669$ & 0 & $8.761\pm0.967$ & 0\\
$233130.08+001631.6$ & $2.659$ & $2$ & $*$ & 0 & $0.573$ & $4.565\pm2.527$ & 0 & $0.457\pm0.253$ & 0\\
$235653.87-010731.5$ & $0.601$ & $2$ & $*$ & 0 & $0.263$ & $35.650\pm5.004$ & 0 & $3.565\pm0.500$ & 0\\

\enddata
\tablenotetext{a}{Sources marked with an asterisk are not known to be radio-loud, but limits were not sensitive enough to guarantee that they were radio-quiet with high confidence.}
\tablenotetext{b}{Blank entries indicate where $F_{2500}$ could not be reliably measured due to bad spectral bins.}
\end{deluxetable}

\clearpage
\begin{landscape}
\clearpage
\begin{deluxetable}{lrrrcrrrr}
\tabletypesize{\scriptsize}
\tablecolumns{9}
\tablewidth{0pc}
\tablecaption{Observations of Each Source\label{epochTable0}}
\tablehead{\colhead{SDSS Name} & \colhead{ObsId}  & \colhead{Exposure} & \colhead{Off-Axis Angle} & \colhead{In HQ} & \colhead{TSTART\tablenotemark{a}} & \colhead{Source Count Rate} & \colhead{Count Luminosity} & \colhead{Total Counts\tablenotemark{b}} \\ \colhead{(J2000)} & \colhead{}  & \colhead{(sec)} & \colhead{(arcmin)} & \colhead{Flag} & \colhead{(sec)} & \colhead{10$^{-6}$ cts cm$^{-2}$ s$^{-1}$} & \colhead{10$^{52}$ cts s$^{-1}$} & \colhead{Full/Soft/Hard Band}}
\startdata
$000622.60-000424.4$ & 4096 & 4451.707 & 13.115 & 0 & 176198559.771 & 192$^{+14.7}_{-14.3}$ & 195.332$^{+14.909}_{-14.506}$ & $274.582^{+19.023}_{-17.827}$ / $192.916^{+16.329}_{-15.099}$ / $85.208^{+11.107}_{-9.888}$\\
$000622.60-000424.4$ & 5617 & 16931.652 & 7.950 & 1 & 238942682.443 & 229$^{+8.25}_{-8.02}$ & 231.930$^{+8.375}_{-8.144}$ & $960.227^{+34.310}_{-33.146}$ / $629.275^{+28.237}_{-27.048}$ / $343.528^{+21.078}_{-19.892}$\\
$004054.65-091526.7$ & 4885 & 9636.413 & 7.774 & 1 & 210608981.312 & 2.73$^{+2.17}_{-2.03}$ & 294.766$^{+233.795}_{-218.609}$ & $6.895^{+4.135}_{-2.756}$ / $5.812^{+3.950}_{-2.534}$ / $1.151^{+2.674}_{-0.997}$\\
$004054.65-091526.7$ & 904 & 38419.192 & 14.503 & 0 & 83056012.160 & 4.82$^{+1.7}_{-1.79}$ & 519.462$^{+183.684}_{-193.478}$ & $119.043^{+12.661}_{-11.495}$ / $58.772^{+9.339}_{-8.131}$ / $63.200^{+9.729}_{-8.502}$\\
$011513.16+002013.1$ & 3203 & 40581.117 & 6.684 & 1 & 126613529.277 & 2.19$^{+0.782}_{-0.772}$ & 18.106$^{+6.477}_{-6.396}$ & $33.976^{+7.413}_{-6.177}$ / $24.072^{+6.492}_{-5.222}$ / $10.361^{+4.746}_{-3.405}$\\
$011513.16+002013.1$ & 3204 & 37616.425 & 3.727 & 1 & 152506743.623 & 4.62$^{+0.943}_{-0.856}$ & 38.268$^{+7.813}_{-7.088}$ & $42.287^{+8.296}_{-7.023}$ / $35.571^{+7.761}_{-6.467}$ / $6.931^{+4.156}_{-2.770}$\\
$014219.01+132746.5$ & 1633 & 1914.645 & 14.098 & 0 & 97032805.579 & 6.04$^{+8.83}_{-8.45}$ & 0.161$^{+0.236}_{-0.226}$ & $4.401^{+3.498}_{-2.130}$ / $0.000^{+2.046}_{-0.000}$ / $4.602^{+3.658}_{-2.228}$\\
$014219.01+132746.5$ & 4010 & 5064.646 & 14.653 & 0 & 161474857.473 & 1.36$^{+5.12}_{-5.37}$ & 0.036$^{+0.137}_{-0.143}$ & $10.006^{+4.583}_{-3.289}$ / $3.427^{+3.355}_{-1.894}$ / $6.912^{+4.145}_{-2.762}$\\
$014320.96+132429.7$ & 1633 & 1914.645 & 12.922 & 0 & 97032805.579 & -3.9$^{+6.83}_{-7.76}$ & -17.960$^{+31.412}_{-35.702}$ & $4.693^{+3.730}_{-2.272}$ / $2.579^{+3.428}_{-1.706}$ / $2.293^{+3.047}_{-1.516}$\\
$014320.96+132429.7$ & 4010 & 5064.646 & 14.251 & 0 & 161474857.473 & -1.05$^{+4.84}_{-5.23}$ & -4.854$^{+22.287}_{-24.066}$ & $19.981^{+5.917}_{-4.677}$ / $8.035^{+4.343}_{-2.982}$ / $12.641^{+5.088}_{-3.768}$\\

\enddata
\tablenotetext{a}{The {\it Chandra} TSTART parameter indicates the time of the start of the observation in seconds since 1~Jan 1998.}
\tablenotetext{b}{The number of full-band counts is not exactly equal to the combination of soft- and hard-band counts due to band-dependent factors in aperture corrections and background estimation.}
\end{deluxetable}
\clearpage
\end{landscape}
\clearpage

\clearpage
\begin{landscape}
\clearpage
\begin{deluxetable}{lrrrrr}
\tabletypesize{\scriptsize}
\tablecolumns{6}
\tablewidth{0pc}
\tablecaption{Observations of Each Source\label{epochTable1}}
\tablehead{\colhead{SDSS Name} & \colhead{ObsId} & \colhead{Source Counts\tablenotemark{a}} & \colhead{BG Counts\tablenotemark{a}} & \colhead{HR} & \colhead{Approx. L$_{0.5-8}$\tablenotemark{b}} \\ \colhead{(J2000)} & \colhead{} & \colhead{Full/Soft/Hard Band} & \colhead{Full/Soft/Hard Band} & \colhead{} & \colhead{10$^{42}$ erg s$^{-1}$}}
\startdata
$000622.60-000424.4$ & 4096 & $254.269^{+19.407}_{-18.883}$ / $186.282^{+16.471}_{-15.695}$ / $71.523^{+11.461}_{-11.251}$ & $20.313^{+6.263}_{-3.777}$ / $6.633^{+4.287}_{-2.152}$ / $13.685^{+5.363}_{-2.833}$ &  & $3216.715^{+245.514}_{-238.886}$\\
$000622.60-000424.4$ & 5617 & $952.258^{+34.388}_{-33.437}$ / $625.993^{+28.266}_{-27.270}$ / $338.754^{+21.144}_{-20.239}$ & $7.969^{+4.337}_{-2.438}$ / $3.282^{+3.374}_{-1.542}$ / $4.774^{+3.714}_{-1.716}$ &  & $3819.410^{+137.927}_{-134.112}$\\
$004054.65-091526.7$ & 4885 & $5.352^{+4.245}_{-3.969}$ / $5.032^{+3.996}_{-3.624}$ / $0.378^{+2.730}_{-2.754}$ & $1.543^{+2.856}_{-0.964}$ / $0.780^{+2.591}_{-0.608}$ / $0.773^{+2.567}_{-0.548}$ &  & $4854.195^{+3850.118}_{-3600.038}$\\
$004054.65-091526.7$ & 904 & $41.771^{+14.770}_{-15.558}$ / $24.029^{+10.651}_{-11.071}$ / $18.773^{+11.109}_{-11.948}$ & $77.272^{+10.485}_{-7.605}$ / $34.743^{+7.515}_{-5.121}$ / $44.427^{+8.395}_{-5.362}$ &  & $8554.465^{+3024.906}_{-3186.192}$\\
$011513.16+002013.1$ & 3203 & $22.319^{+7.984}_{-7.885}$ / $16.675^{+6.914}_{-6.717}$ / $5.941^{+5.026}_{-4.970}$ & $11.657^{+4.900}_{-2.967}$ / $7.397^{+4.226}_{-2.377}$ / $4.420^{+3.620}_{-1.654}$ & -0.582$^{+0.287}_{-0.219}$ & $298.165^{+106.665}_{-105.334}$\\
$011513.16+002013.1$ & 3204 & $40.886^{+8.348}_{-7.572}$ / $34.795^{+7.786}_{-6.972}$ / $6.307^{+4.182}_{-3.725}$ & $1.401^{+2.830}_{-0.933}$ / $0.776^{+2.606}_{-0.623}$ / $0.623^{+2.491}_{-0.461}$ & -0.784$^{+0.106}_{-0.087}$ & $630.201^{+128.666}_{-116.718}$\\
$014219.01+132746.5$ & 1633 & $2.512^{+3.674}_{-3.517}$ / $-0.941^{+2.171}_{-2.460}$ / $3.615^{+3.723}_{-3.408}$ & $1.889^{+2.799}_{-1.126}$ / $0.941^{+2.460}_{-0.727}$ / $0.987^{+2.580}_{-0.695}$ &  & $2.654^{+3.881}_{-3.716}$\\
$014219.01+132746.5$ & 4010 & $1.385^{+5.219}_{-5.473}$ / $1.766^{+3.509}_{-3.427}$ / $-0.342^{+4.658}_{-5.051}$ & $8.621^{+4.374}_{-2.495}$ / $1.661^{+2.855}_{-1.031}$ / $7.254^{+4.228}_{-2.124}$ &  & $0.598^{+2.252}_{-2.361}$\\
$014320.96+132429.7$ & 1633 & $-2.480^{+4.338}_{-4.930}$ / $1.188^{+3.548}_{-3.579}$ / $-3.474^{+3.533}_{-4.297}$ & $7.173^{+4.375}_{-2.214}$ / $1.392^{+3.147}_{-0.914}$ / $5.767^{+4.020}_{-1.787}$ &  & $-295.758^{+517.292}_{-587.938}$\\
$014320.96+132429.7$ & 4010 & $-1.555^{+7.139}_{-7.709}$ / $4.407^{+4.634}_{-4.554}$ / $-5.980^{+6.153}_{-7.003}$ & $21.536^{+6.125}_{-3.992}$ / $3.628^{+3.441}_{-1.615}$ / $18.621^{+5.903}_{-3.459}$ &  & $-79.943^{+367.018}_{-396.323}$\\

\enddata
\tablenotetext{a}{The number of full-band counts is not exactly equal to the combination of soft- and hard-band counts due to band-dependent factors in aperture corrections and background estimation.}
\tablenotetext{b}{Calculated using a power law model with photon index $\Gamma = 2$.}
\end{deluxetable}
\clearpage
\end{landscape}
\clearpage

\begin{deluxetable}{lrrrr}
\tabletypesize{\scriptsize}
\tablecolumns{5}
\tablewidth{0pc}
\tablecaption{Simple Spectral Model for Bright Sources\label{brightFitTable}}
\tablehead{\colhead{SDSS Name} & \colhead{ObsId} & \colhead{Norm} & \colhead{$\Gamma$} & \colhead{Edge Depth ($\tau$)} \\ \colhead{(J2000)} & \colhead{} & \colhead{10$^{-5}$ ph keV$^{-1}$ cm$^{-2}$ s$^{-1}$ at 1~keV} & \colhead{} & \colhead{}}
\startdata
$094745.14+072520.6$ & 7265 & $6.26^{+0.435}_{-0.318}$ & $0.091^{+0.046}_{-0.047}$ & $3.431^{+0.469}_{-0.416}$\\
$094745.14+072520.6$ & 6842 & $6.37^{+0.349}_{-0.258}$ & $0.087^{+0.037}_{-0.037}$ & $3.146^{+0.357}_{-0.322}$\\
$095240.16+515249.9$ & 3195 & $4.95^{+0.146}_{-0.118}$ & $2.335^{+0.038}_{-0.038}$ & $0.000^{+0.193}_{-0.000}$\\
$095240.16+515249.9$ & 7706 & $4.63^{+1.07}_{-0.654}$ & $2.043^{+0.236}_{-0.187}$ & $1.823^{+1.513}_{-1.256}$\\
$095902.76+021906.3$ & 8009 & $5.24^{+0.208}_{-0.163}$ & $1.686^{+0.049}_{-0.048}$ & $0.000^{+0.340}_{-0.000}$\\
$095902.76+021906.3$ & 8015 & $5.86^{+0.446}_{-0.183}$ & $1.874^{+0.074}_{-0.050}$ & $0.021^{+0.367}_{-0.021}$\\
$095924.46+015954.3$ & 8019 & $2.8^{+0.253}_{-0.129}$ & $2.265^{+0.119}_{-0.080}$ & $0.572^{+1.144}_{-0.572}$\\
$095924.46+015954.3$ & 8020 & $2.68^{+0.145}_{-0.114}$ & $2.070^{+0.077}_{-0.073}$ & $0.000^{+0.946}_{-0.000}$\\
$095924.46+015954.3$ & 8025 & $3.67^{+0.209}_{-0.165}$ & $2.291^{+0.088}_{-0.082}$ & $0.000^{+0.199}_{-0.000}$\\
$095924.46+015954.3$ & 8026 & $2.98^{+0.265}_{-0.144}$ & $2.161^{+0.112}_{-0.079}$ & $0.880^{+1.185}_{-0.880}$\\
$105316.75+573550.8$ & 1683 & $5.42^{+0.843}_{-0.713}$ & $2.391^{+0.354}_{-0.222}$ & $1.837^{+3.163}_{-1.837}$\\
$105316.75+573550.8$ & 1684 & $3.52^{+1.16}_{-0.449}$ & $1.587^{+0.258}_{-0.188}$ & $0.663^{+1.326}_{-0.663}$\\
$105316.75+573550.8$ & 4936 & $3.86^{+0.0714}_{-0.0569}$ & $1.762^{+0.024}_{-0.025}$ & $0.000^{+0.073}_{-0.000}$\\
$114656.73+472755.6$ & 767 & $5.59^{+0.299}_{-0.237}$ & $2.068^{+0.083}_{-0.078}$ & $0.000^{+0.246}_{-0.000}$\\
$114656.73+472755.6$ & 768 & $5.79^{+0.252}_{-0.201}$ & $2.204^{+0.070}_{-0.067}$ & $0.000^{+0.166}_{-0.000}$\\
$114656.73+472755.6$ & 1971 & $6.42^{+0.29}_{-0.246}$ & $2.301^{+0.075}_{-0.072}$ & $0.000^{+0.151}_{-0.000}$\\
$022430.60-000038.8$ & 3181 & $5.54^{+0.327}_{-0.207}$ & $2.899^{+0.060}_{-0.058}$ & $0.188^{+0.237}_{-0.188}$\\
$022430.60-000038.8$ & 4987 & $5.91^{+0.097}_{-0.0787}$ & $2.586^{+0.024}_{-0.024}$ & $0.000^{+0.014}_{-0.000}$\\
$090900.43+105934.8$ & 924 & $44.1^{+1.39}_{-0.787}$ & $2.007^{+0.033}_{-0.026}$ & $0.692^{+0.126}_{-0.123}$\\
$090900.43+105934.8$ & 7699 & $43.5^{+2.47}_{-1.71}$ & $1.865^{+0.084}_{-0.060}$ & $0.216^{+0.322}_{-0.216}$\\
$091029.03+542719.0$ & 2452 & $7.96^{+0.241}_{-0.191}$ & $2.065^{+0.064}_{-0.041}$ & $0.043^{+0.327}_{-0.043}$\\
$091029.03+542719.0$ & 2227 & $9.64^{+0.201}_{-0.162}$ & $2.166^{+0.032}_{-0.030}$ & $0.000^{+0.028}_{-0.000}$\\
$100025.24+015852.0$ & 8012 & $13.3^{+0.91}_{-0.291}$ & $2.034^{+0.052}_{-0.036}$ & $0.121^{+0.237}_{-0.121}$\\
$100025.24+015852.0$ & 8011 & $7.13^{+0.285}_{-0.226}$ & $2.018^{+0.057}_{-0.054}$ & $0.000^{+0.072}_{-0.000}$\\
$100025.24+015852.0$ & 8017 & $6.81^{+0.276}_{-0.232}$ & $2.065^{+0.081}_{-0.055}$ & $0.263^{+0.372}_{-0.263}$\\
$100025.24+015852.0$ & 8018 & $5.42^{+0.206}_{-0.163}$ & $1.883^{+0.050}_{-0.048}$ & $0.000^{+0.092}_{-0.000}$\\
$104829.95+123428.0$ & 1587 & $2.99^{+0.191}_{-0.152}$ & $2.307^{+0.105}_{-0.098}$ & $0.000^{+0.206}_{-0.000}$\\
$104829.95+123428.0$ & 7073 & $2.24^{+0.11}_{-0.0877}$ & $2.062^{+0.073}_{-0.069}$ & $0.000^{+0.219}_{-0.000}$\\
$104829.95+123428.0$ & 7076 & $3.01^{+0.374}_{-0.114}$ & $2.011^{+0.099}_{-0.064}$ & $0.101^{+0.509}_{-0.101}$\\
$122511.91+125153.6$ & 803 & $5.16^{+0.141}_{-0.113}$ & $2.168^{+0.039}_{-0.038}$ & $0.000^{+0.084}_{-0.000}$\\
$122511.91+125153.6$ & 5908 & $4.29^{+0.117}_{-0.093}$ & $1.877^{+0.038}_{-0.038}$ & $0.000^{+0.066}_{-0.000}$\\
$122511.91+125153.6$ & 6131 & $4.21^{+0.121}_{-0.0968}$ & $1.846^{+0.040}_{-0.041}$ & $0.000^{+0.122}_{-0.000}$\\
$123759.56+621102.3$ & 580 & $2.02^{+0.106}_{-0.0837}$ & $1.538^{+0.067}_{-0.065}$ & $0.000^{+0.588}_{-0.000}$\\
$123759.56+621102.3$ & 967 & $2.18^{+0.103}_{-0.0819}$ & $1.744^{+0.065}_{-0.063}$ & $0.000^{+0.548}_{-0.000}$\\
$123759.56+621102.3$ & 966 & $2.34^{+0.165}_{-0.0885}$ & $1.776^{+0.086}_{-0.062}$ & $0.205^{+0.551}_{-0.205}$\\
$123759.56+621102.3$ & 2386 & $3.1^{+0.337}_{-0.261}$ & $2.088^{+0.159}_{-0.149}$ & $0.000^{+1.021}_{-0.000}$\\
$123759.56+621102.3$ & 1671 & $3.42^{+0.083}_{-0.0667}$ & $2.059^{+0.036}_{-0.034}$ & $0.000^{+0.022}_{-0.000}$\\
$123759.56+621102.3$ & 2344 & $3.13^{+0.109}_{-0.0871}$ & $2.094^{+0.053}_{-0.050}$ & $0.000^{+0.168}_{-0.000}$\\
$123759.56+621102.3$ & 3293 & $3.45^{+0.104}_{-0.0823}$ & $1.863^{+0.041}_{-0.039}$ & $0.000^{+0.141}_{-0.000}$\\
$123759.56+621102.3$ & 3388 & $4.16^{+0.211}_{-0.167}$ & $1.804^{+0.067}_{-0.065}$ & $0.000^{+0.155}_{-0.000}$\\
$123759.56+621102.3$ & 3408 & $3.91^{+0.175}_{-0.138}$ & $1.832^{+0.060}_{-0.057}$ & $0.000^{+0.598}_{-0.000}$\\
$123759.56+621102.3$ & 3389 & $3.25^{+0.106}_{-0.0843}$ & $1.769^{+0.043}_{-0.042}$ & $0.000^{+0.137}_{-0.000}$\\
$123800.91+621336.0$ & 580 & $2.4^{+0.109}_{-0.0874}$ & $2.318^{+0.076}_{-0.070}$ & $0.000^{+0.076}_{-0.000}$\\
$123800.91+621336.0$ & 967 & $2.14^{+0.152}_{-0.0748}$ & $2.250^{+0.093}_{-0.069}$ & $0.000^{+0.168}_{-0.000}$\\
$123800.91+621336.0$ & 966 & $1.42^{+0.0829}_{-0.0657}$ & $1.953^{+0.085}_{-0.081}$ & $0.000^{+0.148}_{-0.000}$\\
$123800.91+621336.0$ & 957 & $1.29^{+0.174}_{-0.0684}$ & $2.387^{+0.157}_{-0.107}$ & $0.038^{+0.585}_{-0.038}$\\
$123800.91+621336.0$ & 2386 & $3.81^{+0.796}_{-0.535}$ & $2.559^{+0.217}_{-0.194}$ & $2.325^{+1.405}_{-1.219}$\\
$123800.91+621336.0$ & 1671 & $2.86^{+0.076}_{-0.061}$ & $2.534^{+0.047}_{-0.043}$ & $0.000^{+0.071}_{-0.000}$\\
$123800.91+621336.0$ & 2344 & $4.07^{+0.295}_{-0.11}$ & $2.320^{+0.072}_{-0.049}$ & $0.282^{+0.312}_{-0.141}$\\
$123800.91+621336.0$ & 3293 & $1.17^{+0.0616}_{-0.0553}$ & $2.087^{+0.114}_{-0.076}$ & $0.587^{+0.545}_{-0.294}$\\
$123800.91+621336.0$ & 3388 & $2.12^{+0.143}_{-0.108}$ & $2.419^{+0.109}_{-0.099}$ & $0.000^{+0.162}_{-0.000}$\\
$123800.91+621336.0$ & 3408 & $2.86^{+0.137}_{-0.11}$ & $2.426^{+0.080}_{-0.074}$ & $0.000^{+0.079}_{-0.000}$\\
$123800.91+621336.0$ & 3389 & $3.24^{+0.1}_{-0.0805}$ & $2.588^{+0.055}_{-0.050}$ & $0.000^{+0.091}_{-0.000}$\\
$123800.91+621336.0$ & 3409 & $2.02^{+0.104}_{-0.0828}$ & $2.260^{+0.081}_{-0.075}$ & $0.000^{+0.181}_{-0.000}$\\
$123800.91+621336.0$ & 3294 & $2.04^{+0.0722}_{-0.0579}$ & $2.355^{+0.059}_{-0.054}$ & $0.000^{+0.092}_{-0.000}$\\
$123800.91+621336.0$ & 3390 & $1.73^{+0.0682}_{-0.11}$ & $2.332^{+0.084}_{-0.130}$ & $0.173^{+0.286}_{-0.173}$\\
$123800.91+621336.0$ & 3391 & $2.07^{+0.106}_{-0.0631}$ & $2.331^{+0.092}_{-0.055}$ & $0.198^{+0.396}_{-0.198}$\\
$125849.83-014303.3$ & 4178 & $8.12^{+0.196}_{-0.157}$ & $2.271^{+0.033}_{-0.033}$ & $0.000^{+0.095}_{-0.000}$\\
$125849.83-014303.3$ & 6356 & $11.2^{+0.352}_{-0.282}$ & $2.162^{+0.046}_{-0.044}$ & $0.000^{+0.169}_{-0.000}$\\
$125849.83-014303.3$ & 6357 & $11.3^{+0.377}_{-0.299}$ & $2.201^{+0.049}_{-0.046}$ & $0.000^{+0.170}_{-0.000}$\\
$125849.83-014303.3$ & 6358 & $10.9^{+0.374}_{-0.298}$ & $2.142^{+0.050}_{-0.047}$ & $0.000^{+0.136}_{-0.000}$\\
$125849.83-014303.3$ & 5823 & $5.12^{+0.212}_{-0.168}$ & $1.956^{+0.057}_{-0.055}$ & $0.000^{+0.086}_{-0.000}$\\
$125849.83-014303.3$ & 5822 & $5.42^{+0.323}_{-0.254}$ & $1.934^{+0.081}_{-0.078}$ & $0.000^{+0.122}_{-0.000}$\\
$125849.83-014303.3$ & 7242 & $4.82^{+0.213}_{-0.17}$ & $2.004^{+0.062}_{-0.059}$ & $0.000^{+0.546}_{-0.000}$\\
$125849.83-014303.3$ & 7691 & $4.48^{+1.13}_{-0.446}$ & $1.747^{+0.222}_{-0.154}$ & $0.131^{+2.030}_{-0.131}$\\
$142052.43+525622.4$ & 5845 & $3.88^{+0.184}_{-0.146}$ & $2.245^{+0.072}_{-0.067}$ & $0.000^{+0.146}_{-0.000}$\\
$142052.43+525622.4$ & 5846 & $5.37^{+0.197}_{-0.172}$ & $2.333^{+0.060}_{-0.059}$ & $0.000^{+0.182}_{-0.000}$\\
$142052.43+525622.4$ & 6214 & $4.08^{+0.189}_{-0.148}$ & $2.198^{+0.070}_{-0.065}$ & $0.000^{+0.195}_{-0.000}$\\
$142052.43+525622.4$ & 6215 & $3.77^{+0.177}_{-0.14}$ & $2.279^{+0.073}_{-0.067}$ & $0.000^{+0.436}_{-0.000}$\\
$142052.43+525622.4$ & 9450 & $3.31^{+0.232}_{-0.181}$ & $2.061^{+0.100}_{-0.093}$ & $0.000^{+0.195}_{-0.000}$\\
$142052.43+525622.4$ & 9451 & $3.17^{+0.244}_{-0.19}$ & $1.936^{+0.104}_{-0.099}$ & $0.000^{+0.151}_{-0.000}$\\
$142052.43+525622.4$ & 9725 & $4.43^{+0.246}_{-0.194}$ & $2.316^{+0.086}_{-0.080}$ & $0.000^{+0.113}_{-0.000}$\\
$142052.43+525622.4$ & 9843 & $3.67^{+0.571}_{-0.286}$ & $2.245^{+0.198}_{-0.136}$ & $0.250^{+1.000}_{-0.250}$\\
$142052.43+525622.4$ & 9842 & $4.35^{+0.248}_{-0.195}$ & $2.134^{+0.083}_{-0.078}$ & $0.000^{+0.208}_{-0.000}$\\
$142052.43+525622.4$ & 9844 & $3.84^{+0.291}_{-0.227}$ & $2.172^{+0.112}_{-0.104}$ & $0.000^{+0.283}_{-0.000}$\\
$142052.43+525622.4$ & 9866 & $4.23^{+0.281}_{-0.22}$ & $2.060^{+0.094}_{-0.089}$ & $0.000^{+0.267}_{-0.000}$\\
$142052.43+525622.4$ & 9726 & $4.74^{+0.237}_{-0.188}$ & $2.193^{+0.076}_{-0.071}$ & $0.000^{+0.085}_{-0.000}$\\
$142052.43+525622.4$ & 9863 & $5.43^{+0.341}_{-0.269}$ & $2.309^{+0.097}_{-0.091}$ & $0.000^{+0.178}_{-0.000}$\\
$142052.43+525622.4$ & 9870 & $5.91^{+0.512}_{-0.399}$ & $2.029^{+0.121}_{-0.115}$ & $0.000^{+0.182}_{-0.000}$\\
$142052.43+525622.4$ & 9873 & $6.09^{+0.3}_{-0.241}$ & $2.229^{+0.075}_{-0.071}$ & $0.000^{+0.150}_{-0.000}$\\
$142052.43+525622.4$ & 9721 & $5.83^{+0.398}_{-0.354}$ & $2.358^{+0.159}_{-0.105}$ & $0.419^{+0.781}_{-0.419}$\\
$142052.43+525622.4$ & 9722 & $6.66^{+0.424}_{-0.326}$ & $2.161^{+0.125}_{-0.084}$ & $0.164^{+0.722}_{-0.164}$\\
$142052.43+525622.4$ & 9453 & $4.42^{+0.351}_{-0.18}$ & $2.067^{+0.095}_{-0.066}$ & $0.384^{+0.513}_{-0.384}$\\
$142052.43+525622.4$ & 9720 & $3.96^{+0.576}_{-0.206}$ & $2.203^{+0.139}_{-0.091}$ & $0.150^{+0.766}_{-0.150}$\\
$142052.43+525622.4$ & 9723 & $4.58^{+0.434}_{-0.218}$ & $2.116^{+0.112}_{-0.076}$ & $0.575^{+0.656}_{-0.575}$\\
$142052.43+525622.4$ & 9876 & $4.1^{+0.229}_{-0.185}$ & $2.166^{+0.095}_{-0.079}$ & $0.000^{+0.257}_{-0.000}$\\
$142052.43+525622.4$ & 9875 & $6.1^{+0.568}_{-0.292}$ & $2.110^{+0.110}_{-0.077}$ & $0.552^{+0.630}_{-0.552}$\\

\enddata
\end{deluxetable}
\clearpage

\clearpage

\clearpage

\clearpage

\clearpage

\begin{figure} [ht]
  \begin{center}
      \includegraphics[width=5in, angle=-90]{f5.ps}
      \caption{\label{varArchPlotVsSeyfertsFig}Black hole masses, $M_{BH}$, as a function of redshift.  Open circles represent Seyfert AGN and quasars from \citet{pfgkmmmnopvw04}, while filled squares represent masses taken from \citet{sgsrs08} for quasars in our sample.  $M_{BH}$ was estimated by \citet{sgsrs08} using the $H_{\beta}$ line for quasars at $z \le 0.7$, \ion{Mg}{2}$\lambda$2800\AA\ for $0.7 < z < 1.9$, and \ion{C}{4}$\lambda$1549\AA\ for $z > 1.9$.  Vertical, dotted lines are plotted at $z = 0.7, 1.9$ to distinguish the subsamples.  Some well-known local AGN are identified with star symbols on the plot.}
   \end{center}
\end{figure}
\clearpage

\begin{figure} [ht]
  \begin{center}
      \includegraphics[width=5in, angle=-90]{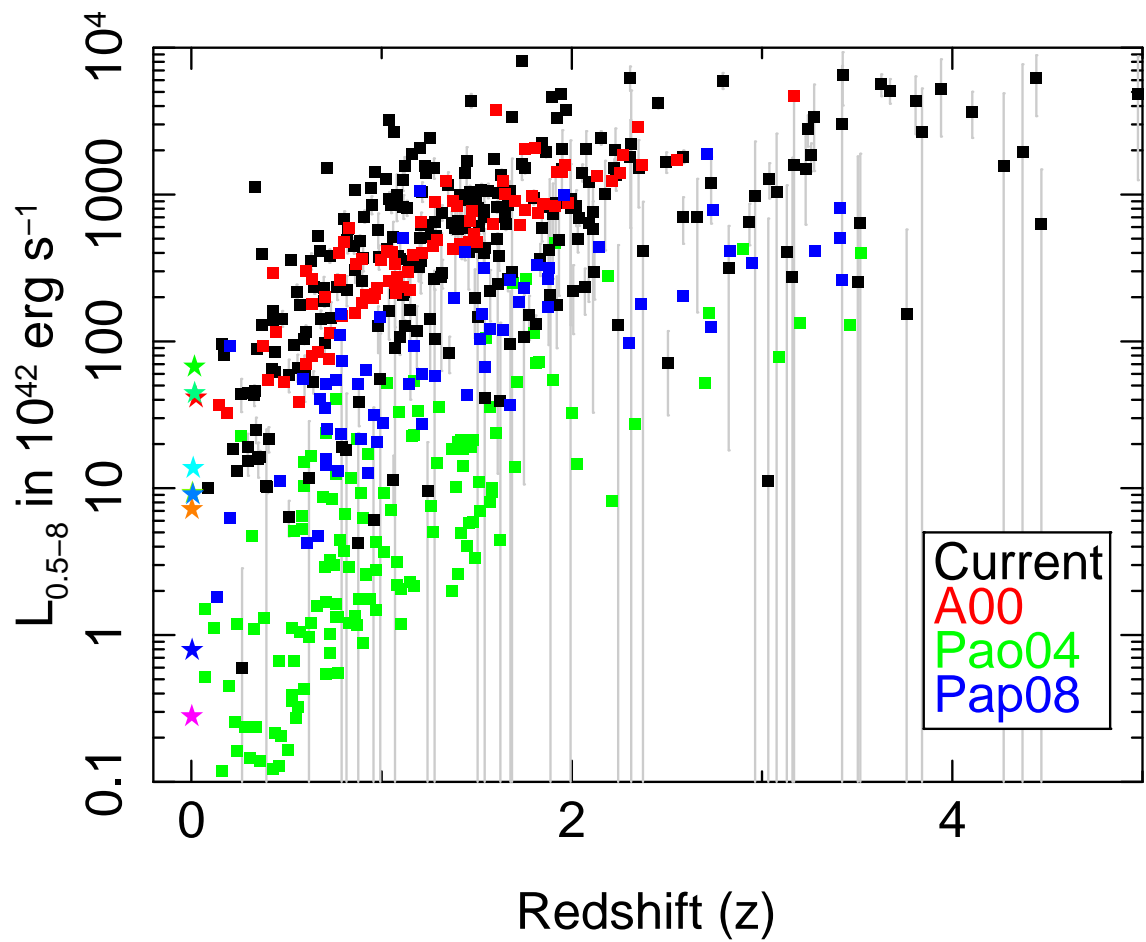}
      \caption{\label{sampleVsOldFig}Median luminosities (estimated as described in \S\ref{chandraRedSec}) and redshifts for sources in our full sample (black squares) compared to the A00 (red squares), Pao04 (green squares), and Pap04 (blue squares) samples.  The $y$-axis is logarithmic, and long error bars stretching downward off the plot indicate sources that were not detected at high significance.  Well-known, local AGN (at low redshifts) are plotted as stars using the same colors as in Figure~\ref{varArchPlotVsSeyfertsFig}.}
   \end{center}
\end{figure}
\clearpage

\begin{figure} [ht]
  \begin{center}
      \includegraphics[width=5in, angle=-90]{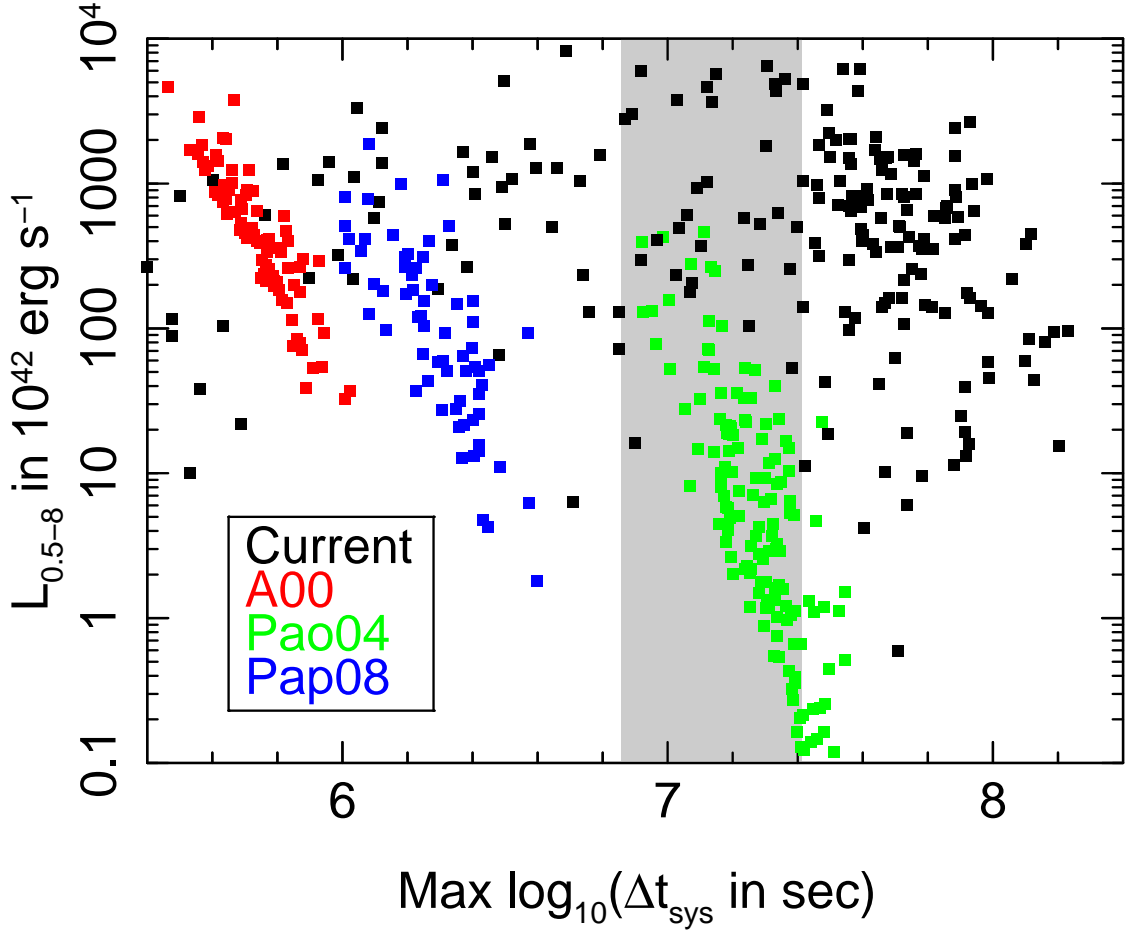}
      \caption{\label{sampledtVsOldFig}Similar to Figure~\ref{sampleVsOldFig}, but showing luminosity as a function of the maximum {\it rest}-frame time between epochs for each source.  We have omitted error bars on luminosity (shown in Figure~\ref{sampleVsOldFig}) for clarity.  For the A00, Pao04, and Pap08 samples, we determined max($\Delta t_{sys}$) using {\it observed}-frame times of 14~d, 435~d, and 52~d to span their observation periods; in practice, they may have binned individual sources to a smaller time scale resolution.  The gray shaded region shows the region of time between power spectrum break time scales for AGN with ($M_{BH}$, bolometric luminosity) = ($10^8 M_{\astrosun}$, $10^{44}$~erg~s$^{-1}$) and ($10^{9.5} M_{\astrosun}$, $10^{47}$~erg~s$^{-1}$), estimated using the relation given in \citet{mkkuf06}.}
   \end{center}
\end{figure}
\clearpage

\clearpage

\clearpage

\clearpage

\clearpage

\clearpage

\clearpage

\clearpage

\clearpage

\clearpage

\clearpage

\clearpage

\clearpage

\clearpage

\clearpage

\clearpage

\clearpage

\end{document}